\documentclass[onecolumn,prd,floats,aps,amsmath,amssymb,nofootinbib, superscriptaddress,preprintnumbers,10pt]{revtex4-2}
\usepackage{comment}
\usepackage[utf8]{inputenc}
\usepackage{mathtools}
\usepackage{bbm}
\usepackage{cancel} 
\usepackage{bm}%
\usepackage{blindtext}
\usepackage{xcolor}
\usepackage{hyperref}%
\usepackage{graphicx}
\usepackage{amsthm}
\usepackage{bm}
\usepackage{verbatim}
\usepackage{amsmath}
\usepackage{siunitx}
\usepackage{amssymb}
\usepackage{slashed}
\usepackage{tikz}
\usetikzlibrary{positioning, shapes, arrows.meta, decorations.pathmorphing}

\newcommand{\Bern}{Institute for Theoretical Physics, Albert Einstein Center for Fundamental Physics,\\University of Bern, Sidlerstrasse 5, CH-3012 Bern, Switzerland}
\newcommand{\hiskp}{HISKP (Theory), Rheinische Friedrich-Wilhelms-Universit\"at Bonn,\\Nussallee 14-16, 53115 Bonn, Germany}
\newcommand{\hpca}{High Performance Computing and Analytics Lab, Rheinische Friedrich-Wilhelms-Universit\"at Bonn,\\ Friedrich-Hirzebruch-Allee 8, 53115 Bonn, Germany}
\newcommand{\CyprusU}{Department of Physics, University of Cyprus, 20537 Nicosia, Cyprus}
\newcommand{\CyprusI}{Computation-based Science and Technology Research Center, The Cyprus Institute,\\20 Konstantinou Kavafi Street, 2121 Nicosia, Cyprus}

\newcommand{\Romadue}{Dipartimento di Fisica and INFN, Universit\`a di Roma ``Tor Vergata",\\Via della Ricerca Scientifica 1, I-00133 Roma, Italy}
\newcommand{\Romatre}{Dipartimento di Matematica e Fisica, Universit\`a Roma Tre and INFN, Sezione di Roma Tre,\\Via della Vasca Navale 84, I-00146 Rome, Italy}
\newcommand{\RomatreINFN}{Istituto Nazionale di Fisica Nucleare, Sezione di Roma Tre,\\Via della Vasca Navale 84, I-00146 Rome, Italy}

\newcommand{\CERN}{Department of Theoretical Physics, European Organization for Nuclear Research, CERN, CH-1211 Genève 23, Switzerland}

\newcommand{\be}{\begin{equation}}
\newcommand{\ee}{\end{equation}}
\newcommand{\bea}{\begin{eqnarray}}
\newcommand{\eea}{\end{eqnarray}}

\begin{document}

\preprint{CERN-TH-2024-197}

\setlength\abovedisplayskip{14pt}
\setlength\belowdisplayskip{14pt}

\setlength{\parskip}{14pt}
\setlength{\parindent}{0pt}

\title{Strange and charm quark contributions
to the muon anomalous magnetic moment\texorpdfstring{\\}{} in lattice QCD with twisted-mass fermions}

\author{C.~Alexandrou}\affiliation{\CyprusU}\affiliation{\CyprusI}
\author{S.~Bacchio}\affiliation{\CyprusI}
\author{A.~De~Santis}\affiliation{\Romadue}
\author{A.~Evangelista}\affiliation{\Romadue}
\author{J.~Finkenrath}\affiliation{\CERN}
\author{R.~Frezzotti}\affiliation{\Romadue} 
\author{G.~Gagliardi}\affiliation{\Romatre}
\author{M.~Garofalo}\affiliation{\hiskp}
\author{N.~Kalntis}\affiliation{\Bern}
\author{B.~Kostrzewa}\affiliation{\hpca}
\author{V.~Lubicz}\affiliation{\Romatre}
\author{F.~Pittler}\affiliation{\CyprusI}
\author{S.~Romiti}\affiliation{\Bern}
\author{F.~Sanfilippo}\affiliation{\RomatreINFN}
\author{S.~Simula}\affiliation{\RomatreINFN}
\author{N.~Tantalo}\affiliation{\Romadue}
\author{C.~Urbach}\affiliation{\hiskp}
\author{U.~Wenger}\affiliation{\Bern}

\begin{abstract}
\vspace{0.05cm}
\centerline{\includegraphics[height=4.7cm]{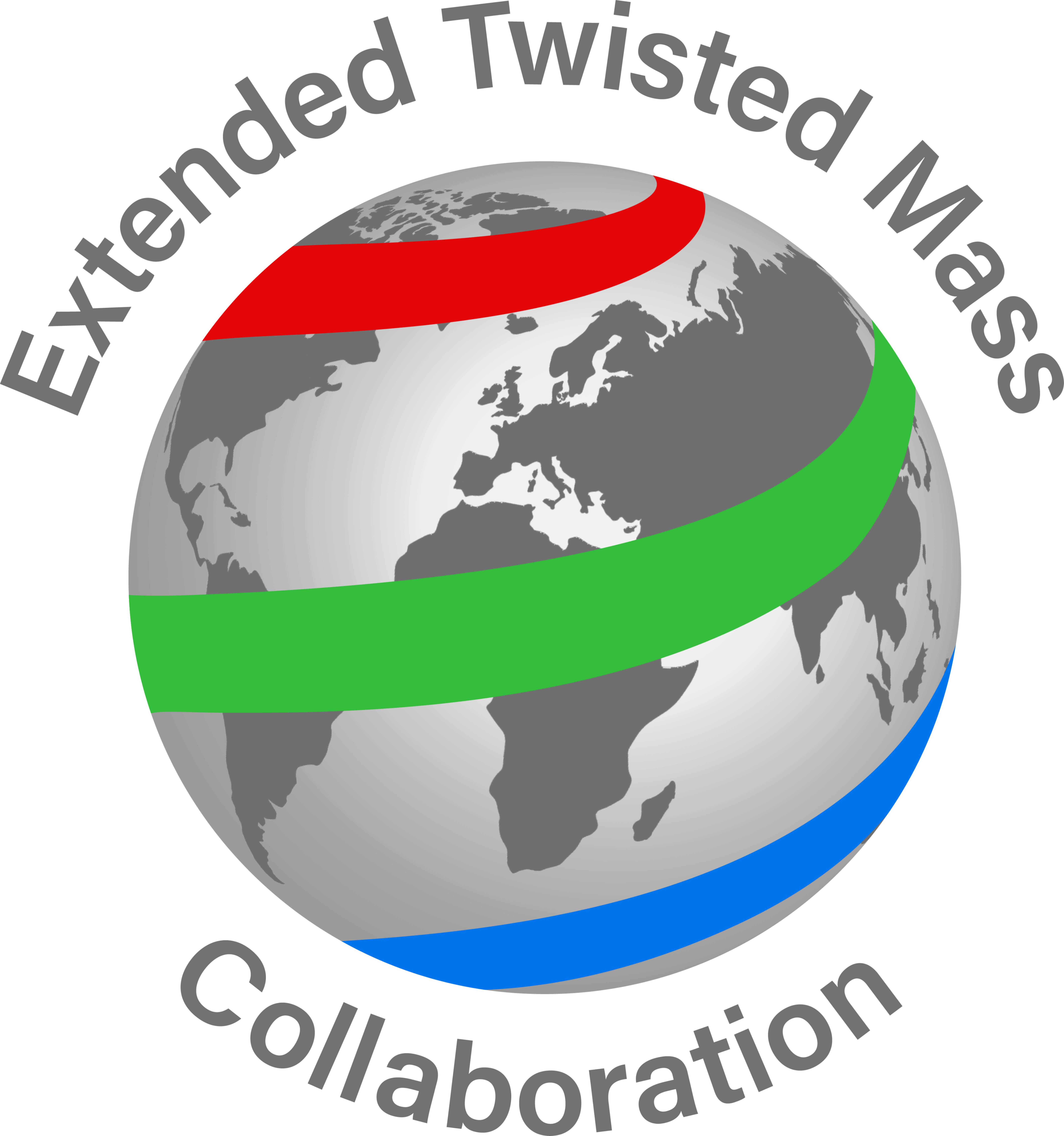}}
\vspace{0.7cm}
We present a lattice calculation of the Hadronic Vacuum Polarization (HVP) contribution of the strange and charm quarks to the anomalous magnetic moment of the muon in isospin symmetric QCD. We employ the gauge configurations generated by the Extended Twisted Mass Collaboration (ETMC) with $N_f = 2 + 1 + 1$ flavors of Wilson-clover twisted-mass quarks at five lattice spacings and
at values of the quark mass parameters that are close and/or include the isospin symmetric QCD point of interest. After computing the small corrections necessary to precisely match this point, and carrying out an extrapolation to the continuum limit based on the data at lattice spacings $a \simeq 0.049, 0.057, 0.068, 0.080$~fm and spatial lattice sizes up to $L \simeq 7.6$~fm, we obtain 
$a_\mu^{\rm HVP}(s) = (53.57 \pm 0.63) \times 10^{-10}$ and 
$a_\mu^{\rm HVP}(c) = (14.56 \pm 0.13) \times 10^{-10}$, for the quark-connected strange and charm contributions, respectively.
Our findings agree well with the corresponding results by other lattice groups.

\end{abstract}

\maketitle

\section{Introduction}
\label{sec:introduction}

The polarization of the vacuum induced by fluctuations of a virtual photon into quarks and gluons, known as the Hadronic Vacuum Polarization (HVP), has recently received a lot of attention and interest due to its importance in the Standard Model (SM) predictions of the anomalous magnetic moment of the muon $a_\mu$.
This quantity is currently investigated at the Fermi National Accelerator Laboratory (FNAL)\,\cite{Muong-2:2021ojo} and at a forthcoming experiment at J-PARC (E34)\,\cite{Abe:2019thb}. The Fermilab Muon $g-2$ experiment (E989), has published the results of the analysis of the Run-1 data collected in 2018\,\cite{Muong-2:2021ojo, Muong-2:2021ovs, Muong-2:2021xzz, Muong-2:2021vma},
where a remarkably good agreement with the previous E821 measurement at BNL\,\cite{Muong-2:2006rrc} is found. More recently, also the results of the analysis of the Run-2 and
Run-3 data, collected in 2019 and 2020, have been published~\cite{Muong-2:2023cdq}, with statistics increased by more than a factor of four and systematic errors reduced by more than a factor of two.

The current experimental world average\,\cite{Muong-2:2023cdq} is $a_{\mu}^\mathrm{exp} = 116\,592\,059 (22) \times 10^{-11}$ with a relative uncertainty of $0.19$ ppm. The ongoing analysis of
the remaining data from three additional years of data collection 
by the Fermilab Muon $g-2$ Collaboration is expected to lead to another factor of two of improvement in statistical precision, while a completely independent cross-check and possibly a further reduction of the total error will come from the forthcoming experiment planned at J-PARC.

From the theoretical side, the dominant source of uncertainty in the determination of $a_{\mu}$ comes from the HVP term at leading order in the electromagnetic (e.m.) coupling, $a_{\mu}^{\rm HVP}$ (see Ref.\,\cite{Aoyama:2020ynm}). 
Presently, there are two approaches for obtaining precise predictions of the HVP contribution. 
The first one makes use of the experimental data on the process $e^+ e^- \to \mbox{hadrons}$, while the other one is represented by numerical simulations of QCD and QCD+QED on the lattice.

The data-driven determination of $a_{\mu}^{\rm HVP}$ quotes~\cite{Davier:2019can, Keshavarzi:2019abf, Aoyama:2020ynm} a precision of $\simeq 0.6 \%$ and it corresponds to a SM prediction for $a_\mu$ that is found\,\cite{Muong-2:2023cdq} to differ by $\simeq 5.0 \sigma$ from the current experimental world average $a_{\mu}^\mathrm{exp}$. However, a recent determination of the cross section $e^+ e^- \to \pi^+ \pi^- (\gamma)$, carried out by the CMD-3 Collaboration\,\cite{CMD-3:2023alj}, shows important tensions with previous measurements, including the one made by the same Collaboration\,\cite{CMD-2:2006gxt} and, if correct, would make nearly negligible the tension with $a_{\mu}^\mathrm{exp}$. Whether radiative corrections can provide an explanation for such tensions in the data-driven determination of $a_{\mu}^{\rm HVP}$ is presently under active investigation (see, e.g., Refs.\,\cite{Abbiendi:2022liz, BaBar:2023xiy, Davier:2023fpl} and also Ref.~\cite{ExtendedTwistedMassCollaborationETMC:2022sta} for a first-principle isoQCD lattice calculation of the $R$-ratio smeared in gaussian energy bins). 

In recent years, impressive progress has been made by the lattice QCD community that enables the evaluation of  $a_{\mu}^{\rm HVP}$ with increasing precision, reaching the goal of a few permille accuracy. A breakthrough concerning the precision achieved came from the lattice calculation performed by the BMW Collaboration in 2020\,\cite{Borsanyi:2020mff}, corresponding to a relative uncertainty of $0.8\%$. This result has been recently updated by the same collaboration\,\cite{Boccaletti:2024guq}, by combining a refinement of their 2020 lattice computation with an experimental data-driven input for the very low energy tail of the $R$-ratio of the process $e^+e^- \to $~hadrons. In this hybrid approach, assuming that the $e^+e^-$ experimental data of input have controlled systematics and are not affected by contributions of physics beyond the SM, they reach a precision of $\simeq 0.5\%$ and obtain a phenomenological prediction for $a_\mu^\mathrm{HVP}$ that, once combined with the SM computations for the QED and electroweak contributions to $a_\mu$, yields a prediction for $a_\mu$ that deviates by only $\simeq 0.9\sigma$ from the current experimental value $a_\mu^\mathrm{exp}$. 
Moreover, very recently, the CLS/Mainz collaboration presented~\cite{Djukanovic:2024cmq} its lattice prediction for $a_\mu^\mathrm{HVP}$ that has an accuracy of $\simeq 1\%$ ($\simeq 0.8\%$ for the isoQCD contribution) and that, once combined with the SM computations for the QED and electroweak contributions, is fully compatible within errors with $a_\mu^\mathrm{exp}$. The CLS/Mainz result exhibits a slightly larger central value than $a_\mu^\mathrm{exp}$ and is in small tension with the BMW 2020 result. Finally, also the RBC/UKQCD collaboration presented~\cite{RBC:2024fic} a lattice result for the dominating light quark contribution to $a_\mu^\mathrm{HVP}$ that is in between the corresponding results of the CLS/Mainz and BMW collaborations and 
compatible with both of them within errors.

Furthermore, the BMW and the CLS/Mainz results are in strong tension (respectively $\simeq 2\sigma$~\cite{Borsanyi:2020mff}, $\simeq 4\sigma$~\cite{Boccaletti:2024guq} and $\simeq 4\sigma$~\cite{Djukanovic:2024cmq}) with the data-driven one of
Ref.~\cite{Aoyama:2020ynm} (i.e. without including the recent CMD3 data). Since 2020, i.e.\ after the appearance of Ref.~\cite{Borsanyi:2020mff}, this issue 
has triggered a lot of investigations of the so-called window contributions to $a_\mu^\mathrm{HVP}$, which were first introduced by the RBC/UKQCD collaboration in Ref.\,\cite{RBC:2018dos}. Such quantities, obtained by introducing suitable weight functions in the Euclidean time-momentum representation of $a_\mu^{\rm HVP}$, have proven to be quite useful since at short and intermediate time distances they can be predicted with high accuracy on the lattice. In particular, in the so-called intermediate window the disagreement between $e^+e^- \to $~hadrons cross-section data-driven results, as quoted by Ref.\,\cite{Aoyama:2020ynm} (i.e. without the recent CMD-3 result), and the lattice determinations has reached the remarkable level of $\simeq 4.5 \sigma$  already in 2022 (see Ref.\,\cite{ExtendedTwistedMass:2022jpw}). However, if one
employs instead the recent determination of the cross section of
$e^+ e^- \to \pi^+ \pi^- (\gamma)$ carried out by the CMD-3 collaboration\,\cite{CMD-3:2023alj}, the experimental results for the intermediate window contribution to $a_\mu^\mathrm{HVP}$ are in agreement with the SM lattice prediction.

In this work, we present a high-precision determination of the
quark-connected contributions to $a_\mu^\mathrm{HVP}$ due to strange and charm flavors that are obtained by the ETMC within the so-called isospin symmetric QCD (isoQCD), where isospin breaking effects, due to different up and down quark masses and quark electric charges, are neglected.

The analysis\footnote{We are currently working on the calculation of the dominating light-quarks quark-connected isoQCD contribution $a_\mu^{\rm HVP}(\ell)$ and, for that quantity, we have implemented a blinded analysis procedure whose details will be explained in a forthcoming publication. In the case of the contributions considered in this paper (as well as in the case of the isoQCD quark-disconnected and of the isospin breaking contributions on which we are also currently working) we have considered unnecessary to implement a blinded analysis.} is performed using the gauge configurations generated by ETMC~\cite{Alexandrou:2018egz, ExtendedTwistedMass:2020tvp, ExtendedTwistedMass:2021qui, Finkenrath:2022eon} with $N_f = 2 + 1 + 1$ flavors of Wilson Clover twisted-mass sea quarks with masses tuned very 
close to the target isoQCD values, i.e. the ones corresponding to our scheme of choice for defining isoQCD which is the so-called Edinburgh/FLAG consensus~\cite{Edinburgh, FlavourLatticeAveragingGroupFLAG:2024oxs}. We have computed the small non-perturbative corrections needed to fine-tune the bare parameters to the target isoQCD values and, after applying these corrections to the simulated ensembles (listed in Table~\ref{tab:simudetails} in the Appendices), we obtain the corrected isoQCD physical-point ensembles listed in Table~\ref{tab:iso_EDI_FLAG}, with lattice spacings $a \simeq 0.049, 0.057, 0.068, 0.080$~fm. Using these isoQCD physical-point ensembles, taking into
account the tiny finite size effects (FSE) and adopting a proper 
Akaike-Information-Criterion (AIC)-based model average for the 
continuum limit extrapolation, the results that we obtain for the quark-connected strange and charm contributions to $a_\mu^\mathrm{HVP}$ are
\bea
    a_\mu^{\rm HVP}(s) & = & (53.57 \pm 0.63) \times 10^{-10} ~ , ~ \\ 
    a_\mu^{\rm HVP}(c) & = & (14.56 \pm 0.13) \times 10^{-10} ~ . ~ \nonumber
\eea
Here the quoted error is the total one resulting from purely statistical, mistuning corrections and continuum-limit extrapolation errors plus the tiny uncertainties related to FSE and to the finite
normalization factors $Z_{V.A}$. 

The results for partial short-distance  (SD) window, intermediate window (W) and long distance (LD) window contributions
are given in Section~III. Our current results for SD and W window contributions are nicely consistent with, and supersede in accuracy, those we published in Ref.\,\cite{ExtendedTwistedMass:2022jpw}. Compared to
that paper, we now employ one physical-point ensemble at finer ($a\simeq 0.049$~fm) lattice spacing which allowed us to exclude the ensembles at unphysical values of the pion mass and at the coarsest lattice spacing used in Ref.\,\cite{ExtendedTwistedMass:2022jpw}  and to achieve a significant reduction of the statistical and continuum extrapolation
errors. Furthermore, precisely because of this error reduction, we carried out a careful analysis of the small mistuning errors affecting our previous results, computed the required corrections and took into account the associated uncertainties, thus improving our control of the total errors.

The paper is organized as follows: In Section~\ref{sec:definitions}, we provide the relevant notations and definitions. 
In Section~\ref{sec:connected} we present our determinations of the strange- and charm-quark connected contributions to the vector correlator, including a detailed analysis of the continuum limit. In Section~\ref{sec:comparison}, we conclude by presenting a comparison
with other available lattice QCD calculations and an outlook.

Further technical information is given in the Appendices as follows:
In Appendix\,\ref{sec:simulations}, we give details about our lattice setup and the bare parameters we used in Monte Carlo simulations, as
a step ``zero'' in our definition of isoQCD. 
In Appendix\,\ref{sec:masses}, we detail the procedure we apply to 
fine-tune the bare parameters of our lattice action in order to implement the chosen definition of isoQCD at a level of accuracy comparable with the statistical errors affecting the hadronic inputs used for theory renormalization.  
In Appendix\,\ref{sec:mass_interpolation_amu} we discuss the evaluation of the strange and charm HVP ($a_\mu^{\rm HVP}(s)$ and $a_\mu^{\rm HVP}(c)$),  along with the corresponding SD, LD and W, contributions at the Edinburgh/FLAG isoQCD point determined in 
Appendix\,\ref{sec:masses}.
In Appendix\,\ref{sec:mistunings}, we present technical details about
our numerical estimate of the systematic effects on both the hadronic renormalization inputs ($F_\pi$, $M_\pi$, $M_K$, $M_{D_s}$) and $a_\mu^{\rm HVP}(s,c)$ themselves stemming from the small mistuning in the bare action parameters used in our Monte Carlo simulations. 
In Appendix\,\ref{sec:renormalization}, we collect the values of the scale-invariant renormalization constants (RCs) of the vector and axial-vector local quark currents, $Z_V$ and $Z_A$, employing the hadronic method of Ref.\,\cite{ExtendedTwistedMass:2022jpw}, which relies on Ward Identities (WIs) and unversality of renormalized matrix elements. Owing to a high statistics determination of the relevant correlators, we achieve a very precise determination of $Z_V$ and $Z_A$, as needed to guarantee a few permille level statistical accuracy of the quark-connected strange and charm  HVP terms at fixed lattice spacing.

\section{Time-momentum representation}
\label{sec:definitions}

Following our previous works\,\cite{Giusti:2017jof,Giusti:2018mdh,Giusti:2019xct, ExtendedTwistedMass:2022jpw}, we adopt the time momentum representation\,\cite{Bernecker:2011gh} and, in continuum notation, evaluate the HVP contribution to the muon anomalous magnetic moment $a_{\mu}^{\rm HVP}$ as
\be
    \label{eq:amu_HVP}
    a_{\mu}^{\rm{HVP}} = 2 \alpha_{em}^2 \int_0^\infty ~ dt \, t^2 \, K(m_\mu t) \,V(t) ~ , ~  
\ee
where $t$ is the Euclidean time and the kernel function $K(m_{\mu} t)$ is defined as\footnote{The leptonic kernel $K(z)$ is proportional to $z^2$ at small values of $z$ and it approaches $1$ as $z \to \infty$.}
\be
    \label{eq:kernel}
    K(z) = 2 \int_0^1 dy ( 1- y) \left[ 1 - j_0^2 \left(\frac{z}{2}\frac{y}{\sqrt{1 - y}} \right) \right]~,\qquad j_{0}(y) = \frac{\sin{(y)}}{y} ~ . ~
\ee
The Euclidean vector correlator $V(t)$ is defined as
 \be
     \label{eq:VV}
     V(t) \equiv \frac{1}{3} \sum_{i=1,2,3} \int d^3{x} ~ \langle J_i(\vec{x}, t) J_i^\dagger(0) \rangle  =
     - \frac{1}{3} \sum_{i=1,2,3} \int d^3{x} ~ \langle J_i(\vec{x}, t) J_i(0) \rangle \; ,
 \ee
with $J_\mu(x)$ being the e.m. current operator
 \be
      \label{eq:Jmu}
     J^\mu(x) \equiv \sum_{f = u, d, s, c, ...} J^\mu_f(x)\;,
     \qquad
     J^\mu_f(x)=q_{\mathrm{em},f} ~ \overline{\psi}_f(x) \gamma_\mu \psi_f(x)\;,
 \ee
 and $q_{\mathrm{em},f}$ the electric charge for the quark flavor $f$ (in units of the positron charge). 

The fermionic Wick contractions appearing in the right hand side (r.h.s.)~of Eq.\,(\ref{eq:VV}) give rise to two distinct topologies of Feynman diagrams, namely to the quark-connected and quark-disconnected contributions.
Connected contributions are flavor diagonal, while the disconnected ones have both diagonal and off-diagonal flavor components. 
In what follows we decompose $a_\mu^{\rm HVP}$ into the following contributions
\be
     \label{eq:amu_HVP_f}
     a_{\mu}^{\rm{HVP}}  = a_{\mu}^{\rm{HVP}}(\ell) + a_{\mu}^{\rm{HVP}}(s) + a_{\mu}^{\rm{HVP}}(c) + a_{\mu}^{\rm{HVP}}({\rm disc.}) + \ldots ~ , ~
\ee
where the first three terms correspond to the quark-connected contributions of mass degenerate up and down ($\ell$) quarks, and a strange ($s$) and a charm ($c$) quark, respectively, while the fourth term represents all quark-disconnected (flavour diagonal and off-diagonal) contributions\footnote{Following Ref.\,\cite{ExtendedTwistedMass:2022jpw}, the separation of quark connected and disconnected contributions to a given correlator can be expressed in terms of local correlators by formally introducing, when needed, a suitable number of extra valence flavours (having the same masses as the physical quarks) and the corresponding ghosts. The different flavor contributions to $a_\mu^{\rm HVP}$, appearing in Eq.\,(\ref{eq:amu_HVP_f}), can be separately extracted from local current-current vector correlators computed within the renormalizable mixed action lattice setup described in detail in Appendix A of Ref.\,\cite{ExtendedTwistedMass:2022jpw}, which is briefly
recalled also in Appendix A of this work.}. 
In Eq.\,(\ref{eq:amu_HVP_f}) the ellipses corresponds to subleading terms, 
namely the isospin breaking effects and the contributions of
quarks heavier than the charm in QCD + QED (i.e.\ the low energy
effective theory of the Standard Model). 

Following the analysis of the RBC/UKQCD collaboration\,\cite{RBC:2018dos}, each of the terms appearing in Eq.~(\ref{eq:amu_HVP_f}) can further be decomposed  by multiplying the integration kernel $K(m_\mu t)$ appearing in Eq.\,(\ref{eq:amu_HVP}) with suitably smoothed Heaviside step-functions, namely
\be
    \label{eq:amu_w}
    a_\mu^{\rm HVP, w} = 2 \alpha_{em}^2 \int_0^\infty ~ dt \, t^2 \, K(m_\mu t) \, \Theta^w(t)\,V(t) ~ \qquad w = \{\rm SD, W, LD \} ~ , ~ 
\ee
where the time-modulating functions $\Theta^w(t)$ are given by
\bea
      \label{eq:Mt_SD}
      \Theta^{\rm SD}(t) & \equiv & 1 -  \frac{1}{1 + e^{- 2 (t - t_0) / \Delta}} ~ , ~ \\[2mm]
      \label{eq:Mt_W}
      \Theta^{\rm W}(t) & \equiv & \frac{1}{1 + e^{- 2 (t - t_0) / \Delta}} -  \frac{1}{1 + e^{- 2 (t - t_1) / \Delta}} ~ , ~ \\[2mm]      
      \label{eq:Mt_LD}
      \Theta^{\rm LD}(t) & \equiv & \frac{1}{1 + e^{- 2 (t - t_1) / \Delta}} ~  ,
\eea
with the parameters $t_0, t_1, \Delta$ chosen\,\cite{RBC:2018dos} to be equal to
\be
    \label{eq:parms}
    t_0 = 0.4 ~ \mathrm{fm} ~ , ~ \qquad t_1 = 1 ~ {\rm fm} ~ , ~ \qquad \Delta = 0.15~{\rm fm} ~ . ~    
\ee
The resulting time-modulating functions $\Theta^{\rm SD, W, LD}(t)$ are shown, e.g., in Fig.\,1 of Ref.\,\cite{ExtendedTwistedMass:2022jpw}.

In this work, together with the full contributions $a_{\mu}^{\rm{HVP}}(s)$ and $a_{\mu}^{\rm{HVP}}(c)$, we also compute the three window observables, 
i.e.\ $a_\mu^{\rm HVP, w} $,~$\rm w = \{\rm SD, W, LD \}$, for the strange and charm (connected) HVP terms. Our results are presented and discussed in the next Section.

The light quark-connected contribution, $a_{\mu}^{\rm HVP}(\ell)$,
and all quark-disconnnected contributions, $a_{\mu}^{\rm{HVP}}({\rm disc.})$, which also include those arising from the product of two strange/charm e.m.\,currents,  will be given in a forthcoming paper devoted to the full HVP contribution
$a_{\mu}^{\rm{HVP}}$ within isoQCD, Ref.~\cite{Latt24_MG}. In a subsequent paper the leading isospin breaking effects  on $a_{\mu}^{\rm{HVP}}$
in QCD+QED with $u$, $d$, $s$ and $c$ active flavours will be included too~\cite{Latt24_AE}.

\section{ The connected contributions to \texorpdfstring{$a_{\mu}^{\rm HVP}(s)$}{amuHVPs} and \texorpdfstring{$a_{\mu}^{\rm HVP}(c)$}{amuHVPc}}
\label{sec:connected}

\begin{table}[t]
\begin{tabular}{lccccccc}
ensemble & $V/a^4$ & $a^{\rm iso}~[\rm fm]$ & $L~[\rm fm]$ &  $am_{\ell}^{\rm iso}$ & $am_{s}^{\rm iso}$ & $am_{c}^{\rm iso}$ & $am_\mathrm{cr}$  \\
\hline
\\
B64    & $64^3\times 128$ & ~ $0.07948(11)$ ~  & ~ 5.09 ~ &~ 0.0006669(28)  ~ & ~0.018267(53)~   &  ~0.23134(52) ~ & ~ -0.4138934(46)	~ \\
B96    & $96^3\times 192$    & ~ $0.07948(11)$ ~  & ~ 7.63 ~ & ~ 0.0006669(28)  ~ & ~0.018267(53)~   &  ~0.23134(52) 	~ & ~ -0.4138934(46)	~ \\[4pt]
C80    & $80^3\times 160$    & ~ $0.06819(14)$ ~   & 5.46 &~ 0.0005864(34) ~
   &  ~0.016053(67)~  &  ~0.19849(64)	~  & ~ -0.3964534(41)  ~ \\
C112    & $112^3\times 224$    & ~ $0.06819(14)$ ~   & 7.64 & ~ 0.0005864(34) ~ 
   &  ~0.016053(67)~  &  ~0.19849(64)	~  & ~ -0.3964534(41)  ~ \\[4pt]
D96    & $96^3\times 192$    & ~ $0.056850(90)$ ~  & 5.46 &~ 0.0004934(24) ~ 
   &  ~0.013559(39)~  &  ~0.16474(44)	~  & ~ -0.3761252(39) ~  \\[4pt]
E112    & $112^3\times 224$    & ~ $0.04892(11)$ ~  & 5.48 &~ 0.0004306(23) ~  &  ~0.011787(55)~  &   ~0.14154(54)	~ & ~  -0.3613136(75) ~  \\[8pt]
\hline
\end{tabular}
\caption{\small \it ETMC gauge ensembles used to compute $a_{\mu}^{\rm HVP}(s)$ and $a_{\mu}^{\rm HVP}(c)$. The values of the lattice spacing and of the bare quark masses are fine-tuned to match our target definition of isoQCD, the one corresponding to the Edinburgh/FLAG consensus\,\cite{Edinburgh, FlavourLatticeAveragingGroupFLAG:2024oxs}. This is done by starting from the simulated gauge ensembles, listed in Table~\ref{tab:simudetails}, and by taking into account the reweighting factors needed to correct the small mistunings of the simulated bare parameters (see Appendices~\ref{sec:simulations} and~\ref{sec:masses}).
\label{tab:iso_EDI_FLAG}}
\end{table}

In this section we present our numerical results for $a_{\mu}^{\rm HVP}(f)$, where $f=\{s,c\}$, and for the corresponding window quantities. The results presented here improve and supersede in accuracy the ones previously obtained in Ref.~\cite{ExtendedTwistedMass:2022jpw}. 

With respect to our previous calculation, in addition to the first five $N_f = 2 + 1 + 1$ isoQCD ensembles listed in Table~\ref{tab:iso_EDI_FLAG} (with lattice spacings $a\simeq \{0.079,0.068,0.057\}$~fm) we included an additional ensemble, the entry E112 in the same table, with the finest lattice spacing ($a\simeq 0.049$~fm) ever simulated by the ETMC. This allowed us to better control the continuum extrapolations and, consequently, to reduce the corresponding systematic errors. Moreover, to further improve the accuracy of our results, in this work we computed the corrections needed to fine-tune the bare parameters of our simulations in order to precisely match our target definition of isoQCD, the Edinburgh/FLAG consensus~\cite{Edinburgh, FlavourLatticeAveragingGroupFLAG:2024oxs}, corresponding to the following hadronic inputs
\begin{flalign}
\label{eq:iso_definition}
M_\pi^\mathrm{iso}=135.0~\mathrm{ MeV}\;,
\quad
M_K^\mathrm{iso}=494.6~\mathrm{ MeV}\;,
\quad
M_{D_s}^\mathrm{iso}=1967~\mathrm{ MeV}\;,
\quad
F_\pi^\mathrm{iso}=130.5~\mathrm{ MeV}\;.
\end{flalign}
The numerical procedure that we used to compute these corrections, on both the input observables and on the target quantities $a_{\mu}^{\rm HVP}(f)$, is described in the Appedices~\ref{sec:simulations},~\ref{sec:masses},~\ref{sec:mass_interpolation_amu} and~\ref{sec:mistunings}  to which we refer for all the technical details of our lattice setup and of our calculation. Here below we present our results by focusing on the main steps of the analysis, i.e.\ the extraction of the observables from the lattice correlators, the estimate of FSE and the continuum extrapolations.

\begin{figure}
    \centering
    \includegraphics[width=0.45\linewidth]{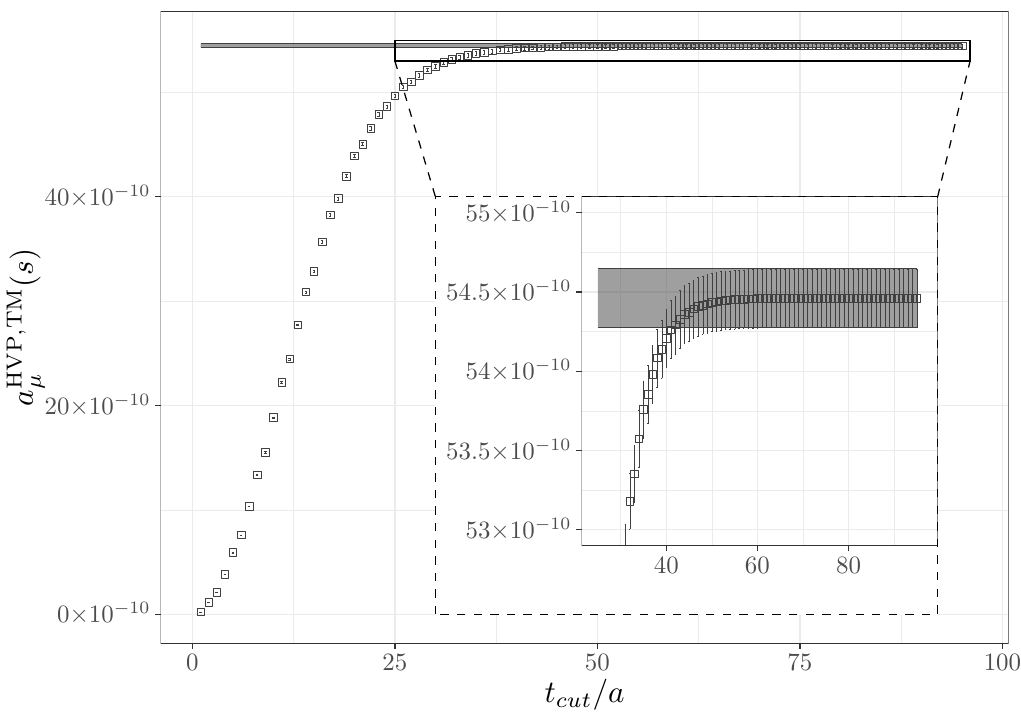}
    \includegraphics[width=0.45\linewidth]{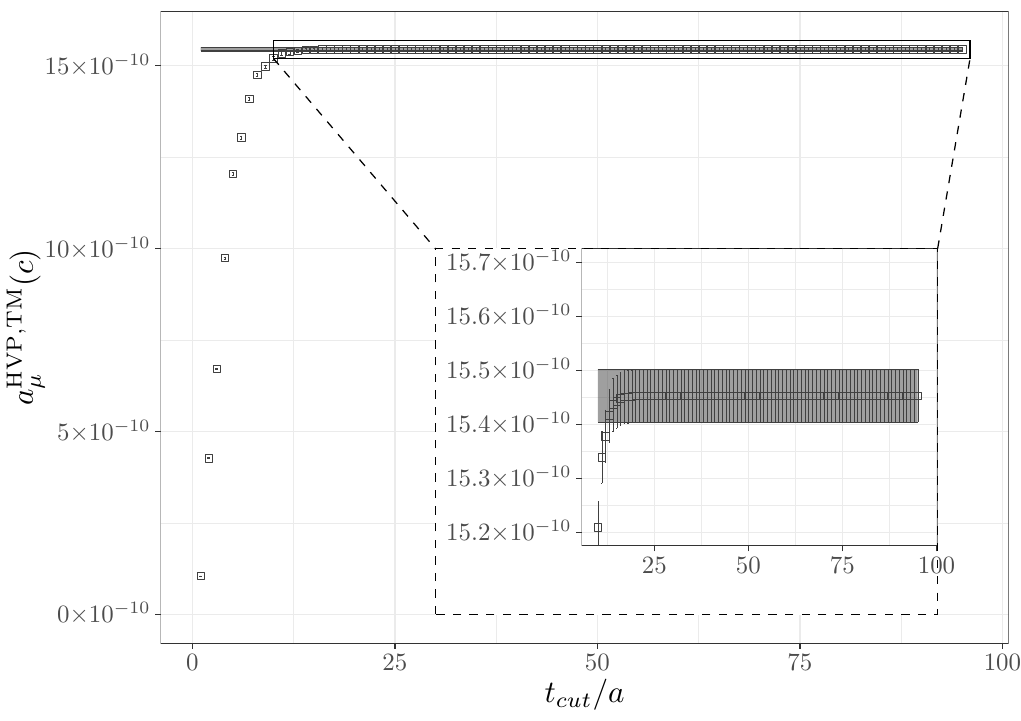}
    \caption{\small \it The left-panel shows $a_\mu^\mathrm{HVP}(s)$ while the right-panel shows $a_\mu^\mathrm{HVP}(c)$ as functions of $t_\mathrm{cut}$. The data correspond to the ensemble D96, to the TM regularization and to the smallest simulated values of the valence strange and charm masses. A detailed study of the dependence upon the valence and see quark masses is presented in the Appendices~\ref{sec:mass_interpolation_amu} and~\ref{sec:mistunings}. We observe that the results are independent of $t_\mathrm{cut}/a$ within the statistical errors for large enough $t_\mathrm{cut}/a$. Similar plots can be shown for the other simulated values of the lattice spacing, of the valence masses and for the OS regularization.  } 
    \label{fig:amu_vs_tcut}
\end{figure}
We consider two lattice discretizations of the e.m. currents $J^{\mu,\mathrm{reg}}_f$ (see Eq.~(\ref{eq:Jmu})), with $\mathrm{reg}=\{\mathrm{TM},\mathrm{OS}\}$, corresponding to the so-called Twisted-Mass (TM) and Osterwalder-Seiler (OS) regularizations, and compute the corresponding connected Wick contractions of the vector correlators $V_f^\mathrm{reg}(t)$ (see Eq.~(\ref{eq:VV})) on each gauge ensemble. The two regularizations become equivalent in the continuum and differ at fixed cutoff by $O(a^2)$ lattice artifacts (see the final part of Appendix~\ref{sec:masses}). From the lattice correlators $V_f^\mathrm{reg}(t)$ we extracted $a_{\mu}^{\rm HVP,reg}(f)$ by using the following discretized version of Eq.~(\ref{eq:amu_HVP}),
\begin{flalign}
&
a_{\mu}^{\rm HVP, reg}(f;an_\mathrm{min}) = 2\alpha_{\rm em}^{2} 
\lim_{t_\mathrm{cut}\mapsto \infty}
a^3\sum_{n=n_\mathrm{min}}^{t_\mathrm{cut}/a} w(n)\, n^{2} K(m_{\mu} an)\, V_{f}^{reg}(an)
\;,
\nonumber \\[8pt]
&
a_{\mu}^{\rm HVP, reg}(f)
\equiv
a_{\mu}^{\rm HVP, reg}(f;0)\;.
\label{eq:defalatt}
\end{flalign}
In Eq.~(\ref{eq:defalatt}), $n=t/a$ is the Euclidean time in lattice units and the lattice spacing $a$ is intended to be fixed at $a=a^{\rm iso}$. We restrict the integral appearing in Eq.~(\ref{eq:amu_HVP}) to the region $[t_\mathrm{min},t_\mathrm{cut}]$, indicate explicitly the limit $t_\mathrm{cut}\mapsto \infty$ and keep track of the dependence upon $t_\mathrm{min}=a n_\mathrm{min}$ to better control our continuum extrapolations (see below). By setting the weights $w(n)$ to $3/8$ at the end-points, to $3/4$ when $n-n_\mathrm{min}$ is a multiple of 3 and to $9/8$ for the remaining points (Simpson-$3/8$ rule), our discretization differs from the corresponding integral by errors of $O(a^2)$.

As customary, we take the $t_\mathrm{cut}\mapsto \infty$ limit of our results by performing a plateaux-analysis of the partial sums as functions of $t_\mathrm{cut}$. Examples of these analyses are shown in Figure~\ref{fig:amu_vs_tcut}.

\begin{figure}[t]
    \centering
    \includegraphics[width=0.48\linewidth]{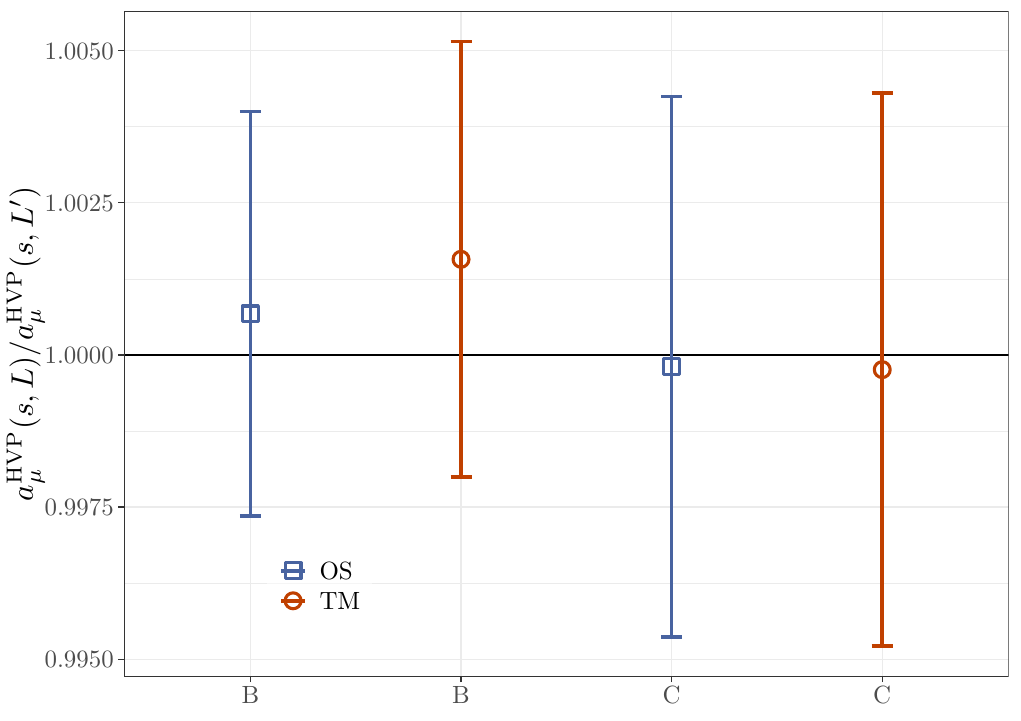}
    \includegraphics[width=0.48\linewidth]{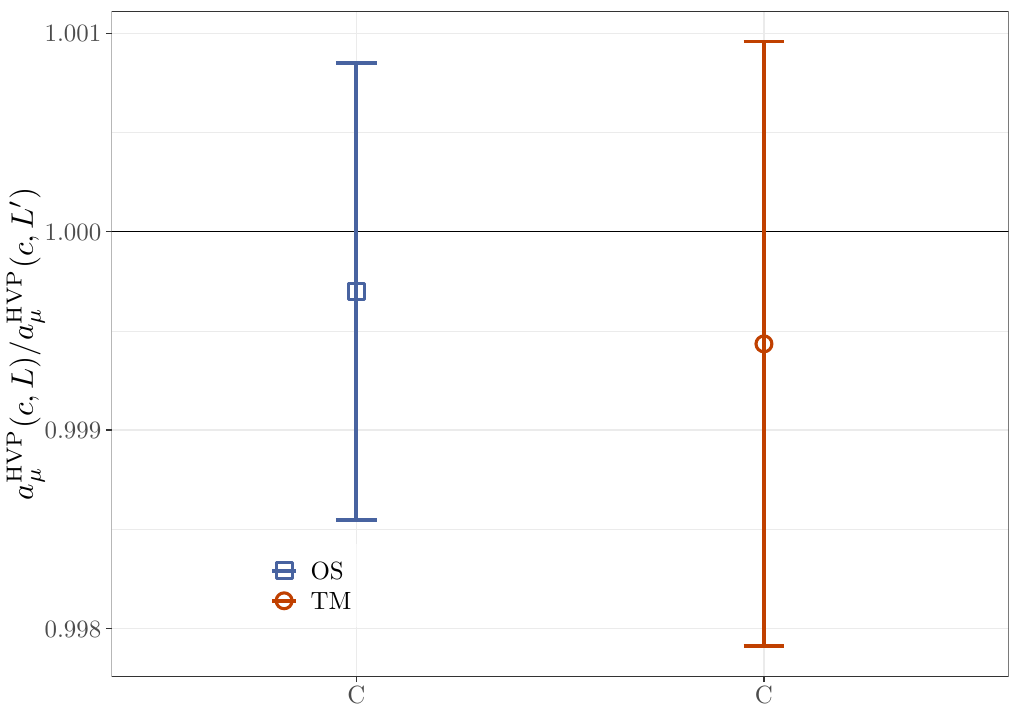}
    \caption{\small \it
    Ratio of $a_\mu^{\rm HVP}(s)$ left and $a_\mu^{\rm HVP}(c)$ for the two 
    regularizations OS and TM, computed at the lattice spacing 
    $a\simeq 0.07$~fm with two linear sizes $L\sim5.4~$fm and $L'\sim7.6~$fm using the ensembles
    C80 and C112. For the strange contribution we plot also the same ratio at
    $a=0.08$~fm with two linear sizes $L\sim5.1$~fm and $L'\sim7.6$~fm using the ensembles
    B64 and B96.}
    \label{fig:FVE_s}
\end{figure}
In order to quantify the FSE on $a_{\mu}^{\rm HVP,reg}(f)$ we perform simulations on two different volumes. More precisely, among the ensembles listed in Table~\ref{tab:iso_EDI_FLAG}, the B64 and the B96 have the same value of the lattice spacing, $a\simeq 0.08$~fm, but different physical volumes, $L\sim5.1~$fm and $L'\sim7.6~$fm. Similarly, the lattice spacing of the ensembles C80 and C112 is $a\simeq 0.07$~fm and the corresponding linear sizes are $L\sim5.4~$fm and $L'\sim7.6~$fm. Figure~\ref{fig:FVE_s} shows the ratio of $a_{\mu}^{\rm HVP,reg}(f)$ computed on the two systems of different linear size (and volume). As it can be seen, the FSE on $a_{\mu}^{\rm HVP,reg}(f)$ are totally negligible w.r.t.\ the statistical errors. Nevertheless, we use these results to estimate a systematic error associated with FSE 
as in \cite{ExtendedTwistedMassCollaborationETMC:2022sta} according to the formula
\begin{flalign}
\Delta_{\mathrm{FSE}}(f)=\max_{\mathrm{reg},a} \left\{ \left|a_{\mu}^{\rm HVP,reg}(f,L)-a_{\mu}^{\rm HVP,reg}(f,L') \right| \mathrm{erf}\left( \frac{\left|a_{\mu}^{\rm HVP,reg}(f,L)-a_{\mu}^{\rm HVP,reg}(f,L') \right|}{\sqrt{2}\sqrt{\Delta^\mathrm{reg}(f,L)^2+\Delta^\mathrm{reg}(f,L')^2}}\right)\right\}\,,
\label{eq:Delta_FVE}
\end{flalign}
where the maximum is computed over the two regularization and over the lattice spacings $a\simeq 0.07$~fm and $a\simeq 0.08$~fm when available. $\Delta^\mathrm{reg}(f,L)$ is the statistical error of $a_{\mu}^{\rm HVP,reg}(f,L)$ and 
$\mathrm{erf}$ is the error function.

By keeping track of the dependence of our results upon $t_\mathrm{min}$ in Eq.~(\ref{eq:defalatt}) we are able to better control our continuum extrapolations and to safely estimate the associated systematic errors. To this end we perform two different analyses. In the main branch of the analysis we fix $t_\mathrm{min}=0$ and perform the continuum extrapolations of our results at fixed cutoff that in this case, with a small abuse of notation, we simply call $a_{\mu}^{\rm HVP, reg}(f)$. In the second branch of the analysis we keep $t_\mathrm{min}$ fixed in physical units by interpolating 
the results $a_{\mu}^{\rm HVP, reg}(f;an_\mathrm{min})$ as functions of the integer variable $n_\mathrm{min}$. Then we extrapolate the results $a_{\mu}^{\rm HVP, reg}(f;t_\mathrm{min})$ to the continuum and add the contributions of the region $[0,t_\mathrm{min}]$ of the integral appearing in Eq.~(\ref{eq:amu_HVP}). These are computed at NNLO in continuum perturbation theory by using the RHAD~\cite{Harlander:2002ur} software package. Finally we study the dependence upon $t_\mathrm{min}$ of the results thus obtained to better quantify the systematic errors associated with our continuum extrapolations. In the case of the full and SD contributions and in both branches of the analysis we compute the tree-level $O(a^2)$ cutoff effects on our results in lattice perturbation theory and remove them before performing the continuum extrapolations.

\begin{figure}[t]
    \centering
    \includegraphics[width=0.45\linewidth]{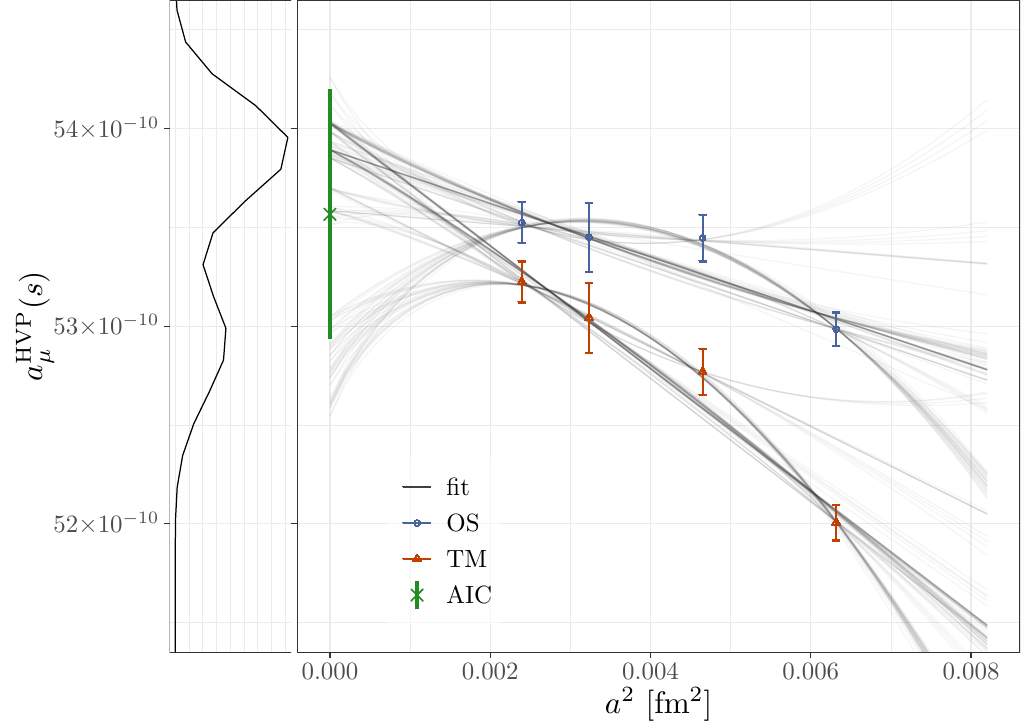}
    \includegraphics[width=0.45\linewidth]{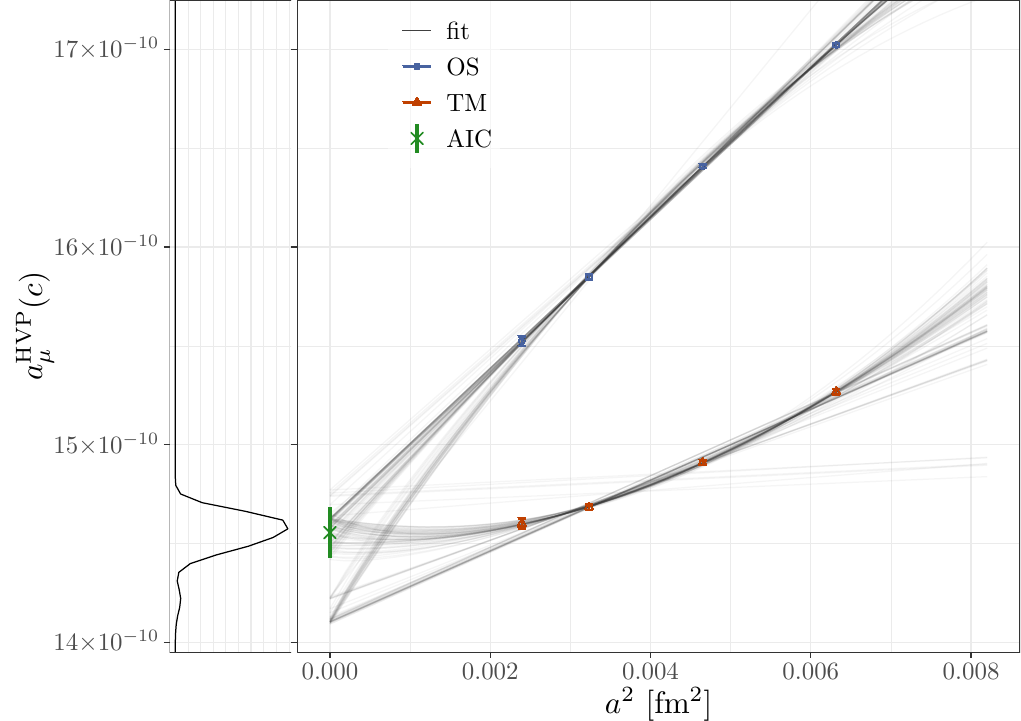}
    \caption{\small \it Continuum extrapolation of the full $a_\mu^\mathrm{HVP}$ for the strange  (left panel) and charm (right panel). For each panel, we show the OS lattice regularization as blue circles, the TM regularization in red triangles, the grey lines represent the various fits, the AIC average of the continuum values is plotted as a green cross and on the left we show the histogram of the continuum values weighted  
    according to the AIC.}
    \label{fig:full}
\end{figure}
We will provide numerical evidence of the branch of the analysis in which we keep track of the dependence upon $t_\mathrm{min}$ in the case of the SD contributions $a_{\mu}^{\rm HVP, SD}(f)$ which are the more sensitive to cutoff effects (see subsection~\ref{sec:SD}). To extrapolate $a_{\mu}^{\rm HVP}(f)\equiv
a_{\mu}^{\rm HVP, reg}(f;t_\mathrm{min}=0)$ to the continuum limit we consider the following Ansatz,
\begin{flalign}
&
a_{\mu}^\mathrm{HVP,reg}(f) = P_0 +P_1^\mathrm{reg} a^2 + P_2^\mathrm{reg} a^4\;, 
\label{eq:fit_amu_continuum}
\end{flalign}
for both the regularizations. We also explore fits in which we remove $P_2^\mathrm{TM}$ and $P_2^\mathrm{OS}$ either together or separately and fits in which the $a^4$ term is replaced with 
$a^2/[\log(a^2/\lambda_0^2)]^n$ for $n=1,2,3$ and $\lambda_0=1$~fm.
We perform all the above fits for three different data sets: (1) the full data set available, (2) the data set excluding the coarsest lattice spacing in both regularizations or only in one of them, and (3) the data set excluding the next-to-coarsest lattice spacing in both regularizations or only in one of them.
To find the average over the different results of the analyses of the  lattice data, we make use of the procedure developed in Ref.~\cite{EuropeanTwistedMass:2014osg}. Namely, starting from $N$ computations with mean values $x_k$ and uncertainties $\sigma_k$ ($k=1,\cdots,N$), based on the same set of input data, their average $x$ and uncertainty $\sigma_x$ are given by
\begin{gather}
    \label{eq:averaging}
    x = \sum_{k=1}^N \omega_k ~ x_k ~ , ~ \qquad
    \sigma_x^2 = \sigma_{x,\mathrm{stat}}^2 + \sigma_{x,\mathrm{syst}}^2 ~ , ~ \qquad
    \sigma_{x,\mathrm{stat}}^2=\sum_{k=1}^N \omega_k~\sigma_k^2 , ~ \qquad
    \sigma_{x,\mathrm{syst}}^2=\sum_{k=1}^N \omega_k ~ (x_k - x)^2~ , ~
\end{gather}
where $\omega_k$ represents the weight associated with the $k$-th determination. The weights $\omega_k$ are based on the Akaike Information Criterion (AIC)\,\cite{Neil:2022joj}, namely 
\be
    \label{eq:AIC}
    \omega_k \propto \mbox{exp}[- (\chi^2 + 2 N_\mathrm{parms} - 2 N_\mathrm{data}) / 2]~ , ~
\ee
where $\chi_k^2$ is the value of the $\chi^2$ variable for the $k$-th computation, $N_\mathrm{parms}$ is the number of free parameters and $N_\mathrm{data}$ the number of data points\footnote{We have verified that the use of the slightly different definition proposed in Ref.\,\cite{Akaike}, namely $\omega_k \propto e^{- (\chi_k^2 + 2 N_\mathrm{parms} - N_\mathrm{data}) / 2} $ leads to very similar averages and errors as compared with those corresponding to the use of Eq.\,(\ref{eq:AIC}).}.
We show the resulting fits in Fig.~\ref{fig:full}. The final results, obtained after averaging all the results of the different analyses
by using Eq.~(\ref{eq:averaging}), are
\begin{align}
     a_{\mu}^{\rm HVP}(s) &=  53.57~(41)_\mathrm{stat}~(48)_\mathrm{cont}~(3)_\mathrm{FSE}\times 10^{-10} =53.57(63)\times 10^{-10}\;,
     \label{eq:values_full} \\[8pt]
     a_{\mu}^{\rm HVP}(c) &= 14.56~(10)_\mathrm{stat}~(9)_\mathrm{cont}~(0)_\mathrm{FSE}\times 10^{-10} =  14.56~(13)\times 10^{-10}\;,
     \label{eq:values_fullc}
\end{align}
where $(.)_\mathrm{stat}$ represents the statistical error resulting from the continuum extrapolation, computed as $\sigma_{x,\mathrm{stat}}$ from Eq.~(\ref{eq:averaging}),  $(.)_\mathrm{cont}$ denotes the systematic error due to the continuum limit, calculated as $\sigma_{x,\mathrm{syst}}$ from Eq.~(\ref{eq:averaging}) and $(.)_\mathrm{FSE}$ denotes the error associated with the finite volume of our lattice simulations, which is calculated using Eq.~(\ref{eq:Delta_FVE}).
All the errors are summed in quadrature to give the total error.

The results for each window contribution are presented in the next subsections.

\subsection{The short-distance window contributions \texorpdfstring{$a_{\mu}^{\rm HVP, SD}(s)$}{amuHVPSDs} and \texorpdfstring{$a_{\mu}^{\rm HVP, SD}(c)$}{amuHVPSDc}}
\label{sec:SD}

\begin{figure}[t]
    \centering
    \includegraphics[width=0.45\linewidth]{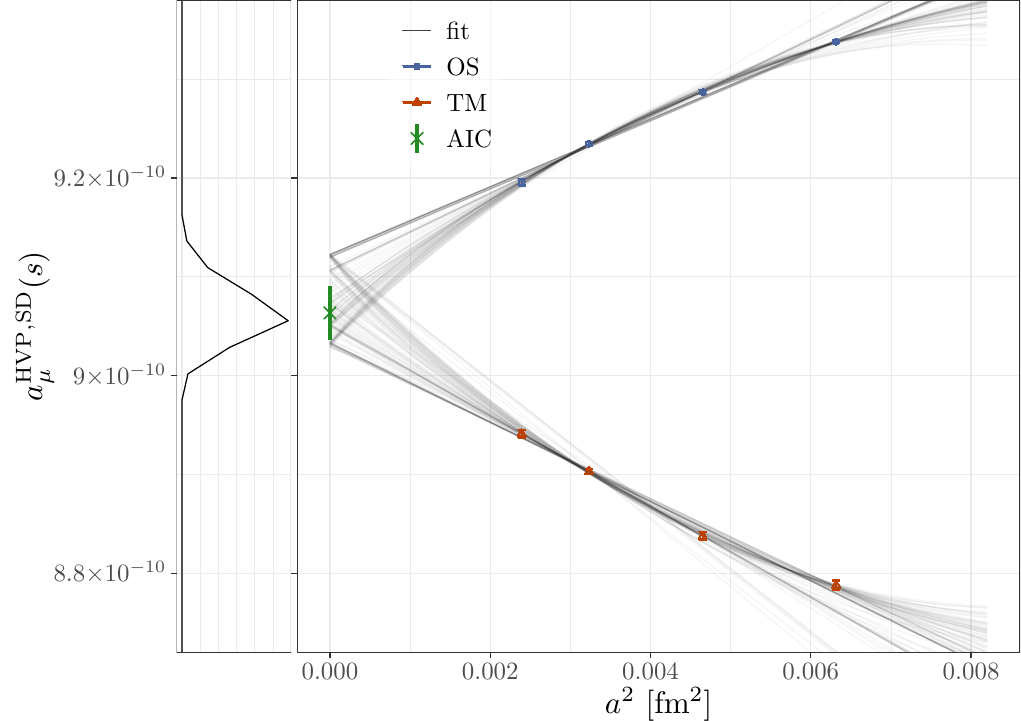}
    \includegraphics[width=0.45\linewidth]{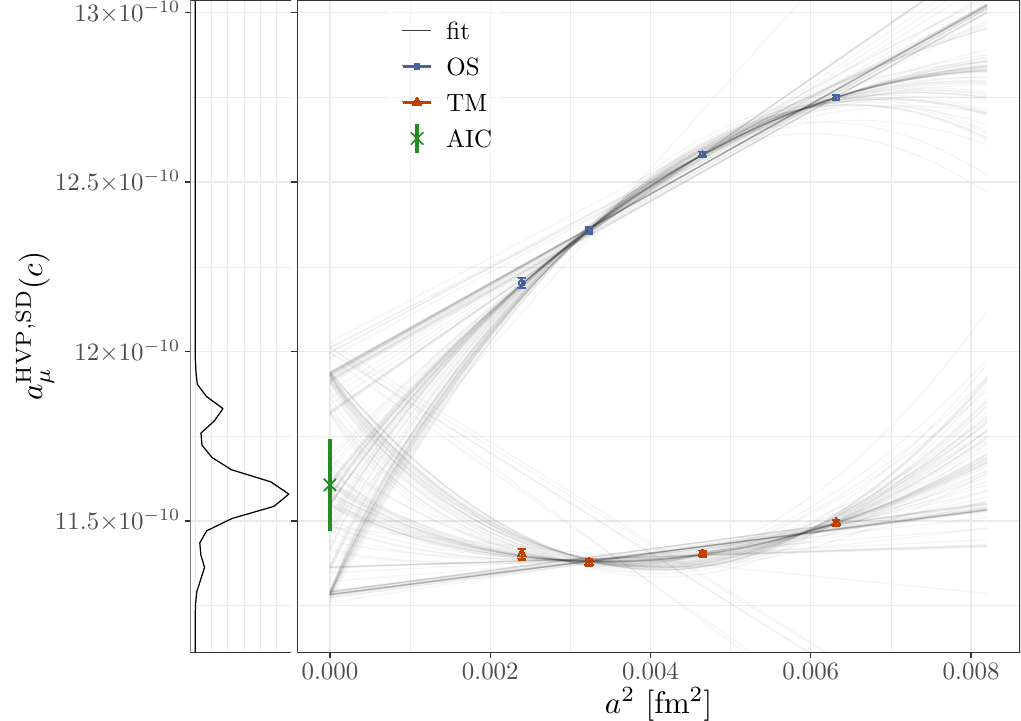}
    \caption{\small \it Continuum extrapolation of the short distance contribution to  $a_\mu^\mathrm{HVP}$ for the strange  (left panel) and charm (right panel). For each panel, we show the OS lattice regularization as blue circles, the TM regularization in red triangles, the grey lines represent the various fits, the AIC average of the continuum values is plotted as a green cross and on the left we show the histogram of the continuum values weighted according to the AIC.}
    \label{fig:SD}
\end{figure}
\begin{figure}[t]
    \centering
    \includegraphics[width=0.45\linewidth]{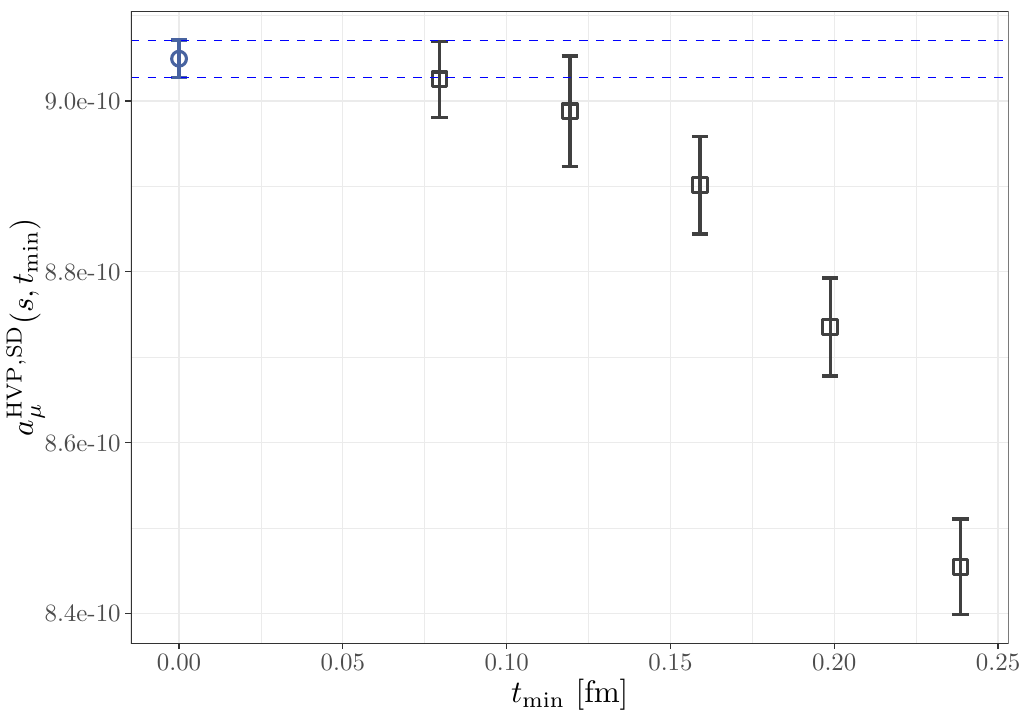}
    \includegraphics[width=0.45\linewidth]{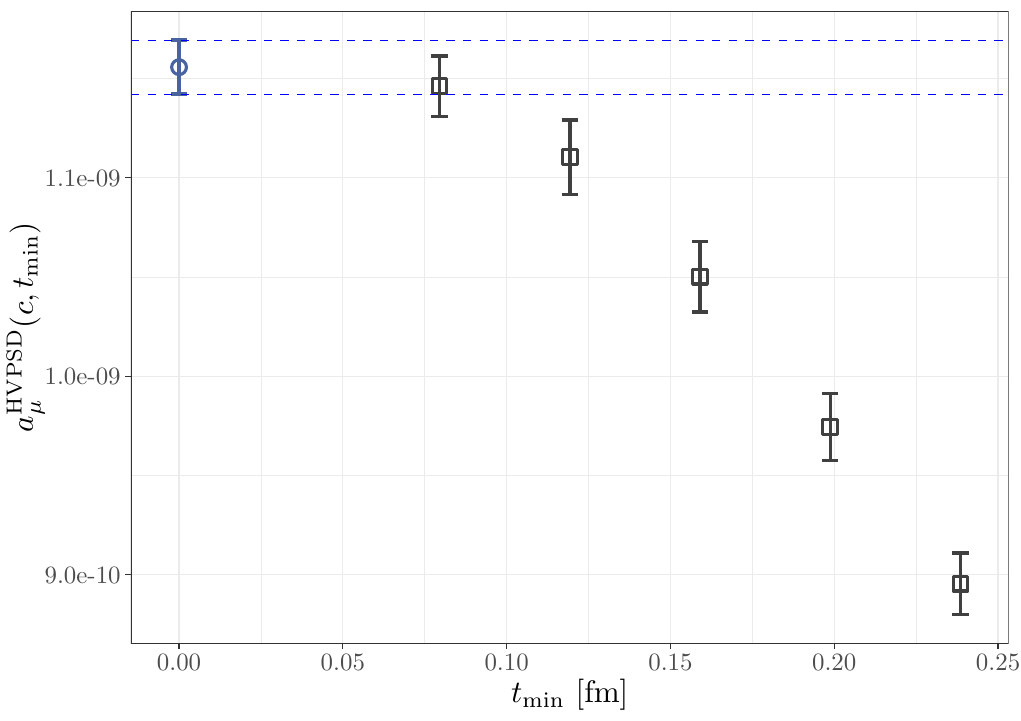}
    \caption{\small \it Continuum values of the short distance contribution to  $a_\mu^{\rm HVP, SD}$ computed as a function of $t_\mathrm{min}$ for the strange  (left panel) and charm (right panel).
    The blue point is computed with the first branch of our analysis, i.e. setting $t_\mathrm{min}=0$ while the black points are computed with the 
    second branch, i.e. keeping $t_\mathrm{min}$ fixed in physical units
    and adding the contribution in the region $[0,t_\mathrm{min}]$ 
    that we have computed at NNLO in continuum perturbation theory by using the RHAD~\cite{Harlander:2002ur} software package.
    }
    \label{fig:SDtmin}
\end{figure}
The SD window contribution is obtained by inserting the kernel $\Theta^\mathrm{SD}(t)$ of Eq.~(\ref{eq:Mt_SD}) in the sum of Eq.~(\ref{eq:defalatt}).
As for the full contribution, we take the $t_\mathrm{cut}\mapsto \infty$ limit of our results performing a plateaux-analysis of the partial sums as functions of $t_\mathrm{cut}$.
Then we subtract from our results the tree-level $O(a^2)$ cutoff effects calculated in lattice perturbation theory (the details of the tree-level calculation can be found
in Appendix~E of Ref.~\cite{ExtendedTwistedMass:2022jpw}).
We first discuss the first branch of our analysis, i.e. taking first $t_\mathrm{min}=0$. The continuum limit and the uncertainty estimate are addressed using the same strategy used in the full contribution. In Fig.~\ref{fig:SD} 
we show the continuum extrapolations and the resulting 
values  are
\begin{align}
     a_{\mu}^{\rm HVP, SD}(s) &= 9.063~(16)_\mathrm{stat}~(22)_\mathrm{cont}~(1)_\mathrm{FSE}\times 10^{-10}=9.063(27)\times 10^{-10}\;,\\[8pt]
     a_{\mu}^{\rm HVP, SD}(c) &= 11.61~(7)_\mathrm{stat}~(11)_\mathrm{cont}~(0)_\mathrm{FSE}\times 10^{-10}=11.61(14)\times 10^{-10}\;.
     \label{eq:values_SD}
\end{align}
The different contributions to the error are estimated as for the full $a_\mu^\mathrm{HVP}$ above.
Our previous determination in Ref.~\cite{ExtendedTwistedMass:2022jpw} was $a_{\mu}^{\rm HVP, SD}(s)=9.074(64)\times 10^{-10}$
and $a_{\mu}^{\rm HVP, SD}(c)=11.61(27)\times 10^{-10}$.
The values given here are compatible with our previous determination and they exhibit a significant reduction of the error.

We now compare the above analysis to our second branch, i.e. when we keep $t_\mathrm{min}$ fixed in physical units by interpolating the results $a_{\mu}^{\rm HVP, SD}(f;an_\mathrm{min})$ as functions
of the integer variable $n_\mathrm{min}$. We then extrapolate those values to the continuum with the same strategy as in the first analysis branch and then add to the integral appearing in Eq.~(\ref{eq:amu_HVP}) the contribution from the region $[0,t_\mathrm{min}]$  which we compute at NNLO in continuum perturbation theory by using the RHAD~\cite{Harlander:2002ur} software package. The results obtained 
from both branches of our analysis are 
plotted in Fig.~\ref{fig:SDtmin}. We observe that for small enough $t_\mathrm{min}$ the two branches of our analysis give compatible results, corroborating the result obtained with the first branch.

\subsection{The intermediate windows \texorpdfstring{$a_{\mu}^{\rm HVP, W}(s)$}{amuHVPWs} and \texorpdfstring{$a_{\mu}^{\rm HVP, W}(c)$}{amuHVPWc}}
\label{sec:amuW}

\begin{figure}[t]
    \centering
    \includegraphics[width=0.45\linewidth]{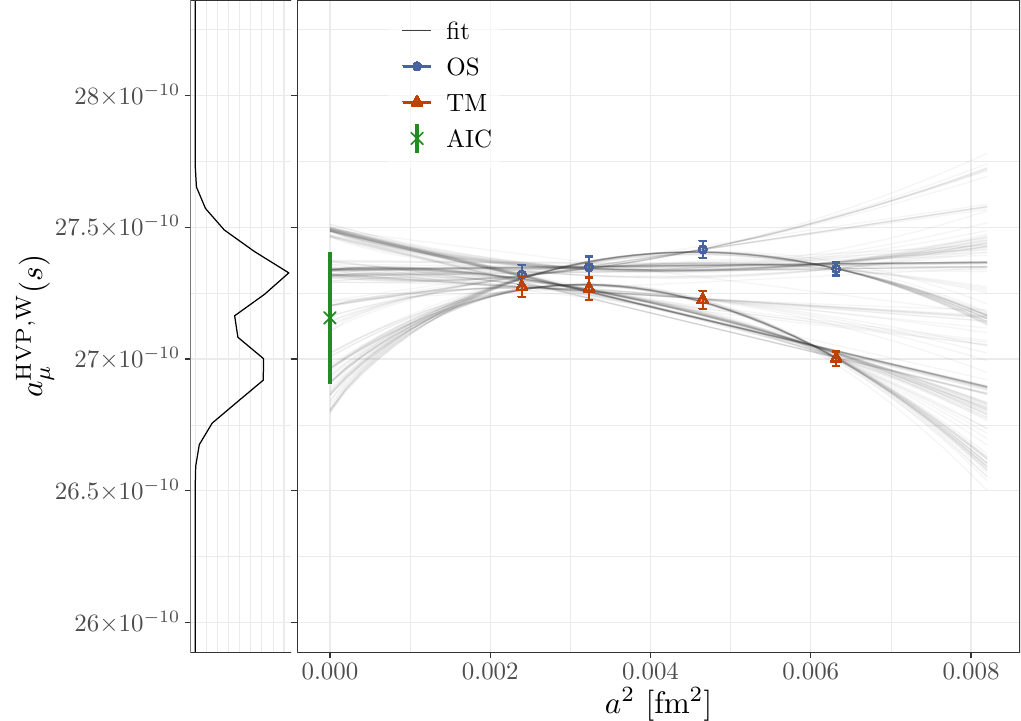}
    \includegraphics[width=0.45\linewidth]{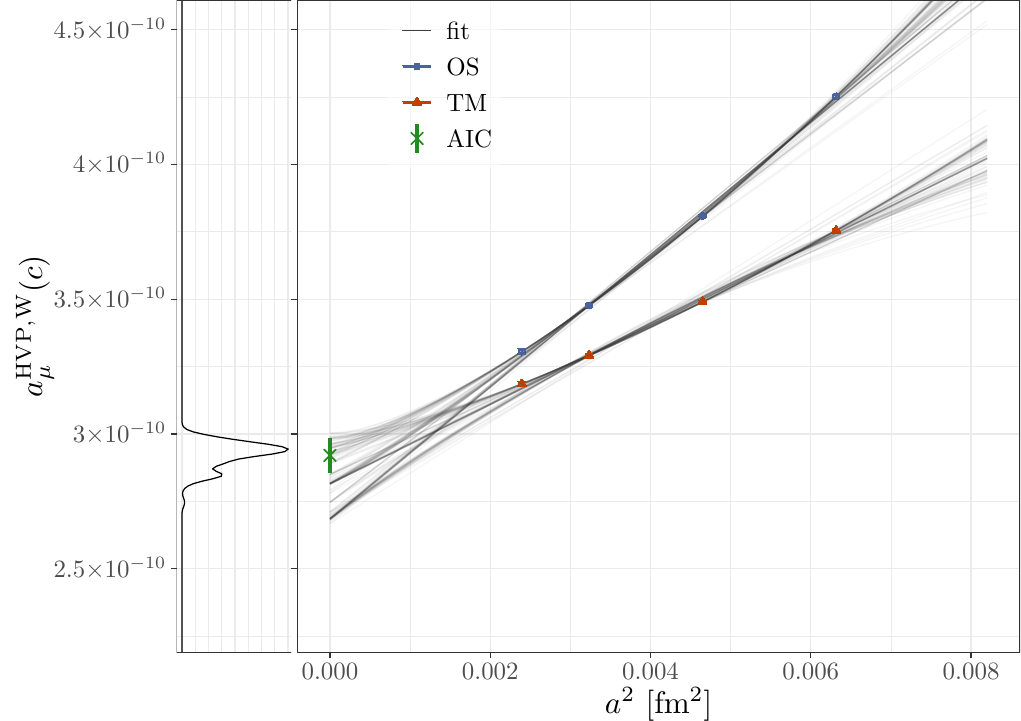}
    \caption{\small \it Continuum extrapolation of the intermediate window contribution to  $a_\mu^{\rm HVP, W}$ for the strange  (left panel) and charm (right panel). For each panel, we show the OS lattice regularization as blue circles, the TM regularization in red triangles, the grey lines represent the various fits, the AIC average of the continuum values is plotted as a green cross and on the left we show the histogram of the continuum values weighted 
        according to the AIC. }
    \label{fig:contW}
\end{figure}
The intermediate window contribution is obtained by inserting the kernel $\Theta^\mathrm{W}(t)$ of Eq.~(\ref{eq:Mt_W}) in the sum of Eq.~(\ref{eq:defalatt}).
As for the full contribution, we take the $t_\mathrm{cut}\mapsto \infty$ limit of our results performing a plateaux-analysis of the partial sums as functions of $t_\mathrm{cut}$.
The continuum extrapolation is done as in the case of the full contributions 
and the various fits are shown in Fig.~\ref{fig:contW}.
The values obtained are 
\begin{align}
     a_{\mu}^{\rm HVP, W}(s) &= 27.16~(15)_\mathrm{stat}~(20)_\mathrm{cont}~(2)_\mathrm{FSE}\times 10^{-10}=27.16(25)\times 10^{-10}\;,\\[8pt]
     a_{\mu}^{\rm HVP, W}(c) &= 2.920~(43)_\mathrm{stat}~(48)_\mathrm{cont}~(0)_\mathrm{FSE}\times 10^{-10}=2.920(64)\times 10^{-10}\;.
     \label{eq:values_W}
\end{align}
Our previous determination in Ref.~\cite{ExtendedTwistedMass:2022jpw} was $a_{\mu}^{\rm HVP, W}(s)=27.28(20)\times 10^{-10}$
and $a_{\mu}^{\rm HVP, W}(c)=2.90(12)\times 10^{-10}$.
The values given here are compatible with our previous determination.
We note a reduction of the error for the charm contribution 
only. For the strange contribution the total error slightly increased w.r.t.\ our previous determination. This is due to the contribution to the statistical errors coming from the uncertainties on the fine-tuning corrections that, at this level of precision, must be taken into account (see Appendix~\ref{sec:mistunings} and, in particular, Fig.~\ref{fig:amu_tuning}).

\subsection{The long distance windows \texorpdfstring{$a_{\mu}^{\rm HVP, LD}(s)$}{amuHVPLDs} and \texorpdfstring{$a_{\mu}^{\rm HVP, LD}(c)$}{amuHVPLDc}}
\label{sec:amuLD}

The long-distance window contribution is computed by inserting the kernel 
\(\Theta^\mathrm{LD}(t)\) from Eq.~(\ref{eq:Mt_LD}) into the summation of Eq.~(\ref{eq:defalatt}).
As for the full contribution, we take the $t_\mathrm{cut}\mapsto \infty$ limit of our results performing a plateaux-analysis of the partial sums as functions of $t_\mathrm{cut}$.
The continuum extrapolation is done as in the case of the full contributions and the values obtained are 
\begin{align}
     a_{\mu}^{\rm HVP, LD}(s) &=17.32~(29)_\mathrm{stat}~(24)_\mathrm{cont}~(1)_\mathrm{FSE}\times 10^{-10} =17.32(38) \times 10^{-10}\;, 
     \label{eq:values_LDs}\\[8pt]
     a_{\mu}^{\rm HVP, LD}(c) &= 0.01352~(37)_\mathrm{stat}~(68)_\mathrm{cont}~(14)_\mathrm{FSE}\times 10^{-10}=0.01352(79)\times 10^{-10}\;.
     \label{eq:values_LD}
\end{align}
By subtracting the values in the continuum of the short distance, Eq.~(\ref{eq:values_SD}), and of the intermediate window, Eq.~(\ref{eq:values_W}),
from the full contribution,
Eq.~(\ref{eq:values_fullc}), and by propagating the error in quadrature we get
$a_{\mu}^{\rm HVP, LD}(s)=17.35(73)\times 10^{-10}$ and $a_{\mu}^{\rm HVP, LD}(c)=0.03(20)\times 10^{-10}$,
which are consistent with Eq.~(\ref{eq:values_LD}) but with larger errors.

\section{Comparison with other lattice QCD results and Outlook}
\label{sec:comparison}
\begin{figure}[t]
    \centering
    \includegraphics[width=0.3\linewidth]{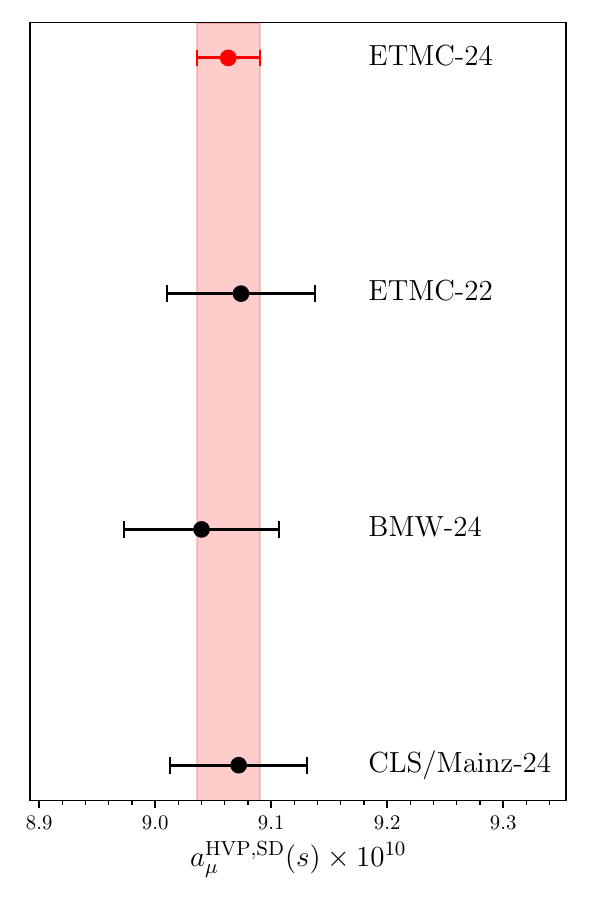}
    \includegraphics[width=0.3\linewidth]{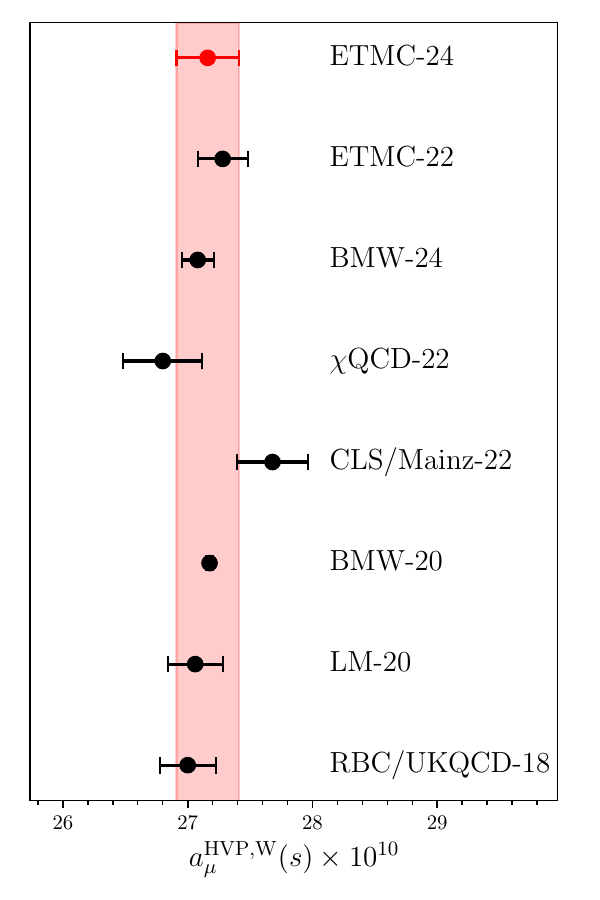}
    \includegraphics[width=0.3\linewidth]{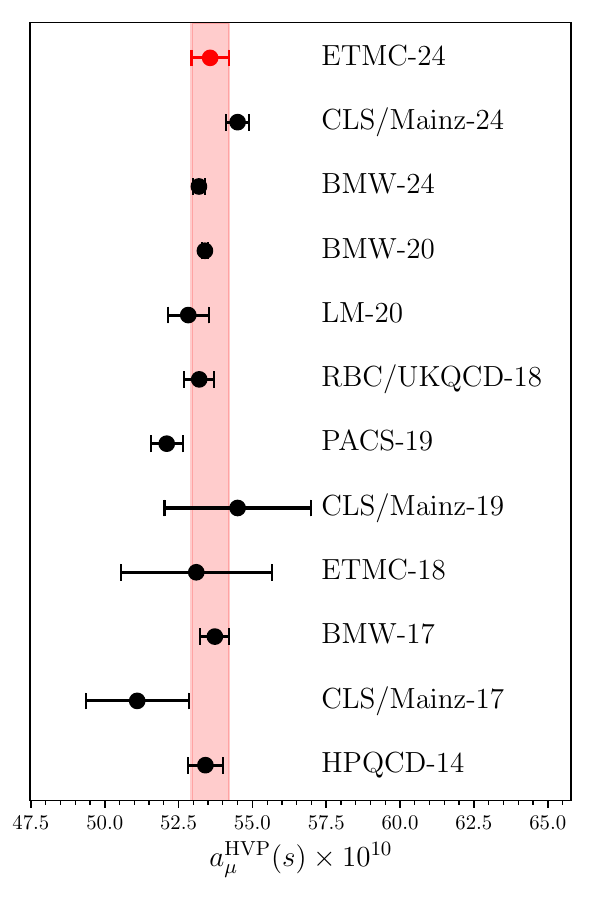}    
    \caption{\small \it We show the lattice QCD determinations of $a_\mu^\mathrm{HVP,SD}(s)$ (left panel), of $a_\mu^{\mathrm{HVP,W}}(s)$ (central panel) and of $a_\mu^\mathrm{HVP}(s)$ (right panel) obtained in this work (in red) and in Refs.~\cite{Boccaletti:2024guq,Kuberski:2024bcj,ExtendedTwistedMass:2022jpw,Wang:2022lkq,Ce:2022kxy,Borsanyi:2020mff,Lehner:2020crt,Shintani:2019wai,Gerardin:2019rua,RBC:2018dos,Giusti:2017jof,Budapest-Marseille-Wuppertal:2017okr,DellaMorte:2017dyu,Chakraborty:2014mwa,Djukanovic:2024cmq}.  In all  panels the vertical red band corresponds to our determination and it is displayed to ease the comparison. Note that the BMW-24 result of Ref.~\cite{Boccaletti:2024guq} for $a_\mu^\mathrm{HVP}(s)$ refers to a time window defined in the range [0,2.8]fm. }
    \label{fig:comparison_s}
\end{figure}
\begin{figure}[t]
    \begin{center}
    \includegraphics[width=0.3\linewidth]{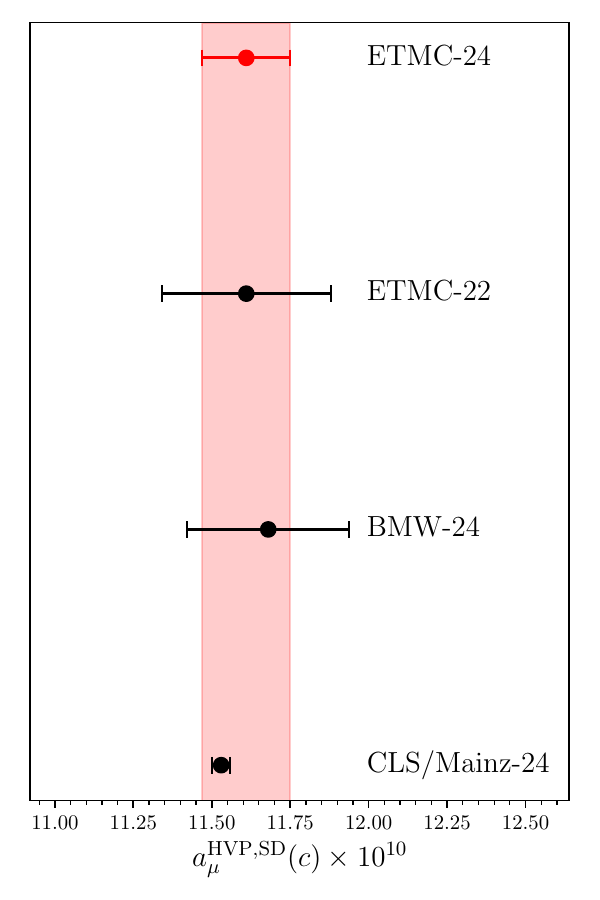}
    \includegraphics[width=0.3\linewidth]{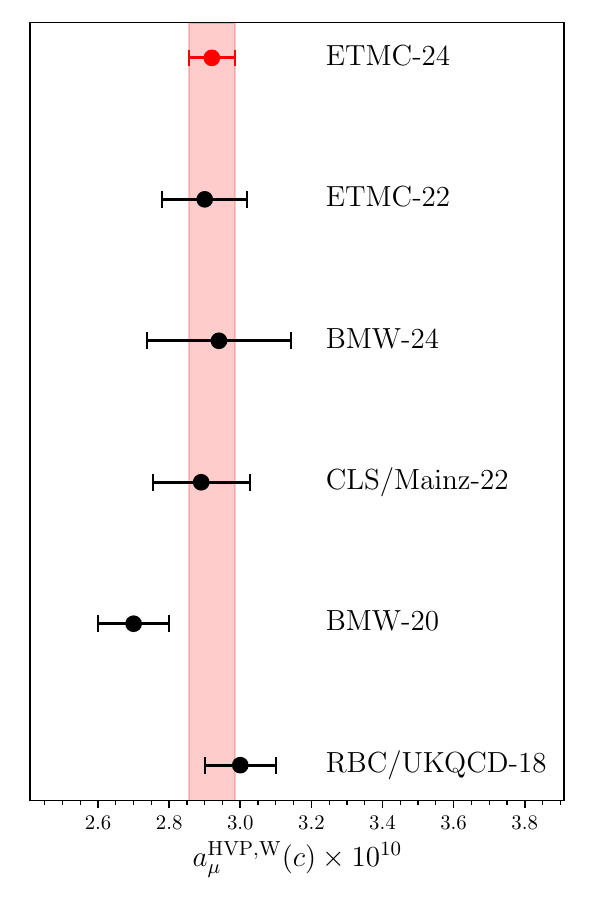}
    \includegraphics[width=0.3\linewidth]{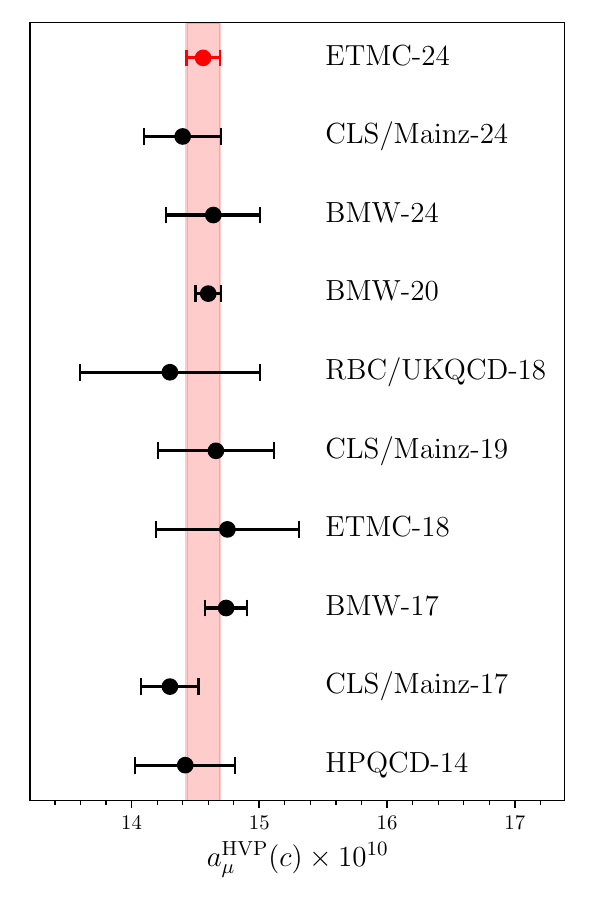}    
    \end{center}
    \caption{\small \it
     We show the lattice QCD determinations of $a_\mu^\mathrm{HVP,SD}(c)$ (left panel), of $a_\mu^{\mathrm{HVP,W}}(c)$ (central panel) and of $a_\mu^\mathrm{HVP}(c)$ (right panel) obtained in this work (in red) and in Refs.~\cite{Boccaletti:2024guq,Kuberski:2024bcj,ExtendedTwistedMass:2022jpw,Wang:2022lkq,Ce:2022kxy,Borsanyi:2020mff,Lehner:2020crt,Shintani:2019wai,Gerardin:2019rua,RBC:2018dos,Giusti:2017jof,Budapest-Marseille-Wuppertal:2017okr,DellaMorte:2017dyu,Chakraborty:2014mwa,Djukanovic:2024cmq}.    In all panels the vertical red band corresponds to our determination and it is displayed to ease the comparison. Note that the BMW-24 result of Ref.~\cite{Boccaletti:2024guq} for $a_\mu^\mathrm{HVP}(c)$ refers to a time window defined in the range [0,2.8]fm.}
    \label{fig:comparison_c}
\end{figure}
In the isospin symmetric limit of QCD, according to the Edinburgh/FLAG consensus as specified in Eq.~(\ref{eq:iso_definition}),  we obtain results for the total, short-distance, intermediate window and long-distance contributions to $a_\mu^\mathrm{HVP}$ coming from the strange and charm quark-connected diagrams that are listed 
in Eqs.~(\ref{eq:values_full})--(\ref{eq:values_LD}).
In Fig.~\ref{fig:comparison_s} and Fig.~\ref{fig:comparison_c}, to the best of our knowledge,
we compare\footnote{The comparison of our results with the other lattice determinations presented in Figs.~\ref{fig:comparison_s} and~\ref{fig:comparison_c} has been done by directly using the results quoted in Refs.~\cite{Aoyama:2020ynm,Boccaletti:2024guq,Kuberski:2024bcj,ExtendedTwistedMass:2022jpw,Wang:2022lkq,Ce:2022kxy,Borsanyi:2020mff,Lehner:2020crt,Shintani:2019wai,Gerardin:2019rua,RBC:2018dos,Giusti:2017jof,Budapest-Marseille-Wuppertal:2017okr,DellaMorte:2017dyu,Chakraborty:2014mwa,Djukanovic:2024cmq} without taking into account the small differences associated with the fact that, in some cases, slightly different definitions of isoQCD have been adopted. Moreover, the BMW-24~\cite{Boccaletti:2024guq} results correspond to the window $[0.2.8]$~fm, obtained by setting $t_0=2.8$~fm in Eq.~(\ref{eq:Mt_SD}).} our results in this work with the corresponding ones by other lattice studies, see Refs.~\cite{Aoyama:2020ynm,Boccaletti:2024guq,Kuberski:2024bcj,ExtendedTwistedMass:2022jpw,Wang:2022lkq,Ce:2022kxy,Borsanyi:2020mff,Lehner:2020crt,Shintani:2019wai,Gerardin:2019rua,RBC:2018dos,Giusti:2017jof,Budapest-Marseille-Wuppertal:2017okr,DellaMorte:2017dyu,Chakraborty:2014mwa,Djukanovic:2024cmq}.

The results for the short-distance and intermediate window contributions to $a_\mu^\mathrm{HVP}(s)$ and $a_\mu^\mathrm{HVP}(c)$ presented in this work exhibit in most cases a significant reduction of the uncertainty compared to our previous determination in Ref.~\cite{ExtendedTwistedMass:2022jpw}. This is more notable in the case of $a_\mu^\mathrm{HVP,SD}(c)$ as it can be appreciated in the left panel of Fig.~\ref{fig:comparison_c}. Our new determinations of
$a_\mu^\mathrm{HVP,SD}(s)$ and $a_\mu^\mathrm{HVP,W}(c)$ are presently the most precise among the available lattice results. 

In addition to this update, we also provide the total and the long-distance quark-connected contributions for $a_\mu^\mathrm{HVP}(s)$ and $a_\mu^\mathrm{HVP}(c)$, see 
Eqs.~(\ref{eq:values_full})--(\ref{eq:values_fullc})
and
Eqs.~(\ref{eq:values_LDs})--(\ref{eq:values_LD}).
Our results for the total quark-connected $a_\mu^\mathrm{HVP}(s)$ and $a_\mu^\mathrm{HVP}(c)$ are in excellent agreement with all the other lattice determinations. The result we quote for $a_\mu^\mathrm{HVP}(c)$ is among the most accurate ones.

Being concerned with analyses that give results with typical total errors at the 0.5--1.0\% level, we have chosen a definite prescription for the isosymmetric QCD theory, the Edinburgh/FLAG consensus, and made an effort to consider and quantify as well as possible all the uncertainties related to the
necessary hadronic renormalization inputs and their feedback on the
observables of interest here. 
The subpercent accuracy level achieved for
the observable estimators at finite lattice spacing has required considering a set of continuum fit Ansatz that go well beyond the basic fit linear in $a^2$ and are averaged using the AIC, as discussed in Sect.~III. A similar kind of analyses will likely be necessary for several other physical observables that are obtained in lattice studies at comparable accuracy level.

At the level of precision reached now for $a_\mu^\mathrm{HVP}(s)$ and $a_\mu^\mathrm{HVP}(c)$ and their partial window contributions, the evaluation of the leading isospin breaking contributions within QCD + QED with $u$, $d$, $s$ and $c$ active flavours is of course mandatory. This work is currently in progress by our collaboration using the RM123 
method\,~\cite{deDivitiis:2013xla}. Similarly in progress is the evaluation of the light quark-connected contribution, $a_\mu^\mathrm{HVP}(\ell)$, and the all-flavours quark-disconnected contribution, $a_\mu^\mathrm{HVP}({\rm disc.})$, as well as of the corresponding leading isospin breaking corrections in QCD+QED.

\section*{Acknowledgments}

We thank all members of ETMC for the most enjoyable collaboration. We thank Giancarlo Rossi for a careful
reading of the draft of this paper. 

C.A.~acknowledges partial support from the European Joint Doctorate AQTIVATE that received funding from the European Union’s research and innovation programme under the Marie Sklodowska-Curie Doctoral Networks action under the Grant Agreement No 101072344 and from the European Union’s Horizon 2020 Research and Innovation Programme ENGAGE under the Marie Sklodowska-Curie COFUND scheme with grant agreement No.~101034267.
N.K., S.R., and U.W.~acknowledge support by the Swiss National Science Foundation (SNSF) project No. 200020\_208222. 
V.L., F.S., G.G., R.F., N.T., A.E.~and A.D.S.~are supported by the Italian Ministry of University and Research (MUR) under the grant PNRR-M4C2-I1.1-PRIN 2022-PE2 Non-perturbative aspects of fundamental interactions, in the Standard Model and beyond F53D23001480006 funded by E.U.-NextGenerationEU.
S.B., J.F.~and F.P.~received financial support from the Inno4scale project, which received funding from the European High-Performance Computing Joint Undertaking (JU) under Grant Agreement No.~101118139. The JU receives support from the European Union's Horizon Europe Programme.
J.F.~received support by the DFG research unit FOR5269 ”Future methods for studying
confined gluons in QCD” and acknowledges financial support by the Next Generation Triggers project (https://nextgentriggers.web.cern.ch). M.G. and C.U. are supported by the Deutsche Forschungsgemeinschaft (DFG,
German Research Foundation) as part of the CRC 1639 NuMeriQS – project no.~511713970.
F.S. is supported by ICSC – Centro Nazionale di Ricerca in High Performance Computing, Big Data and Quantum Computing, funded by European Union -NextGenerationEU and by Italian  Ministry of University and Research (MUR) projects FIS\_00001556 and PRIN\_2022N4W8WR

The open-source packages tmLQCD~\cite{Jansen:2009xp,Abdel-Rehim:2013wba,Deuzeman:2013xaa,Kostrzewa:2022hsv}, LEMON~\cite{Deuzeman:2011wz}, DD-$\alpha$AMG~\cite{Frommer:2013fsa,Alexandrou:2016izb,Bacchio:2017pcp,Alexandrou:2018wiv}, QPhiX~\cite{joo2016optimizing,Schrock:2015gik} and QUDA~\cite{Clark:2009wm,Babich:2011np,Clark:2016rdz} have been used in the ensemble generation.

The authors gratefully acknowledge the Gauss Centre for Supercomputing e.V.~(www.gauss-centre.eu) for funding this project by providing computing time on the GCS Supercomputers SuperMUC-NG at Leibniz Supercomputing Centre and JUWELS~\cite{JUWELS, JUWELS-BOOSTER} at Juelich Supercomputing Centre.
The authors acknowledge the Texas Advanced Computing Center (TACC) at The University of Texas at Austin for providing HPC resources (Project ID PHY21001).
The authors gratefully acknowledge PRACE for awarding access to HAWK at HLRS within the project with Id Acid 4886.
We acknowledge the Swiss National Supercomputing Centre (CSCS) and the EuroHPC Joint Undertaking for awarding this project access to the LUMI supercomputer, owned by the EuroHPC Joint Undertaking, hosted by CSC (Finland) and the LUMI consortium through the Chronos programme under project IDs CH17-CSCS-CYP and CH21-CSCS-UNIBE as well as the EuroHPC Regular Access Mode under project ID EHPC-REG-2021R0095.
We are grateful to CINECA and EuroHPC JU
for awarding this project access to Leonardo 
supercomputing facilities hosted at CINECA. 
We gratefully acknowledge EuroHPC JU for the computer
time on Leonardo-Booster provided to us through the 
Extreme Scale Access Call grant EHPC-EXT-2024E01-027.
We gratefully acknowledge
CINECA for the provision of GPU time under the specific initiative INFN-LQCD123 and IscrB S-EPIC.

\appendix

\section{Lattice setup and simulation details}
\label{sec:simulations}

In this work, we compute correlation functions, and from these extract the physical observables of interest, on the gauge ensembles produced by ETMC in isoQCD with $N_f = 2 + 1 + 1$ flavors of Wilson-Clover twisted-mass quarks as described in Refs.\,\cite{Alexandrou:2018egz, ExtendedTwistedMass:2020tvp, ExtendedTwistedMass:2021qui, Finkenrath:2022eon}. As already done in Ref.\,\cite{ExtendedTwistedMass:2022jpw}, the correlation functions are evaluated in the mixed-action lattice theory corresponding to the following renormalizable action
\begin{flalign}
S=S_\mathrm{YM}(g_0) + 
S_\mathrm{TM}(\mu_i^\mathrm{sim},m_\mathrm{cr}^\mathrm{sim}) + 
S_\mathrm{ghost}(m_f^\mathrm{sim},m_\mathrm{cr}^\mathrm{sim}) + 
S_\mathrm{OS}(m_f,m_\mathrm{cr})\;.
\label{eq:Sfull}
\end{flalign}

The gluon action is the mean-field improved Iwasaki one~\cite{Aoki:1998qd},
\begin{flalign}
S_\mathrm{YM}(g_0) =
\frac{\beta}{3}\sum_x\sum_{\mu<\nu}\left(
b_0\left\{
1-\mathrm{ReTr}\left[U^{1\times 1}_{\mu\nu}(x)\right]
\right\}
+
b_1\left\{
1-\mathrm{ReTr}\left[U^{1\times 2}_{\mu\nu}(x)\right]
\right\}
\right)\;,
\end{flalign}
where $\beta=6/g_0^2$ is the inverse bare QCD gauge coupling, $b_1=-0.331$, $b_0=1-8b_1$, $U^{1\times 1}_{\mu\nu}(x)$ is the square plaquette and $U^{1\times 2}_{\mu\nu}(x)$  the rectangular one, see Ref.~\cite{Iwasaki:1985we}. 

The Twisted Mass (TM) quark action depends on the bare mass parameters $\mu_i^\mathrm{sim}$ with $i=\{\ell,\sigma,\delta\}$ and on the critical mass $m_\mathrm{cr}^\mathrm{sim}$ and is given by
\begin{flalign}
S_\mathrm{TM}(\mu_i^\mathrm{sim},m_\mathrm{cr}^\mathrm{sim})
&=
\sum_x \bar{\Psi}_\ell\left\{
\gamma_\mu \bar{\nabla}_\mu[U]-i\tau^3\gamma_5 \left(
W^\mathrm{cl}[U]+m_\mathrm{cr}^\mathrm{sim} 
\right) +\mu_\ell^\mathrm{sim}
\right\}\Psi_\ell
\nonumber\\[8pt]
&+
\sum_x \bar{\Psi}_h\left\{
\gamma_\mu \bar{\nabla}_\mu[U]-i\tau^1\gamma_5 \left(
W^\mathrm{cl}[U]+m_\mathrm{cr}^\mathrm{sim} 
\right) +\mu_\sigma^\mathrm{sim}
+\tau^3\mu_\delta^\mathrm{sim}
\right\}\Psi_h \;,
\end{flalign}
where $\Psi_\ell^T=(u,d)$ is the light-quark TM doublet, $\Psi_h^T=(c,s)$ is the heavy TM doublet. We refer to Ref.~\cite{Sheikholeslami:1985ij} for the explicit expression of the Wilson-Clover term $W^\mathrm{cl}[U]$. The TM quark action is non-diagonal in the charm-strange flavor sector.

The Osterwalder-Seiler (OS) action is flavor-diagonal also in the heavy sector and is given by
\begin{flalign}
\label{eq:OS_valence_quark_action}
S_\mathrm{OS}(m_f,m_\mathrm{cr})
&=
\sum_{f}
\sum_x \bar{q}_f\left\{
\gamma_\mu \bar{\nabla}_\mu[U]-ir_f\gamma_5 \left(
W^\mathrm{cl}[U]+m_\mathrm{cr} 
\right) +m_f
\right\}q_f \;,
\end{flalign}
where the flavor index $f=\{u,d,s,c\}$ runs over the four lightest quarks and, in isoQCD, we take $m_u=m_d=m_\ell$ and set $r_{u,c}=1$ and $r_{d,s}=-1$. 

Finally, we have the {\em ghost} action, which is given by
\begin{flalign}
S_\mathrm{ghost}(m_f^\mathrm{sim},m_\mathrm{cr}^\mathrm{sim})
&=
\sum_{f}
\sum_x \bar{\phi}_f\left\{
\gamma_\mu \bar{\nabla}_\mu[U]-ir_f\gamma_5 \left(
W^\mathrm{cl}[U]+m_\mathrm{cr}^\mathrm{sim} 
\right) +m_f^\mathrm{sim}
\right\}\phi_f \;,
\end{flalign}
where the pseudo-quarks $\phi_f$ are the bosonic fields associated with the OS quark fields $q_f$ and, therefore,  in this case we set $r_{u,c}=1$ and $r_{d,s}=-1$. 

By integrating out all the quarks and pseudo-quarks fields appearing in $S$, one gets the weight that should be used to generate the gluon field gauge configurations corresponding to the action given in Eq.~(\ref{eq:Sfull}). This, up to its normalization, is given by
\begin{flalign}
P[U]
&=
e^{-S_\mathrm{YM}(g_0)}
\times
\mathcal{D}_\mathrm{TM}^\ell(\mu_\ell^\mathrm{sim},m_\mathrm{cr}^\mathrm{sim})
\mathcal{D}_\mathrm{TM}^h(\mu_\sigma^\mathrm{sim},\mu_\delta^\mathrm{sim},m_\mathrm{cr}^\mathrm{sim})
\times
\prod_f 
\frac{
\mathcal{D}_\mathrm{OS}^f(m_f,m_\mathrm{cr})
}{
\mathcal{D}_\mathrm{OS}^f(m_f^\mathrm{sim},m_\mathrm{cr}^\mathrm{sim})
}
\label{eq:Sweight1}
\\[12pt]
&=
e^{-S_\mathrm{YM}(g_0)}
\times
\frac{
\mathcal{D}_\mathrm{TM}^h(\mu_\sigma^\mathrm{sim},\mu_\delta^\mathrm{sim},m_0^\mathrm{sim})
}{
\mathcal{D}_\mathrm{OS}^s(m_s^\mathrm{sim},m_\mathrm{cr}^\mathrm{sim})\
\mathcal{D}_\mathrm{OS}^c(m_c^\mathrm{sim},m_\mathrm{cr}^\mathrm{sim})
}
\times
\prod_f \mathcal{D}_\mathrm{OS}^f(m_f,m_0)\;,
\label{eq:Sweight2}
\end{flalign}
where
\begin{flalign}
&
\mathcal{D}_\mathrm{TM}^h(\mu_\sigma^\mathrm{sim},\mu_\delta^\mathrm{sim},m_\mathrm{cr}^\mathrm{sim})
=
\det\left\{
\gamma_\mu \bar{\nabla}_\mu[U]-i\tau^1\gamma_5 \left(
W^\mathrm{cl}[U]+m_\mathrm{cr}^\mathrm{sim} 
\right) +\mu_\sigma^\mathrm{sim}
+\tau^3\mu_\delta^\mathrm{sim}
\right\}\;,
\nonumber \\[8pt]
&
\mathcal{D}_\mathrm{OS}^f(m_f^\mathrm{sim},m_\mathrm{cr}^\mathrm{sim})
=
\det\left\{
\gamma_\mu \bar{\nabla}_\mu[U]-ir_f\gamma_5 \left(
W^\mathrm{cl}[U]+m_\mathrm{cr}^\mathrm{sim} 
\right) +m_f^\mathrm{sim}
\right\}\;,
\end{flalign}
and where, in passing from Eq.~(\ref{eq:Sweight1}) to Eq.~(\ref{eq:Sweight2}), we  use the relation
\begin{flalign}
\mathcal{D}_\mathrm{TM}^\ell(\mu_\ell^\mathrm{sim},m_\mathrm{cr}^\mathrm{sim})
&=
\det\left\{
\gamma_\mu \bar{\nabla}_\mu[U]-i\tau^3\gamma_5 \left(
W^\mathrm{cl}[U]+m_\mathrm{cr}^\mathrm{sim} 
\right) +\mu_\ell^\mathrm{sim}
\right\}
=
\mathcal{D}_\mathrm{OS}^u(\mu_\ell^\mathrm{sim},m_\mathrm{cr}^\mathrm{sim})\
\mathcal{D}_\mathrm{OS}^d(\mu_\ell^\mathrm{sim},m_\mathrm{cr}^\mathrm{sim})
\;.
\end{flalign}
Some important remarks are in order. 
Our gluon gauge field configurations are generated with the action $S_\mathrm{YM}(g_0) + S_\mathrm{TM}(\mu_i^\mathrm{sim},m_\mathrm{cr}^\mathrm{sim})$ which, at the price of introducing charm-strange flavor mixing at fixed cutoff, is automatically ${\cal O}(a)$-improved. In addition, it has a real and positive weight, which can thus be interpreted as a probability density and which is given by 
\begin{flalign}
P^\mathrm{sim}[U]=
e^{-S_\mathrm{YM}(g_0)}
\times
\mathcal{D}_\mathrm{TM}^\ell(\mu_\ell^\mathrm{sim},m_\mathrm{cr}^\mathrm{sim})
\mathcal{D}_\mathrm{TM}^h(\mu_\sigma^\mathrm{sim},\mu_\delta^\mathrm{sim},m_\mathrm{cr}^\mathrm{sim})
\;.
\label{eq:Ssimweight}
\end{flalign}

The OS quarks and the corresponding ghosts are introduced in our setup for two reasons: 
\begin{itemize}
\item to avoid the technical complications associated with the heavy-flavor mixing at finite cutoff in the calculation of physical observables which we define in terms of the OS quarks: this is possible because the TM and OS actions can be matched by relying on the following, renormalization scale and scheme independent, relations
\begin{flalign}
m_\ell=\mu_\ell\;,
\qquad
m_s=\mu_{\sigma}-\frac{Z_P}{Z_S}\mu_\delta\;,
\qquad
m_c=\mu_{\sigma}+\frac{Z_P}{Z_S}\mu_\delta\;,
\label{eq:mummatching}
\end{flalign}
and/or matching physical quantities computed with both the TM and OS actions (see below).
Once the matching is performed, the ratio of the TM and the ghost determinants appearing in Eq.~(\ref{eq:Sweight2}) is a mere lattice artifact of $O(a^2)$,
\begin{flalign}
\frac{
\mathcal{D}_\mathrm{TM}^h(\mu_\sigma^\mathrm{sim},\mu_\delta^\mathrm{sim},m_0^\mathrm{sim})
}{
\mathcal{D}_\mathrm{OS}^s(m_s^\mathrm{sim},m_\mathrm{cr}^\mathrm{sim})\
\mathcal{D}_\mathrm{OS}^c(m_c^\mathrm{sim},m_\mathrm{cr}^\mathrm{sim})
}
=1+O(a^2)\;.
\end{flalign}
This implies that our mixed-action lattice theory, which is fully unitary in the light-quarks sector, can also be considered unitary in the heavy-quarks sector up to tiny $O(a^2)$ violations;
 
 \item to improve the precision  of the tuning of the critical mass counterterm $m_\mathrm{cr}$ and to match our target definition of isoQCD  by fine-tuning the OS quark masses at the values $m_f=m_f^\mathrm{iso}$:
this is accomplished by evaluating the re-weighting factors appearing in Eq.~(\ref{eq:Sweight1}), 
\begin{flalign}
W_f(m_f,m_\mathrm{cr})
=
\frac{
\mathcal{D}_\mathrm{OS}^f(m_f,m_\mathrm{cr})
}{
\mathcal{D}_\mathrm{OS}^f(m_f^\mathrm{sim},m_\mathrm{cr}^\mathrm{sim})
}\;,
\label{eq:reSweight2}
\end{flalign}
which are identically equal to one when $m_f=m_f^\mathrm{sim}$ and $m_\mathrm{cr}=m_\mathrm{cr}^\mathrm{sim}$.
\end{itemize}
In the remaining part of this appendix, we explain how the simulated values of the bare parameters $\mu_i^\mathrm{sim}$ and $m_\mathrm{cr}^\mathrm{sim}$ are fixed during the Monte Carlo simulations. 
The fine-tuning of the bare parameters, i.e. the determination of the parameters $m_f^\mathrm{iso}$ and $m_\mathrm{cr}$, will be the subject of Appendix~\ref{sec:masses}.

The critical mass counter-term $m_\mathrm{cr}^\mathrm{sim} \sim 1/a$ is set to a unique value for all flavors\,\cite{Frezzotti:2004wz} and is tuned, at each simulated value of the bare gauge coupling $g_0$, in order to guarantee automatic ${\cal{O}}(a)$-improvement of physical observables\,\cite{Frezzotti:2003ni,Frezzotti:2005gi}. 
This is done, as explained in detail in Ref.~\cite{Alexandrou:2018egz}, by computing the Partially Conserved Axial Current (PCAC) quark mass of the TM light doublet,
\begin{flalign}
2 m_\mathrm{PCAC}(m_f^{sim}, m_0) = \frac{
\sum_{\vec x}
\left\langle 
[\partial_0 \bar\chi_\ell \gamma_5 \gamma_0 \tau^1 \chi_\ell ](t,\vec x) [ \bar\chi_\ell \gamma_5  \tau^1 \chi_\ell ](0) 
\right\rangle
}{
\sum_{\vec x}
\left\langle 
[\bar\chi_\ell \gamma_5 \gamma_0 \tau^1 \chi_\ell ](t,\vec x)[\bar\chi_\ell \gamma_5 \gamma_0 \tau^1 \chi_\ell ](0) 
\right\rangle
}\; \, 
\label{eq:mpcaccondition}
\end{flalign}
\begin{flalign}
\chi_\ell(x)= \exp \left(-i \frac{\pi}{4} \gamma_5\tau^3\right) \Psi_\ell(x)\;,
\qquad
\bar\chi_\ell(x) =\bar \Psi_\ell(x) \exp\left(-i \frac{\pi}{4} \gamma_5\tau^3\right) \ ,
\label{eq:mpcaccondition-bis}
\end{flalign} 
and by determining $m_0=m_{\rm cr}^\mathrm{sim}$ through the condition $am_\mathrm{PCAC}(m_f^\mathrm{sim},m_0)<0.1 am_\ell^\mathrm{sim}/Z_A$, where $m_\ell^\mathrm{sim}= \mu_\ell^\mathrm{sim}$ is the bare mass of the light TM  doublet
and $Z_A$ is an estimate, for which even a modest accuracy of several percent is enough here, of the renormalization constant of the operator $\bar\chi_\ell \gamma_5 \gamma_0 \tau^1 \chi_\ell$. 

In the early stages of the Monte Carlo simulations of the various
ensembles, at each $\beta=6/g_0^2$, the value of $m_\ell^\mathrm{sim}$ is chosen so as to obtain $M_\pi$ as close as possible to the reference value $M_\pi= 135$ MeV. The simulated values of $\mu_\sigma^\mathrm{sim}$ and $\mu_\delta^\mathrm{sim}$, i.e.\ the bare mass parameters of the heavy TM doublet, are tuned in order to reproduce the renormalization group invariant (RGI) values $M_{D_s}/f_{D_s} = 7.9(0.1)$ and $m_c^\mathrm{sim}/m_s^\mathrm{sim} = 11.8(0.2)$.
As detailed in Ref.~\cite{Alexandrou:2018egz}, the two 
conditions above are first imposed on the mass parameters
$m_s$ and $m_c$ of the valence quark action (A.4), for
some trial values of the sea quark mass parameters $\mu_\sigma$
and $\mu_\delta$. We then use Eq.~(A.11) to set
$\mu_\sigma = \frac{1}{2}(m_s+ m_c)$ while $\mu_\delta$ is
determined by the equivalent to Eq.~(\ref{eq:mummatching}), but statistically more precise condition of matching the mass of the unitary Kaon, i.e.\
it is obtained from the appropriate two-point correlators with interpolating operators 
$\bar\Psi_\ell\gamma_5\Psi_h$ and  $\bar\Psi_h \gamma_5 \Psi_\ell$ made out the fields entering in the TM action,
to the mass of the valence Kaon evaluated using the OS 
quark lattice action with $r_s= - r_u$. Note that by following this strategy the renormalization scale and scheme independent ratio $Z_P/Z_S$ appearing in Eq.~(\ref{eq:mummatching}) is not needed. After this matching step, the Monte Carlo simulation is then repeated at the selected values of $\mu_\sigma$ and $\mu_\delta$  and a set of corresponding gauge configurations is used to re-evaluate $m_s$ and $m_c$. After convergence of this iterative procedure %
we get the values of 
$m_s^\mathrm{sim}$ and $m_c^\mathrm{sim}$ as well as of
$\mu_\sigma^\mathrm{sim}$ and $\mu_\delta^\mathrm{sim}$.

\begin{table}[t!]
\begin{center}
    \begin{tabular}{ccccccc}
    ~~~ ensemble ~~~ & ~~~ $\beta$ ~~~ & ~~~ $V/a^{4}$ ~~~ & ~~~ $a^{\rm sim}$ (fm) ~~~ & ~~~ $a\mu_{\ell}^{\rm sim}$ ~~~ & ~ $M_{\pi}$ (MeV) ~ &  ~ $M_{\pi}L$ ~ \\
  \hline
  \\
  cB211.072.64 & $1.778$ & $64^{3}\times 128$ & $0.08$ & $0.00072$ & $140$ & $3.6$ \\
  
  cB211.072.96 & $1.778$ & $96^{3}\times 192$ & $0.08$ & $0.00072$ & $140$ &  $5.4$ \\
  
  cC211.060.80 & $1.836$ & $80^{3}\times 160$ & $0.07$ & $0.00060$ & $137$ &  $3.8$ \\

  cC211.060.112 & $1.836$ & $112^{3}\times 224$ & $0.07$ & $0.00060$ & $137$  & $5.3$ \\
  
  cD211.054.96 & $1.900$ & $96^{3}\times 192$ & $0.06$ & $0.00054$ & $141$ & $3.9$ \\
  
  cE211.044.112 & $1.960$ & $112^{3}\times 224$ & $0.05$ & $0.00044$ & $136$ & $3.8$ \\[8pt]
  \hline
\\

 cA211.53.24 & $1.726$ & $24^{3}\times 48$ & $0.09$ & $0.00530$ & $360$ &  $4.0$  \\
 cA211.40.24 & $1.726$ & $24^{3}\times 48$ & $0.09$ & $0.00400$ & $315$ & $3.5$ \\
 cA211.30.32 & $1.726$ & $32^{3}\times 64$ & $0.09$ & $0.00300$ & $272$ &  $4.0$ \\ 
 cA211.12.48 & $1.726$ & $48^{3}\times 96$ & $0.09$ & $0.00120$ & $174$ & $3.8$ \\ 
 cB211.25.48 & $1.778$ & $48^{3}\times 96$ & $0.08$ & $0.0025$ & $260$ & $5.0$ \\
 cB211.14.64 & $1.778$ & $64^{3}\times 112$ & $0.08$ & $0.0014$ & $194$ & $5.0$ \\[8pt]
 \hline
    \end{tabular}
\end{center}
\caption{We provide the full list of the ETMC gauge ensembles used in this work. These are produced by performing Monte Carlo simulations with the action $S_\mathrm{YM}(g_0) + S_\mathrm{TM}(\mu_i^\mathrm{sim},m_0^\mathrm{sim})$ and, therefore, with the probabilistic weight $P^\mathrm{sim}[U]$ given in Eq.~(\ref{eq:Ssimweight}), see Refs.\,\cite{Alexandrou:2018egz, ExtendedTwistedMass:2020tvp, ExtendedTwistedMass:2021qui, Finkenrath:2022eon}. The bare parameters of these simulations (reported in the table) are slightly different from the ones (determined in this work) corresponding to our target definition of isoQCD which, among the other inputs, prescribes $M_\pi^\mathrm{iso}=135$~MeV. The ensembles at heavier pion masses, listed in the last six lines of the table, are only used to check the determination of the fine-tuned isoQCD bare parameters (see Eq.~(\ref{eq:global_fit}) and the related discussion). The large volumes ensembles with $M_\pi\simeq M_\pi^\mathrm{iso}$, listed in the first six lines of the table, are corrected for the small mistunings of the bare parameters by applying the reweighting technique thoroughly discussed in Appendix~\ref{sec:masses}. The values of the light, strange and charm quark masses corresponding to our definition of isoQCD ($m_{\ell,s,c}^{\rm iso}$), along with the values of the critical mass counterterm ($m_{\rm cr}$) and the lattice spacing $a^{\rm iso}$, which we use for the present calculation of the strange and charm HVP, are reported in Table~\ref{tab:iso_EDI_FLAG}. Note the different naming conventions used in Table~\ref{tab:iso_EDI_FLAG} to distinguish the reweighted ensembles from the simulated ones that are listed here.    
}
\label{tab:simudetails}
\end{table}

 The essential information on 
 the ETMC ensembles that are relevant for this work are collected in Table\,\ref{tab:simudetails}. With respect to Ref.\,\cite{ExtendedTwistedMass:2022jpw} two new dedicated gauge ensembles, the cE211.044.112 and the cC211.060.112, are included in the current analysis %
 to improve the control of cutoff and finite-size effects. 
The cE211.044.112 ensemble corresponds to our finest lattice spacing $a \approx 0.05$~fm.
We remind that the cB211.074.96 and the cC211.060.112 ensembles, which have a spatial lattice size $L \approx 7.6$ fm, are used to estimate FSEs by comparing to  the cB211.074.64 and cC211.060.80 ensembles of smaller spatial size, respectively. 
Note that for the ensembles listed in the upper part of Table\,\ref{tab:simudetails}, which are the only ones that are used for the calculation of $a_{\mu}^{\rm HVP}(s)$ and $a_{\mu}^{\rm HVP}(c)$, the pion mass is simulated very close to the reference value $M_\pi= 135$~MeV. For the evaluation of the quark connected contribution to $a_{\mu}^{\rm HVP}(s)$ ($a_{\mu}^{\rm HVP}(c)$), the inversions of the Dirac operator are performed using up to $N_\mathrm{hits}=112$ ($N_\mathrm{hits}=24$) spin-diluted spatial stochastic sources per gauge configuration.

\section{Scheme defining isospin symmetric \texorpdfstring{$N_f=2+1+1$}{Nf211} QCD}
\label{sec:masses}

In this appendix, we describe in detail the procedure that we use to fine-tune the bare parameters of our lattice action, i.e.\ the determination of the parameters $m_\mathrm{cr}$,
$m_\ell^\mathrm{iso}$,  $m_s^\mathrm{iso}$ and $m_c^\mathrm{iso}$.

Our target definition of isoQCD is the one corresponding to the Edinburgh/FLAG consensus\,\cite{Edinburgh, FlavourLatticeAveragingGroupFLAG:2024oxs}and is implemented by using the hadronic inputs given in Eq.~(\ref{eq:iso_definition}) to determine, at any fixed value of the strong bare coupling $g_0$, the bare quark masses $m_f^\mathrm{iso}$ and the lattice spacing $a^\mathrm{iso}$. 

In appendix~\ref{sec:simulations} (see also Ref.~\cite{Alexandrou:2018egz} for more details) we discussed the strategy used to set the bare parameters $\mu_i^\mathrm{sim}$ of the Monte Carlo simulations and, therefore, also the corresponding matched parameters $m_f^\mathrm{sim}$ (see Eq.~(\ref{eq:mummatching}) and text below it). Since the conditions that we used to fix $m_f^\mathrm{sim}$ correspond to an alternative possible definition of isoQCD, that in fact differs from the Edinburgh/FLAG one for corrections that are of the order of isospin breaking effects on hadronic quantities, it turns out that the differences $m_f^\mathrm{iso}-m_f^\mathrm{sim}$ are very small. Moreover, the tuning of the chiral symmetry breaking bare parameter $m_\mathrm{cr}^\mathrm{sim}$ achieved at the time in which the Monte Carlo simulations have been performed is very accurate and, consequently, also the difference $m_\mathrm{cr}-m_\mathrm{cr}^\mathrm{sim}$ is very small, of the order of the statistical errors on $m_\mathrm{PCAC}$ (see Eq.~(\ref{eq:mpcaccondition})). By relying on these observations, we split the re-weighting factors $W_f(m_f,m_\mathrm{cr})$ defined in Eq.~(\ref{eq:reSweight2}) according to
\begin{flalign}
W_f(m_f,m_\mathrm{cr})=\hat W_f(m_\mathrm{cr})\, \bar W_f(m_f) \;,
\qquad
\hat W_f(m_\mathrm{cr})
=
\frac{
\mathcal{D}_\mathrm{OS}^f(m_f^\mathrm{sim},m_\mathrm{cr})
}{
\mathcal{D}_\mathrm{OS}^f(m_f^\mathrm{sim},m_\mathrm{cr}^\mathrm{sim})
}\;,
\qquad
\bar W_f(m_f)
=
\frac{
\mathcal{D}_\mathrm{OS}^f(m_f,m_\mathrm{cr})
}{
\mathcal{D}_\mathrm{OS}^f(m_f^\mathrm{sim},m_\mathrm{cr})
}\;,
\end{flalign}
and expanded them in powers of the differences
\begin{flalign}
\Delta m_f=m_f-m_f^\mathrm{sim}\;,
\qquad
\Delta m_\mathrm{cr}=m_\mathrm{cr}-m_\mathrm{cr}^\mathrm{sim}\;,
\label{eq:mdiffnot}
\end{flalign}
which are treated as being of equal order, $O(\Delta m)$, and we neglect $O(\Delta m^2)$ corrections. Therefore, the formulae used to evaluate the re-weighting factors are
\begin{flalign}
\hat W_f(m_\mathrm{cr}) = 1
+
\Delta m_\mathrm{cr}
\mathrm{Tr}
\left[
(-ir_f\gamma_5)
\frac{1}{D_\mathrm{OS}^\mathrm{sim}}
\right] + O(\Delta m)^2\;,
\qquad
\bar W_f(m_f) = 1
+
\Delta m_f
\mathrm{Tr}
\left[
\frac{1}{D_\mathrm{OS}^\mathrm{sim}}
\right] + O(\Delta m)^2\;,
\label{eq:Wexpansions}
\end{flalign}
where we use the compact operator notation
\begin{flalign}
D_\mathrm{OS}^\mathrm{sim}=
\gamma_\mu \bar{\nabla}_\mu[U]-ir_f\gamma_5 \left(
W^\mathrm{cl}[U]+m_\mathrm{cr}^\mathrm{sim} 
\right) +m_f^\mathrm{sim} \;.
\end{flalign}

These formulae have been employed to evaluate the sea quark contribution to the derivatives of physical observables w.r.t. the critical mass and the bare quark masses, 
as we are now going to explain. 

Any physical quantity $O$ is calculated by taking its path-integral expectation value according to
\begin{flalign}
O=\frac{
\left\langle O[U]\ \prod_f \hat W_f(m_\mathrm{cr})\, \bar W_f(m_f) \right\rangle^\mathrm{sim}
}{
\left\langle \prod_f \hat W_f(m_\mathrm{cr})\, \bar W_f(m_f) \right\rangle^\mathrm{sim}
}\;,
\end{flalign}
where the expectation value $\langle \cdot \rangle^\mathrm{sim}$ includes the simulated probabilistic weight $P^\mathrm{sim}[U]$ given in Eq.~(\ref{eq:Ssimweight}). By using Eq.~(\ref{eq:Wexpansions}) and by expanding the previous formula at first order w.r.t. the mass differences we have
\begin{flalign}
\label{eq:der_sea}
O=O^\mathrm{sim} 
+ \Delta m_\mathrm{cr}\ \partial_{cr}^\mathrm{sea} O
+
\sum_f \Delta m_f\ \partial_f^\mathrm{sea} O\;,
\end{flalign}
where $O^\mathrm{sim}=\langle O[U] \rangle^\mathrm{sim}$ while the derivative w.r.t. the sea critical mass is given by
\begin{flalign}
\label{eq:der_sea_mc}
\partial_{cr}^\mathrm{sea} O
=
\left\langle O[U]\ 
\sum_f
\mathrm{Tr}
\left[
(-ir_f\gamma_5)
\frac{1}{D_\mathrm{OS}^\mathrm{sim}}
\right]
\right\rangle^\mathrm{sim}
-
\left\langle O[U]
\right\rangle^\mathrm{sim}
\
\sum_f
\left\langle
\mathrm{Tr}
\left[
(-ir_f\gamma_5)
\frac{1}{D_\mathrm{OS}^\mathrm{sim}}
\right]
\right\rangle^\mathrm{sim}\;,
\end{flalign}
and the derivative w.r.t. the sea mass of the $f$-flavour is given by
\begin{flalign}
\label{eq:der_sea_mf}
\partial_f^\mathrm{sea} O
=
\left\langle O[U]\ 
\mathrm{Tr}
\left[
\frac{1}{D_\mathrm{OS}^\mathrm{sim}}
\right]
\right\rangle^\mathrm{sim}
-
\left\langle O[U]
\right\rangle^\mathrm{sim}
\
\left\langle
\mathrm{Tr}
\left[
\frac{1}{D_\mathrm{OS}^\mathrm{sim}}
\right]
\right\rangle^\mathrm{sim}\;.
\end{flalign}
We use Eqs.~(\ref{eq:der_sea})-(\ref{eq:der_sea_mf}) to calculate the sea-quark contribution to the variation of a physical quantity \( O \) under a small change \(\Delta m_{f}\) (or \(\Delta m_{\rm cr}\)) in the quark mass \( m_f \) (or the critical mass \( m_{\rm cr} \)).\footnote{For the charm sea-quark mass corrections, the results obtained using leading-order reweighting turn out to be too noisy to be used. We estimate the charm sea-quark mass derivative $\partial_{c}^{\rm sea}$ from the strange sea-quark mass derivative $\partial_{s}^{\rm sea}$, assuming the scaling $m_{c}\partial_{c}^{\rm sea} \sim m_{s}\partial_{s}^{\rm sea}$ (see the discussion in Appendix~\ref{sec:mistunings} for more details on this point).\label{charm_der_note}}
So far, our discussion has centered on the variation of \( O \) as the sea-quark masses and critical mass vary, with the valence quark masses entering \( O \) held constant.

To explain the procedure we implemented for determining \( m_{f}^{\rm iso} \) (for $f=\ell, s, c)$ and \( m_{\rm cr} \), it is helpful to separate the dependencies of the observable \( \mathcal{O} \) on the flavour quark masses ($m_f$) and the critical mass parameter $m_0$ that arise, after use of the Wick theorem at fixed gauge configuration, from the fermionic
determinant (sea quark effects) and from the relevant quark propagators (valence quark effects) and to show them explicitly.
In fact, in the intermediate steps for tuning \( m_{f}^{\rm iso} \) and \( m_{\rm cr} \) (see below), a given mass parameter, for practical reasons, can temporarily assume different values in the quark determinants (sea quark mass parameters) and in the quark propagators (valence quark mass parameters, which we label with a superscript ``${\rm val}$''). 
For this reason, we introduce the following notation:
\begin{align}
\label{eq:def_O_4args}
O( m_{f}, m_{0} | m^{\rm val}_{f}, m^{\rm val}_{0} ) \equiv \mathcal{O}^{\rm sim}(m^{\rm val}_{f}, m^{\rm val}_{0}) + (m_{0}-m_{0}^{\rm sim})\, \partial_{\rm cr}^{\rm sea}\,O + \sum_{f} (m_{f}-m_{f}^{\rm sim}) \, \partial_{f}^{\rm sea}O ~,
\end{align}
where on the l.h.s.\ the first set of variables in the argument
denotes the sea quark mass parameters and the second set refers to the valence quark mass parameters, while on the r.h.s.\
\( \mathcal{O}^{\rm sim}(m^{\rm val}_{f}, m^{\rm val}_{0}) \) denotes \( O \) computed with valence quark mass parameters 
\( m^{\rm val}_{f} \) and \( m^{\rm val}_{0} \) at sea quark
mass parameters \( m_{f}^{\rm sim} \) and \( m_{\rm cr}^{\rm sim} \) .

We are now in position to discuss the procedure that in practice we follow to match the Edinburgh/FLAG definition of isosymmetric QCD.
In order to determine the dependence of \( \mathcal{O}^{\rm sim}(m^{\rm val}_{f}, m^{\rm val}_{0}) \) on the {\em valence} quark masses \( m^{\rm val}_{f} \), in general we perform the inversion of the Dirac operator using a few values of \( m^{\rm val}_{f} \) for all quark flavors, while keeping the valence critical mass parameter fixed at $m^{\rm val}_{0}=m_{\rm cr}^{\rm sim}$. Instead, in order to determine the $m^{\rm val}_{0}$-dependence, we explicitly evaluate the derivative 
\begin{align}
\partial_{\rm cr}^{\rm val} O \equiv \partial_{m^{\rm val}_{0}}\mathcal{O}^{\rm sim}(m^{\rm val}_{f},m^{\rm val}_{0})~,
\end{align}
which, after setting $m_{0}^{\rm val}=m_{0}$, allows us to write
\begin{align}
\label{eq:final_O}
O( m_{f}, m_{0} | m^{\rm val}_{f}, m_{0} )  = \mathcal{O}^{\rm sim}(m^{\rm val}_{f}, m_{\rm cr}^{\rm sim}) + (m_{0}-m_{0}^{\rm sim})\,\left[  \partial_{\rm cr}^{\rm sea}\,O + \partial_{\rm cr}^{\rm val}\, O \right] + \sum_{f} (m_{f}-m_{f}^{\rm sim}) \, \partial_{f}^{\rm sea}O ~. 
\end{align}

The fine-tuning of $m_{\rm cr}$, i.e. the determination of the small mismatch $m_{\rm cr} - m_{\rm cr}^{\rm sim}$, can be carried out independently of the tuning of the quark masses $m_{f} \to m_{f}^{\rm iso}$. Indeed,
 on one hand $m_{\rm cr} = a^{-1} w_{\rm cr}(g_0^2, am_f)$,
 as determined from the condition~(\ref{eq:mpcaccondition}) in twisted mass lattice QCD, depends very weakly (actually only at O($a$) level for $m_{\rm cr}$) on the values of the individual quark masses $m_f$.
 On the other hand the values of $m_f^{\rm sim}$ at which the condition~(\ref{eq:mpcaccondition}) is imposed are quite close to the target 
 values $m_f^{\rm iso}$ and the value of $m_0^{\rm sim}$ determined in the early stages of the Monte Carlo simulation
 differ from zero by a small amount, which is always (in modulus) below $0.06 m_\ell^{\rm sim}$ and typically different from zero by 
 two to six standard deviations \cite{ETMCsim24}. For these reasons, within the linear approximation approach we follow here, the differences $\Delta m_f$, $f=\ell,s,c$ (see
 Eq.~(\ref{eq:mdiffnot}))
 can safely be treated as negligible second order effects in the determination of $\Delta m_{\rm cr} = m_{\rm cr} - m_{\rm cr}^{\rm sim}$. 
 
\begin{figure}
\centering
\includegraphics[scale=0.4]{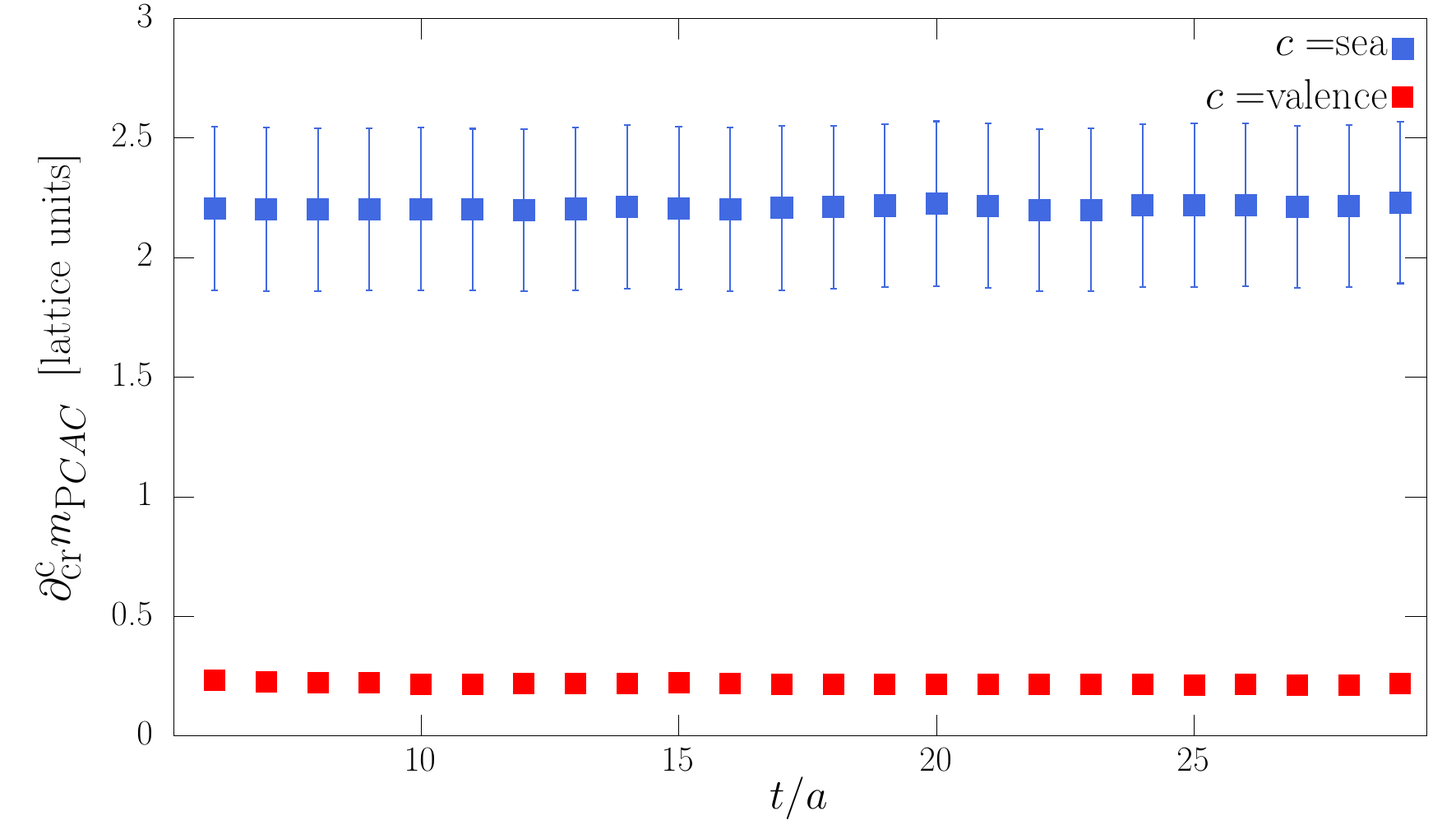}
\caption{\small\it Sea-quark (blue) and valence-quark (red) contribution to the derivative of the $m_{\rm PCAC}$ mass defined in Eq.~(\ref{eq:mpcaccondition}), as a function of the Euclidean time $t/a$ on the cB211.072.64 ensemble. \label{fig:mpcac_der} }
\end{figure}

Considering $O = m_{\rm PCAC}$, see Eq.~(\ref{eq:mpcaccondition}),
expressed in the notation of Eq.(\ref{eq:def_O_4args}), and making use of Eq.~(\ref{eq:final_O}),
we determine $m_{\rm cr}$ by solving the equation
\begin{align} 
m_{\rm PCAC}(m_{f}^{\rm sim} ; m_{\rm cr} | m_{f}^{\rm sim}, m_{\rm cr}) = m_{\rm PCAC}(m_{f}^{\rm sim}, m_{\rm cr}^{\rm sim} | m_{f}^{\rm sim} , m_{\rm cr}^{\rm sim}) +   (m_{\rm cr} - m_{\rm cr}^{\rm sim})\left[ \partial_{\rm cr}^{\rm sea}  + \partial_{\rm cr}^{\rm val}\right] m_{\rm PCAC} = 0 ~.
\end{align}
In Figure~\ref{fig:mpcac_der}, we present the derivatives of \( m_{\rm PCAC} \) with respect to the valence quark (\( \partial_{\rm cr}^{\rm val} \, m_{\rm PCAC} \)) and sea quark (\( \partial_{\rm cr}^{\rm sea} \, m_{\rm PCAC} \)) critical mass for the cB211.072.64 ensemble. As illustrated, both valence- and sea-quark contributions are of the same sign. However, the magnitude of the valence-quark contribution is approximately an order of magnitude smaller than that of the sea-quark contribution. The cB211.072.64 ensemble is the only one for which we have computed the valence-quark mass derivative \( \partial_{\rm cr}^{\rm val} \, m_{\rm PCAC} \). For the C-type and D-type ensembles, we focused on the dominant sea-quark derivative alone, introducing an additional \( 10\% \) uncertainty to account for the uncomputed valence-quark term. For the cE211.044.112 ensemble, where neither the valence- nor sea-quark contributions to the derivative are available, we estimated \( \left[ \partial_{\rm cr}^{\rm val} + \partial_{\rm cr}^{\rm sea} \right] m_{\rm PCAC} \) by using the D-type ensemble derivative as a central value. We then included a systematic uncertainty based on the observed difference between the C-type and D-type ensemble derivatives.

Having fine-tuned the critical mass by determining the small difference $m_{\rm cr}- m_{\rm cr}^{\rm sim}$ for all lattice spacings employed in the present analysis, we now proceed to discuss the conditions that determine the quark masses $m_{f}^{\rm iso}$ as needed to match our definition of isosymmetric QCD given in Eq.~(\ref{eq:iso_definition}). To simplify the notation, we set from now on
\begin{align}
O(m_{f}| m^{\rm val}_{f}) \equiv O(m_{f}, m_{\rm cr} | m^{\rm val}_{f}, m_{\rm cr})~.
\end{align}
From the theoretical perspective, the solution $m_{f=\ell,s,c}^{\rm iso}$ of our tuning problem is obtained by solving the following system of equations
\begin{flalign}
O^i( m_{f}^{\rm iso} \vert m_{f}^{\rm iso}) =  O^i(m_f^{\rm sim}\vert m_f^{\rm iso})
+
\sum_{f}\left[m_{f}^{\rm iso}- m_f^\mathrm{sim}\right]
\partial_f^\mathrm{sea}
O^i(m_f^\mathrm{sim}\vert m_{f}^{\rm iso})
=[O^i]^\mathrm{\rm iso}
\;,
\end{flalign}
where
\begin{flalign}
O=\left\{ 
\frac{aM_\pi}{aF_\pi},
\frac{aM_K}{aF_\pi},
\frac{aM_{D_s}}{aF_\pi} \right\}\;,
\end{flalign}
and then by defining the lattice spacing according to
\begin{flalign}
a^\mathrm{\rm iso} = \frac{aF_\pi(m_f^\mathrm{\rm iso}\vert m_f^\mathrm{\rm iso})}{F_\pi^\mathrm{\rm iso}}\;.
\end{flalign}

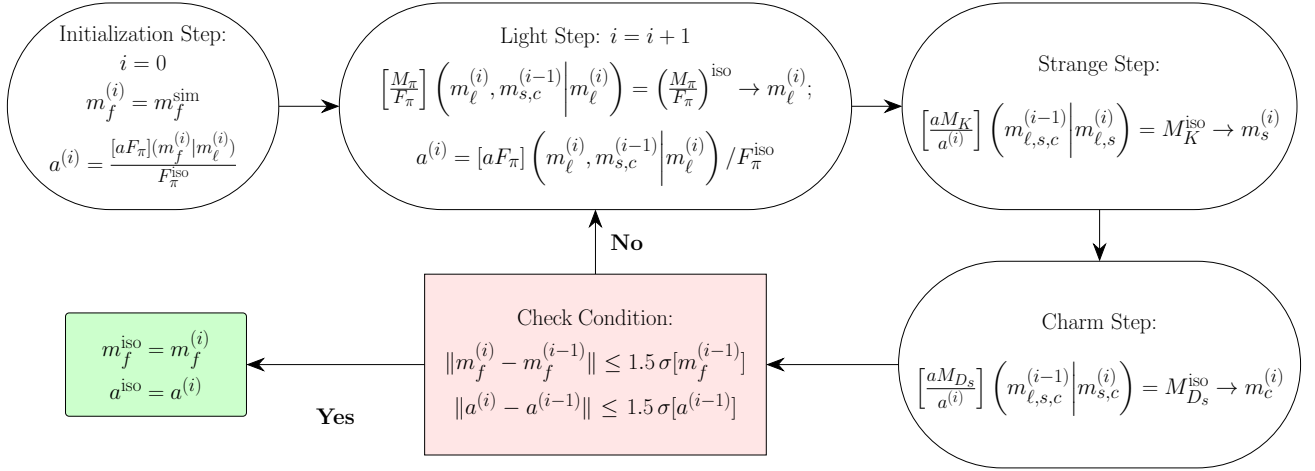
\begin{figure}[h!]
\centering
\resizebox{17.1cm}{!}{\hspace{-0.5cm}
\begin{tikzpicture}[
    node distance=20cm and 12cm, %
    every node/.style={draw, minimum height=8.0cm , minimum width=2cm, align=center, font=\large}, %
    process/.style={rounded rectangle, fill=white},
    lightstep/.style={rounded rectangle, fill=white, draw, minimum height=8.0cm , minimum width=2cm, align=center, font=\Huge}, %
    decision/.style={rectangle, aspect=2, fill=red!10, text width=13cm, minimum height=7cm, align=center, font=\huge},
    line/.style={draw, -{Stealth[scale=3]}, thick}, %
    iso/.style={draw, rectangle, rounded corners, minimum height=4cm, minimum width=7cm, fill=green!20, font=\large} %
]

\node[process] (init) {\Huge Initialization Step: \\[10pt] \Huge $i = 0$ \\ [12pt]  \Huge $m_f^{(i)} = m^{\rm sim}_{f}$ \\[10pt] \Huge $a^{(i)} = \frac{ [a {F_\pi}](m_f^{(i)}|m_\ell^{(i)})}{F_{\pi}^{\rm iso}}$};

\node[lightstep, right of=init, node distance=17.5cm] (light) { \Huge Light Step: $i = i + 1$ \\[15pt]  \Huge $\left[\frac{{M_\pi}}{F_{\pi}}\right]\left(m_{\ell}^{(i)}, m_{s,c}^{(i-1)} \bigg| m_{\ell}^{(i)} \right) = \left(\frac{M_\pi}{F_{\pi}}\right)^{\rm iso} \rightarrow m_\ell^{(i)}$; \\[15pt]
\Huge $a^{(i)} =  [a {F_\pi}]\left(m_{\ell}^{(i)}, m_{s,c}^{(i-1)}\bigg|m_\ell^{(i)}\right)/F_{\pi}^{\rm iso}$};

\node[process, right of=light, node distance=19.5cm] (strange) {\Huge Strange Step: \\[30pt] \Huge $\left[\frac{a{M_K}}{a^{(i)}}\right]\left( m_{\ell,s,c}^{(i-1)} \bigg| m_{\ell,s}^{(i)}  \right) = M_K^{\rm iso} \rightarrow m_s^{(i)}$};

\node[process, below of=strange, node distance=10cm] (charm) { \Huge Charm Step: \\[30pt] \Huge $\left[ \frac{a{M_{D_s}}}{a^{(i)}}\right]\left( m_{\ell,s,c}^{(i-1)} \bigg| m_{s,c}^{(i)} \right) = M_{D_s}^{\rm iso} \rightarrow m_c^{(i)}$};

\node[decision, left of=charm, node distance=19.5cm] (check) {\Huge Check Condition: \\[18pt] \Huge $\,\,\,\, \|m_f^{(i)} - m_f^{(i-1)}\| \leq 1.5  \, \sigma[ m_f^{(i-1)}]\,\,\,\,$ \\[10pt] \Huge $\,\,\,\, \|a^{(i)} - a^{(i-1)}\| \leq 1.5 \,\sigma[a^{(i-1)}]\,\,\,\,$};

\node[iso, left of=check, node distance = 17.0cm] (yes) {\Huge $m_f^{\rm iso} = m_f^{(i)}$ \\[8pt] \Huge $a^{\rm iso} = a^{(i)}$};

\path[line] (init) -- (light);
\path[line] (light) -- (strange);
\path[line] (strange) -- (charm);
\path[line] (charm) -- (check);
\path[line] (check) -- node[midway, left=0.1cm, below=-2cm,  font=\bfseries\Huge, draw=none, fill=none] {Yes} (yes); %
\path[line] (check.north) to[out=90, in=90, looseness=0] node[midway, above right=-5.5cm, right=0.2cm, font=\bfseries\Huge, draw=none, fill=none] {No} (light.south); %

\end{tikzpicture}}
\caption{\small\it Schematic description of the iterative algorithm used to determine $m_{f}^{\rm iso}$, which is described in detail in the text below. At each iteration, we update the values of the quark masses and of the lattice spacing until convergence is achieved. In the convergence condition, shown in the pink box, $\sigma[X]$ denotes the full (mostly statistical) error on $X$, for $X = m_f^{(i-1)},
a^{(i-1)}$.  \label{fig:iterative_procedure}}
\end{figure}
In practice, we solve the system by implementing an iterative procedure that we are now going to explain in detail. An illustrative
sketch of this procedure is also shown in 
Fig.~(\ref{fig:iterative_procedure}).

Before starting the iteration we have the following
\begin{flalign}
&
\mathtt{Initialization\ Step:}\qquad i=0~, \qquad
m_{f}^{(i)}=m_f^\mathrm{sim}\;,\qquad a^{(i)} \equiv \frac{aF_\pi\left( 
m_{\ell,s,c}^{(i)}
\vert 
m_\ell^{(i)} 
\right)
}{F_\pi^\mathrm{\rm iso}}\;,
\end{flalign}
where $i$ is the iteration index, and we are using the fact that $aF_{\pi}$ does not depend on the strange and charm valence masses. The iteration then starts and runs on the flavour index that we order from lighter to heavier, namely $f=\ell, s , c$. To improve the tuning of the light quark mass and of the lattice spacing we implement the following
\begin{align}
\label{eq:light_step}
\mathtt{Light\ Step:} \qquad & i=i+1\;,\qquad  \left[\frac{aM_\pi}{aF_\pi}\right] 
\left( 
m_{\ell}^{(i)}, m_{s,c}^{(i-1)}
\vert 
m_\ell^{(i)} 
\right)
=
\left[\frac{M_\pi}{F_\pi}\right]^\mathrm{\rm iso}
\quad
\longrightarrow 
\quad
m_\ell^{(i)}~,\nonumber\\[10pt]
& a^{(i)} \equiv \frac{aF_\pi\left( 
m_{\ell}^{(i)}, m_{s,c}^{(i-1)}
\vert 
m_\ell^{(i)} 
\right)
}{F_\pi^\mathrm{\rm iso}}\;,
\end{align}
where again we are using the fact that $aM_\pi$ and $aF_{\pi}$ do not depend upon the strange and charm valence masses. 

In all steps, as already mentioned in the text above, the valence quark masses are changed by performing the needed valence-quark propagator inversions required to compute the input observables. 

To improve the tuning of the strange quark mass we then implement the following
\begin{flalign}
&
\mathtt{Strange\ Step:} \qquad \left[\frac{aM_K}{a^{(i)}}\right]
\left( 
m_{\ell,s,c}^{(i-1)}
\vert 
m_{\ell,s}^{(i)}
\right)
=
M_K^\mathrm{\rm iso
}
\quad
\longrightarrow 
\quad
m_s^{(i)}\;.
\end{flalign}
Here we are using the fact that $aM_K$ does not depend upon the charm valence masses. 

To improve the tuning of the charm quark mass we then implement the following %
\begin{flalign}
&
\mathtt{Charm\ Step:} \qquad \left[\frac{aM_{D_s}}{a^{(i)}}\right]
\left( 
m_{\ell,s,c}^{(i-1)}, 
\vert 
m_{s,c}^{(i)} 
\right)
=
M_{D_s}^\mathrm{\rm iso}
\quad
\longrightarrow 
\quad
m_c^{(i)}\;.
\end{flalign}
At the end of the \texttt{charm-step} the lattice spacing and all quark masses are updated,
and we check whether we reach the target precision given by the following convergence condition
\begin{flalign}
\label{eq:convergence}
&
\mathtt{Check\ Condition:}
\nonumber \\[8pt]
&
\mathtt{if}
\qquad \left\|m_f^{(i)}-m_f^{(i-1)}\right\| \le \frac{3}{2}\,\sigma[ m_f^{(i-1)} ] \,\,\,\mathtt{and}\,\,\, \left\|a^{(i)}-a^{(i-1)}\right\| \le \frac{3}{2}\,\sigma [a^{(i-1)}]
\nonumber \\[8pt]
&
m_f^\mathrm{\rm iso}=m_f^{(i)}\;,
\quad
a^\mathrm{\rm iso}=a^{(i)}\;,
\nonumber \\[8pt]
&
\mathtt{end}
\nonumber \\[8pt]
&
\mathtt{else\quad goto\quad light-step}
\end{flalign}
where $\sigma [m_f^{(i-1)}]$ and $\sigma [a^{(i-1)}]$  are the statistical errors on the tuned quark masses and on the lattice spacing.

It turns out that at the end of the first iteration, $i=1$, the convergence condition is not satisfied. The values of the simulated sea-quark masses $m_{f}^{\rm sim}= m_{f}^{(0)}$ differ from the quark masses $m_{f}^{(1)}$ by a few percent. The difference is between $2-7\%$ for $m_{\ell}^{(1)}-m_{\ell}^{(0)}$, and between $2-4\%$ for $m_{s}^{(1)}-m_{s}^{(0)}$ and $m_{c}^{(1)}-m_{c}^{(0)}$. We therefore continue to $i=2$, incorporating the sea-quark mass corrections through the linear-reweighting approximation of Eq.~(\ref{eq:def_O_4args}). The sea-quark mass corrections to $aM_{\pi}, aF_{\pi}, aM_{K}$ and $aM_{D_{s}}$ turn out to be extremely tiny. Compared to the statistical errors of the uncorrected quantities, they amount to at most $1.5\sigma$ for $aF_{\pi}$, about $1\sigma$ for $aM_{\pi}$ and are completely negligible for $aM_{K}$ and $aM_{D_{s}}$. This allows us to verify the condition in Eq.~(\ref{eq:convergence}) and to exit the loop at the end of the iteration $i=2$. 

We now give a separate description of the tuning steps described above and present our final results for the quark masses $m_{f}^{\rm iso}$. 

\begin{itemize}
\item \textsc{Light Step}: To determine $m_{\ell}^{(i)}$ and $a^{(i)}$ at each step $i$, we employ two methods: the \textit{direct} and \textit{global fit} approaches. In the \textit{direct} approach, starting from the values of $aF_{\pi}$ and $aM_{\pi}$ on the nearly-physical ensembles of Table~\ref{tab:simudetails}, we apply Eq.~(\ref{eq:def_O_4args}) to vary the light sea-quark mass, while in order to be able to perform the corresponding change in the valence sector we have produced data for $aF_{\pi}$ and $aM_{\pi}$ at a second value of $m_{\ell}^{\rm val} < m_{\ell}^{\rm sim}$. Since the nearly-physical ensembles of Table~\ref{tab:simudetails} have pion masses which differ from $M_{\pi}^{\rm iso}$ at most by $5~{\rm MeV}$, the valence- and sea-quark mass corrections needed to match the Edinburgh/FLAG conditions are small, making the linear reweighting approximation reliable. To confirm this, we carry out a second type of analysis, based on a global fit of $aF_{\pi}$ and $M_{\pi}/F_{\pi}$, exploiting all the ensemble in Table~\ref{tab:simudetails}, included those with larger-than-physical pion mass. The \textit{global-fit} analysis has been already discussed in Appendix A (Section 2) of Ref.~\cite{ExtendedTwistedMass:2022jpw}, to which we refer for additional details. In a nutshell, for this analysis we perform a  global fit of $aF_{\pi}$ and $am_{\ell}$, using all the ensembles of Table~\ref{tab:simudetails}, according to the following ChPT-inspired Ansatz
\begin{align}
\label{eq:global_fit}
aF_{\pi}(\xi_{\pi}, \beta) &=  a\overline{F}_{\pi}(\beta) \cdot \left\{1-2\xi_{\pi}\log(\xi_{\pi}/\xi_{\pi}^{\rm iso})+ [P+P_{disc} (aF_\pi(\xi_{\pi},\beta))^2] (\xi_{\pi}-\xi^{\rm iso}_{\pi})\right\} \nonumber \\[10pt]
am_{\ell}(\xi_{\pi},\beta) &= a\overline{m}_{\ell}(\beta) \frac{\xi_{\pi}}{\xi_{\pi}^{\rm iso}} \cdot \left\{ 1 + 5\xi_{\pi}\log(\xi_{\pi}/\xi_{\pi}^{\rm iso}) + (B+a^2B_{disc})(\xi_{\pi} -\xi_{\pi}^{\rm iso}) \right\}^{-1}~,
\end{align}
where $\beta = 6/g_0^2$ is the inverse QCD gauge coupling and 
\begin{align}
\xi_{\pi} = \frac{ M_{\pi}^{2}}{16\pi^{2} F_{\pi}^{2}}~.
\end{align}
In Eq.~(\ref{eq:global_fit}), 
$a\overline{F}_{\pi}(\beta)$, $a\overline{m}_{\ell}(\beta)$, $P$, $P_{\rm disc}$, $B$ and $B_{disc}$ are free-fit parameters (for $a\overline{F}_{\pi}(\beta)$ and $a\overline{m}_{\ell(\beta)}$ there is a different fit parameter for each $\beta$). In the global fit method, at each iteration step, the (bare) light quark mass and the lattice spacing corresponding to the Edinburgh/FLAG definition of isosymmetric QCD are given, for each value of the coupling $\beta$,  by
\begin{align}
a^{(i)}(\beta) = \frac{ a\overline{F}_{\pi}(\beta)}{F_{\pi}^{\rm iso}}~, \qquad  m_{\ell}^{(i)}(\beta) = \overline{m}_{\ell}(\beta)~.
\end{align}
The hadronic observables $F_{\pi}$ and $M_{\pi}$ (and hence $\xi_{\pi}$) entering Eq.~(\ref{eq:global_fit}) are intended as infinite-volume quantities. The needed infinite-volume extrapolation of our lattice data is carried out by employing next-to-leading-order ChPT, i.e. making use of the Gasser-Leutwyler formulae~\cite{Gasser:1983yg}. We have checked that next-to-leading-order ChPT nicely describes the spread between the $F_{\pi}$ and $M_{\pi}$ values on the ensembles of Table~\ref{tab:simudetails} produced at different values of the spatial volume, but equal values of $\beta$ and quark masses. 

The advantage of the global fit approach is that it captures the dependence of both \( F_{\pi} \) and \( M_{\pi}/F_{\pi} \) on \( m_{\ell} \) without relying on leading-order reweighting, instead utilizing gauge ensembles generated away from the physical point (we emphasize that all data contributing to the global fit in Eq.~(\ref{eq:global_fit}) were produced at different values of $m_{\ell}$, but for each ensemble, the valence light-quark mass is equal to the sea light-quark mass).

\begin{figure}[t!]
    \centering
    \includegraphics[width=0.45\linewidth]{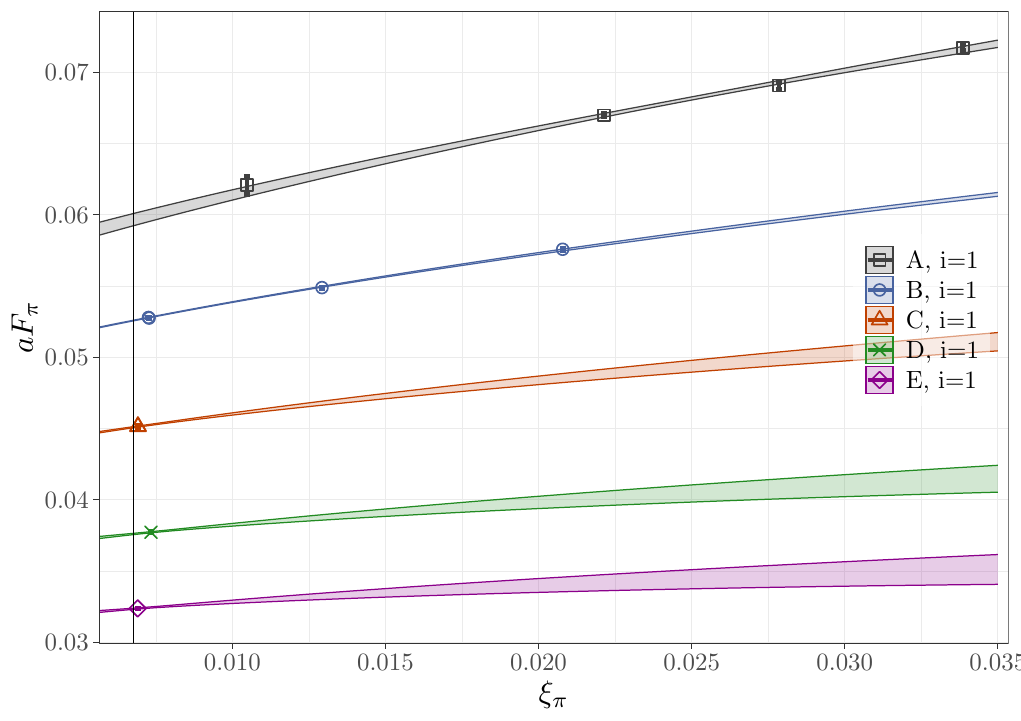}
    \includegraphics[width=0.45\linewidth]{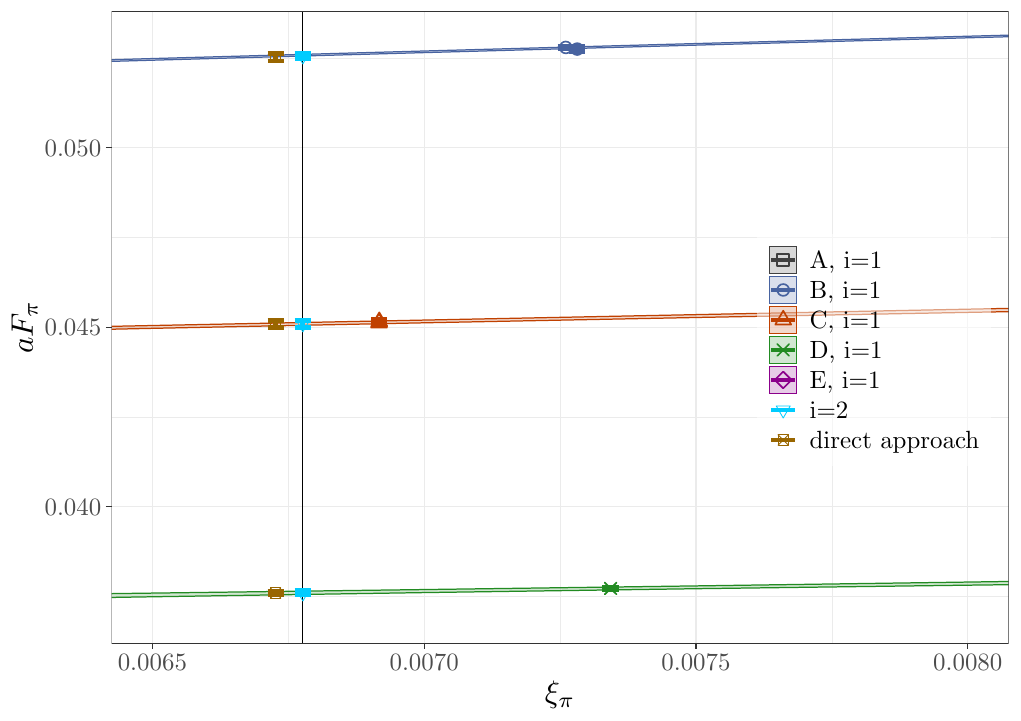} \\
    \hspace{0.15cm}
    \includegraphics[width=0.444\linewidth]{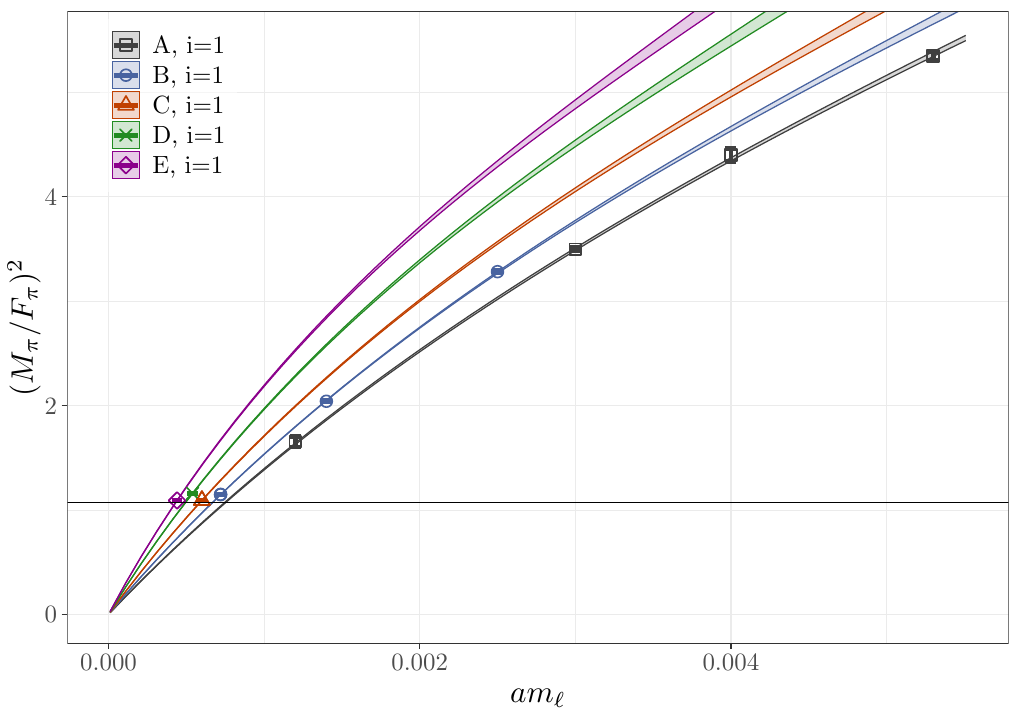}
    \includegraphics[width=0.444\linewidth]{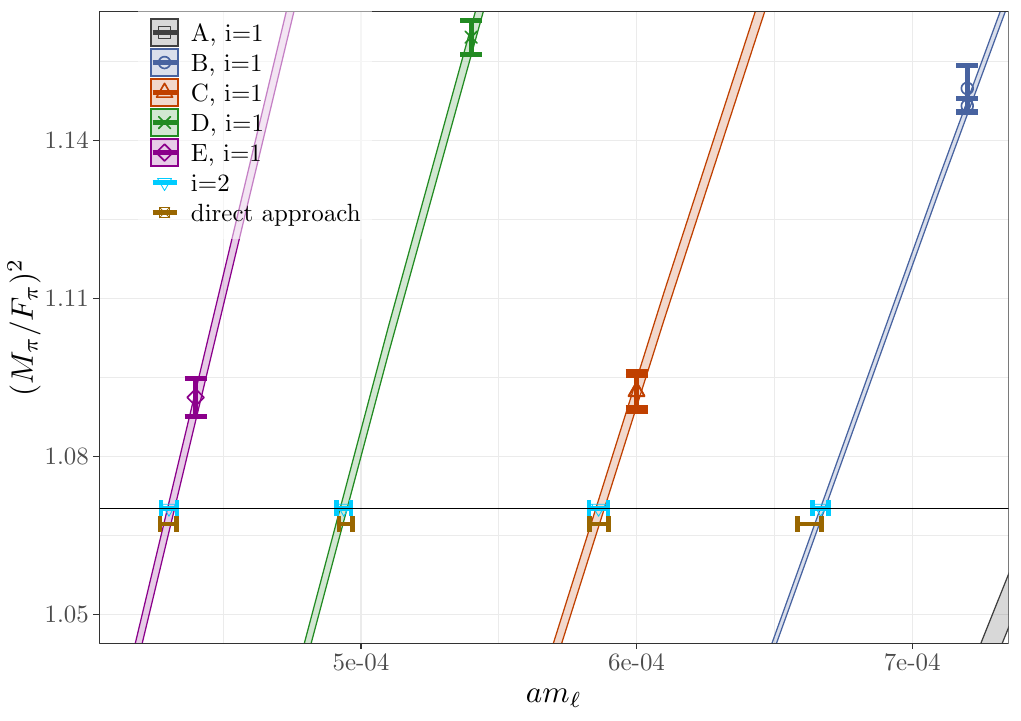}
    \caption{\small\it Results of the global fit to $a F_{\pi}$ (top panels) and $am_{\ell}$ (bottom panels). The coloured bands correspond to the best fit curve obtained in the global fit at each value of $\beta$, while the datapoints to our determination of $aF_{\pi}$ and $(M_{\pi}/F_{\pi})^{2}$ on each of the gauge ensembles of Table~\ref{tab:simudetails}. The rightmost plots show a zoom in the region close to the physical point, where, for comparison, we include the result obtained using the direct approach at the iteration $i=1$ (datapoints in brown), and the final result after the iteration $i=2$ (datapoints in sky-blue), i.e. after inserting the charm and strange sea-quark mass corrections making use of Eq.~(\ref{eq:def_O_4args}). In these plots, the vertical and horizontal lines correspond to the point $\xi_{\pi}= \xi_{\pi}^{\rm iso}$. In the bottom-right (top-right) plot the datapoints in brown, corresponding to the results of the direct approach at the iteration $i=1$, are slightly shifted vertically (horizontally) for better visualization.  
    \label{fig:scale_setting}}
\end{figure}

In Figure~\ref{fig:scale_setting}, we show the result of the global fit and compare it with that of the direct approach, which as already stressed, only uses the nearly-physical ensembles of Table~\ref{tab:simudetails}, and relies on leading-order reweighting to describe the light sea-quark mass dependence of $M_{\pi}$ and $F_{\pi}$. The reduced $\chi^{2}$ of the global fit to $aF_{\pi}$ and $am_{\ell}$ is very good, about $0.5$. The results in the two figures correspond to the global fit performed during the first light-step iteration ($i=1$), where the strange and charm sea-quark masses are held fixed to their simulation value $m_{s,c}^{\rm sim}$. As the figures show, the agreement between the two approaches is excellent; the difference between the two determinations of $m_{\ell}^{(1)}$ and $a^{(1)}$ is smaller than the statistical uncertainty. 

Two remarks are relevant here. First, as it is shown by Eq.~(\ref{eq:light_step}), during the $i$-th light-step the strange and charm quark masses are held fixed to the value obtained at the $(i-1)$-th iteration, and therefore for $i=1$ they are set to the simulated values $m_{s,c}^{(i=1)}= m_{s,c}^{\rm sim}$. At the end of the first iteration we do not reach convergence, and therefore at the second iteration ($i=2$), we proceed by adding the strange and charm sea-quark corrections to $F_{\pi}$ and $M_{\pi}$ according to Eq.~(\ref{eq:def_O_4args}). These corrections have basically the only effects of increasing the uncertainty on the determination of $m_{\ell}$ and of the lattice spacing. In the left panels of Figure~\ref{fig:scale_setting} we add, for comparison, the results obtained for $m_{\ell}^{(2)}$ and $a^{(2)}F_{\pi}^{\rm iso}$, which then also correspond to our final results for $m_{\ell}^{\rm iso}$ and  $a^{\rm iso}$. Indeed, after the second iteration we achieve convergence, and hence exit the loop. 

The second comment concerns the corrections to $F_{\pi}$ and $M_{\pi}$ that stem from the small-mistuning, $m_{\rm cr}- m_{\rm cr}^{\rm sim}$, of the critical mass. These corrections 
may of course be included by adding the second term 
in Eq.~(\ref{eq:final_O}). 
In the specific case of $M_{\pi}$ and $F_{\pi}$, however, to include such corrections, one can also profit from the existence of the following analytic expression~\cite{ExtendedTwistedMass:2021qui}, for the leading, $\mathcal{O}(a^{0})$, dependence of $M_{\pi}$ and $F_{\pi}$ on the $m_{\rm PCAC}$ mass
\begin{align}
\label{eq:an_corr_mpcac_fpi_mpi}
 F_{\pi}(m_{\rm PCAC}=0) &=F_{\pi}(m_{\rm PCAC})\sqrt{1+(Z_A m_{\rm PCAC}/ m_\ell)^2} \nonumber \\[12pt]
    M_{\pi}^2(m_{\rm PCAC}=0) &= \frac{M_{\pi}^2(m_{\rm PCAC})}{\sqrt{1+(Z_A m_{\rm PCAC}/m_\ell)^2}}\,,
\end{align}
where $Z_{A}$ (see Appendix~\ref{sec:renormalization}) is the renormalization constant of the axial current. We have employed the analytic expressions~(\ref{eq:an_corr_mpcac_fpi_mpi}) to correct $F_{\pi}$ and $M_{\pi}$ on all the ensembles of Table~\ref{tab:simudetails}, included those with higher-than-physical pion masses, for which a first-principle estimate of the critical mass derivative in Eq.~(\ref{eq:final_O}) is not available to us. We have explicitly checked, in the case of the nearly-physical ensembles of Table~\ref{tab:simudetails}, that the corrections produced by the analytic expressions above, agree within uncertainties with the numerical results
from the second term in the r.h.s. of Eq.~(\ref{eq:final_O}).

\item \textsc{Strange Step}:
The determination of the strange-quark mass $m_{s}^{\rm iso}$ (as well as of the charm quark mass to be discussed in the next bullet-point), turns out to be substantially less involved than the determination of $m_{\ell}^{\rm iso}$ and of the lattice spacing.

For this analysis, we have employed only the nearly-physical ensemble of Table~\ref{tab:simudetails}, which as already remarked, are the only ones entering
the analysis of $a_{\mu}^{\rm HVP}(s)$ and $a_{\mu}^{\rm HVP}(c)$. At each iteration, to be able to vary the valence light and strange quark masses we have performed the inversion of the two-point pseudoscalar light-strange correlator employing two values of $m_{s}^{\rm val}$ and two values of $m_{\ell}^{\rm val}$ (one of these two values is $m_{\ell}^{\rm val}= m_{\ell}^{\rm sim}$). The critical mass corrections due to the small difference $m_{\rm cr}- m_{\rm cr}^{\rm sim}$ have been included by adding the second term of Eq.~(\ref{eq:def_O_4args}) for $O= M_{K}$. This correction, however, turns out to be much smaller than the statistical errors and can safely be neglected. During the first iteration $i=1$, the light, strange, and charm quark masses are fixed to their simulation value and therefore no correction has been applied. During the second iteration, $i=2$, we have applied these corrections adding the last term in Eq.~(\ref{eq:def_O_4args}). Moreover, all the 
sea-quark mass corrections turn out to be extremely tiny and negligible. The only (small) difference between $m_{s}^{(1)}$ and $m_{s}^{(2)}$, comes from the slightly increased uncertainties on the lattice spacing. As already pointed out, at the end of the second iteration, we achieve convergence and exit the loop. 

\begin{figure}[t!]
\centering
\includegraphics[scale=0.355]{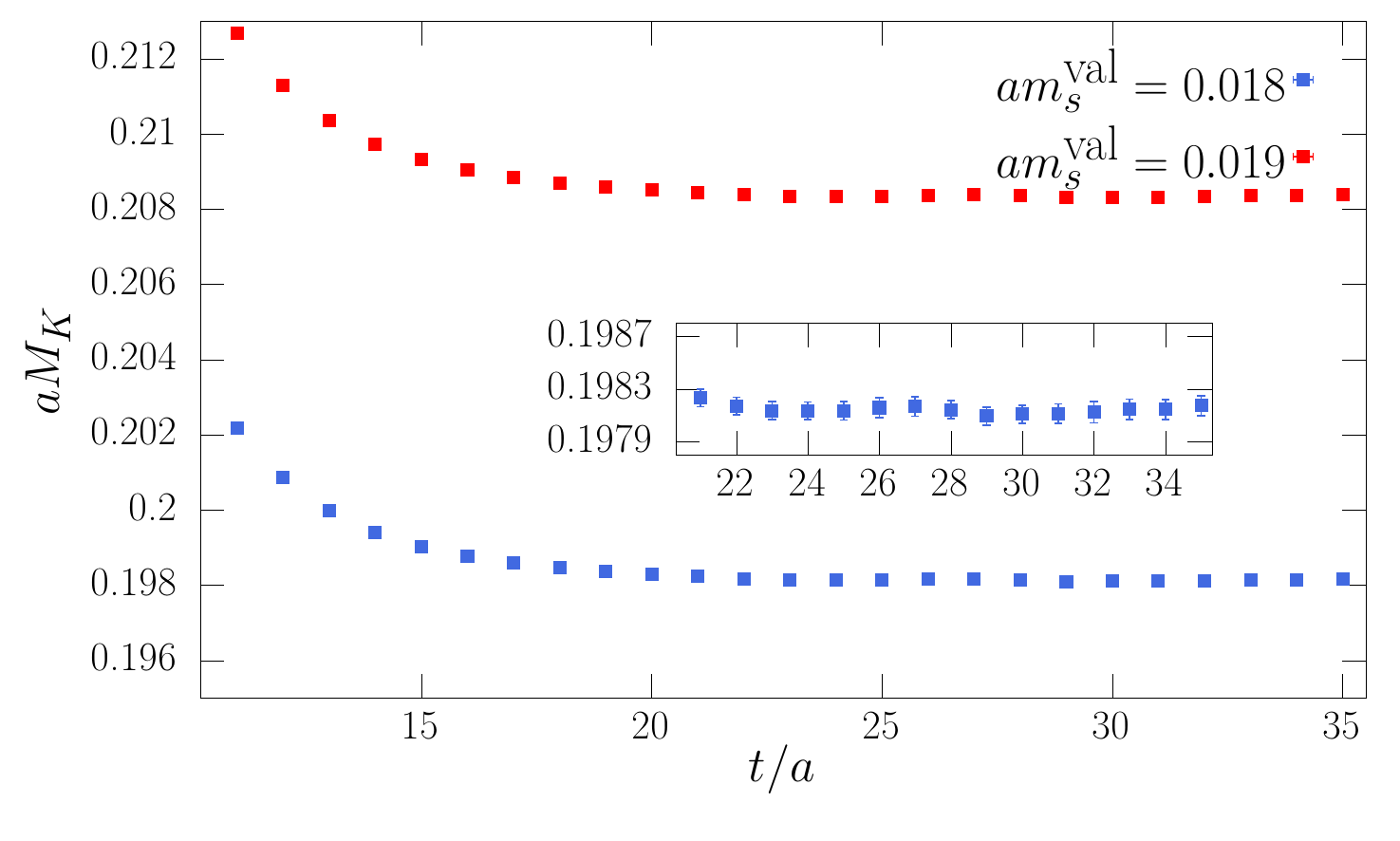}
\includegraphics[scale=0.355]{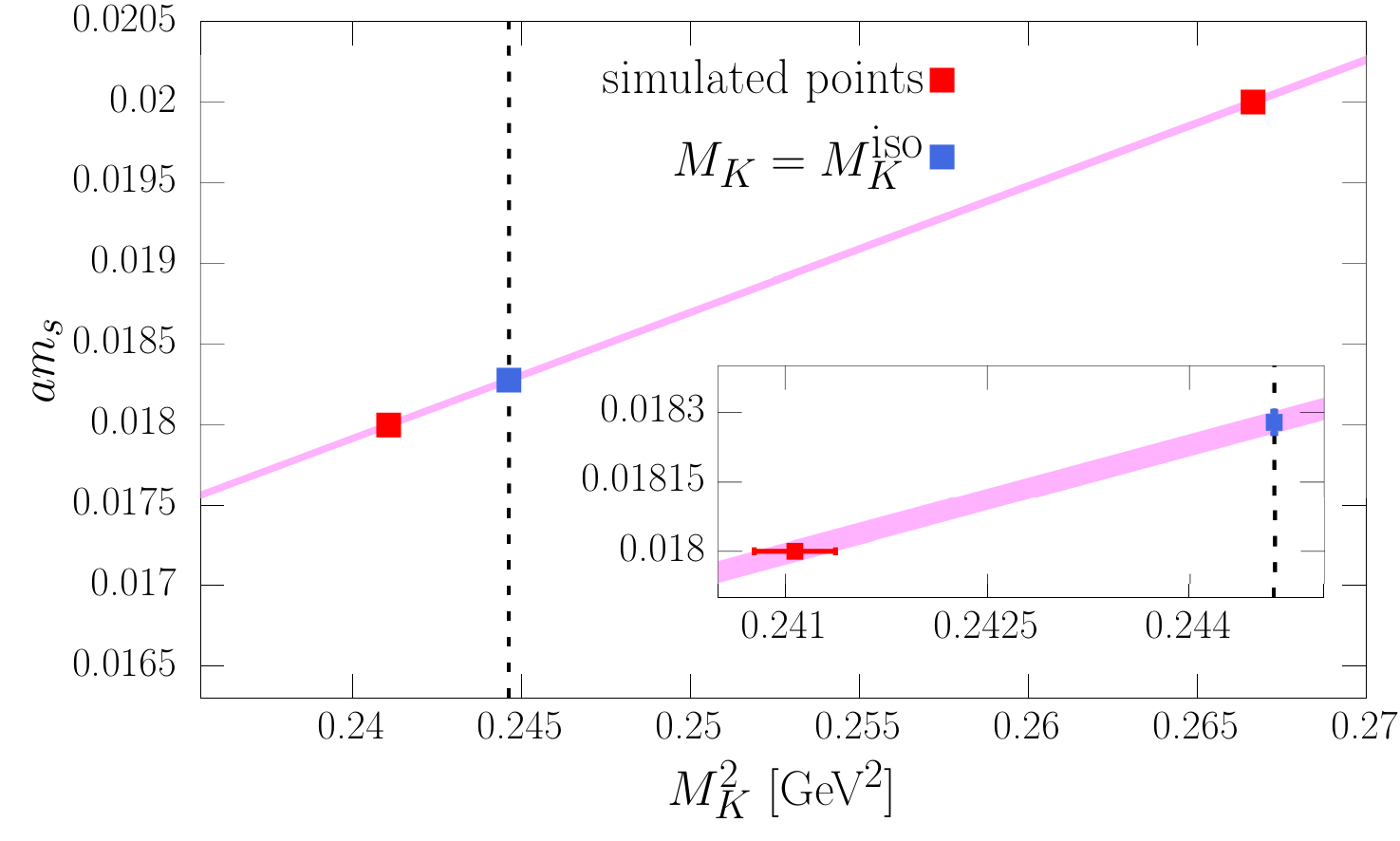}
\caption{\small\it Left: effective mass of the kaon for the two simulated values of the valence strange quark mass $am_{s}^{\rm val} = 0.018,0.019$ on the cB211.072.64 ensemble. Right: interpolation of the strange quark mass $am_{s}$ to the physical point defined by $M_{K}=M_{K}^{\rm iso}=494.6~{\rm MeV}$, indicated in the plot by the vertical dashed line. The interpolation has been performed according to the Ansatz of Eq.~(\ref{eq:ams_ansatz}).  \label{fig:ams}}
\end{figure}

To illustrate the quality of the effective-mass plateaus, we show in the left panel of Figure~\ref{fig:ams} the effective mass of the kaon, determined on the cB211b.072.64 ensemble, for the two different values of the valence strange quark mass employed for this calculation and for $m_{\ell}^{\rm val}= m_{\ell}^{\rm sim}$. The panel on the right shows instead the result of the final strange-mass interpolation needed to impose $M_{K} = M_{K}^{\rm iso}$, which we performed according to the following ChPT-inspired Ansatz
\begin{align}
\label{eq:ams_ansatz}
m_{s} = A + B M_{K}^{2}~,
\end{align}
where  $A$ and $B$ are fit parameters.

\item \textsc{Charm Step}: For charm mass tuning we have included in the analysis only the nearly-physical ensembles of Table~\ref{tab:simudetails}. At each iteration, we vary the valence strange and charm quark masses and evaluate the two-point pseudoscalar strange-charm correlator employing up to two values of $m_{s}^{\rm val}$ and up to three values of $m_{c}^{\rm val}$.\footnote{On the cC211.060.112, cD211.054.96 and cE211.044.112 ensembles, we have considered a single value of $m_{s}^{\rm val}$, carefully chosen to be very close to $m_{s}^{(1)} \sim m_{s}^{(2)} = m_{s}^{\rm iso}$. We have checked, using our determination of the slope $\partial M_{D_s} /\partial m_{s}^{\rm val}$ on the ensembles where results at two different values of the valence strange-quark mass are available, that the small difference $m_{s}^{\rm val} -m_{s}^{(i)}$, produces a completely negligible impact on the determination of $m_{c}^{(i)}$.}

For $aM_{D_{s}}$, all the sea-quark mass and critical mass corrections, which have been included during the second iteration step $i=2$ making use of Eq.~(\ref{eq:def_O_4args}), turn out to be negligible within statistical uncertainty, and, as in the case of the strange-quark mass, the only difference between $m_{c}^{(1)}$ and $m_{c}^{(2)}$ comes from the slightly increased lattice spacing uncertainties. 

\begin{figure}
\centering
\includegraphics[scale=0.355]{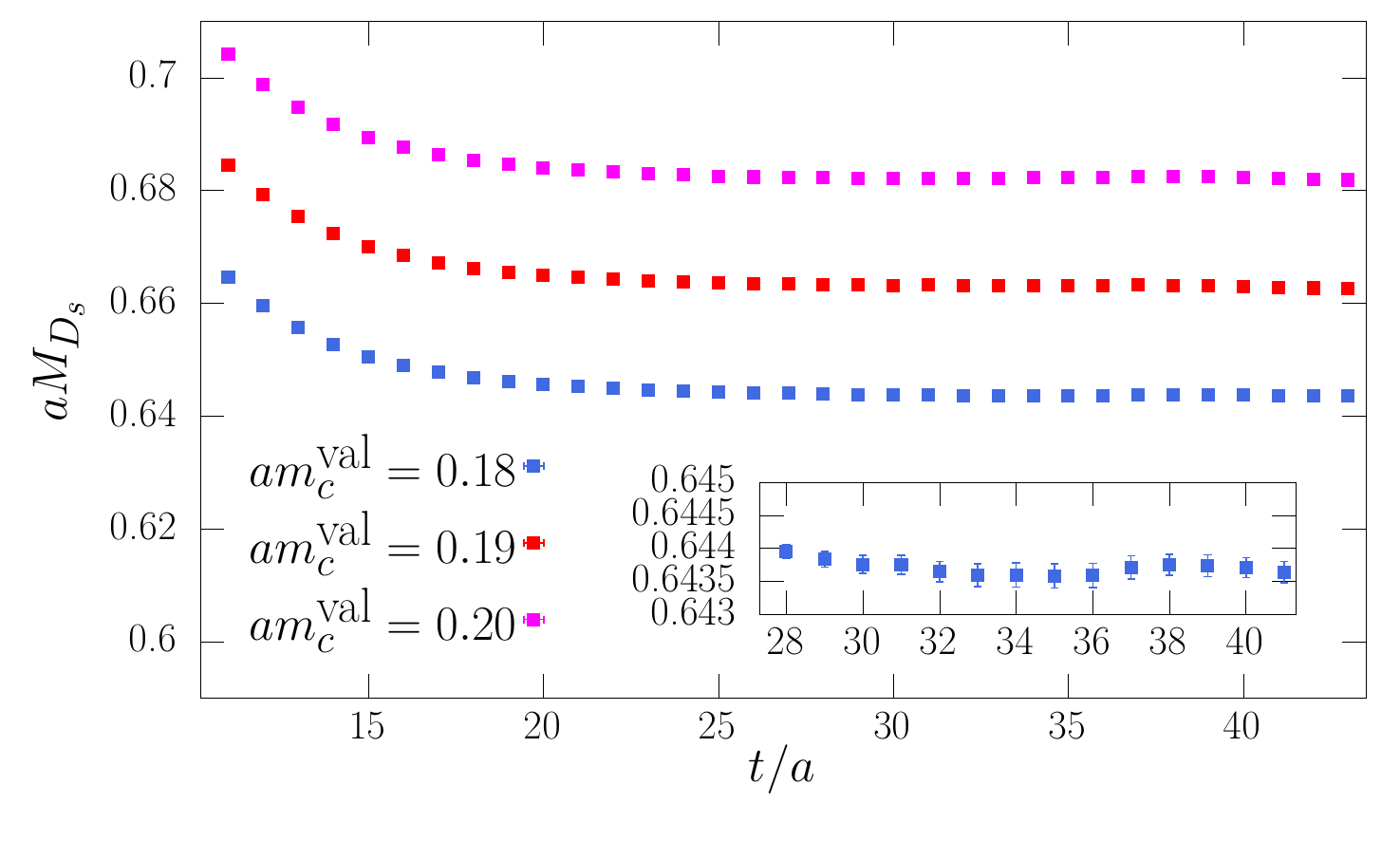}
\includegraphics[scale=0.355]{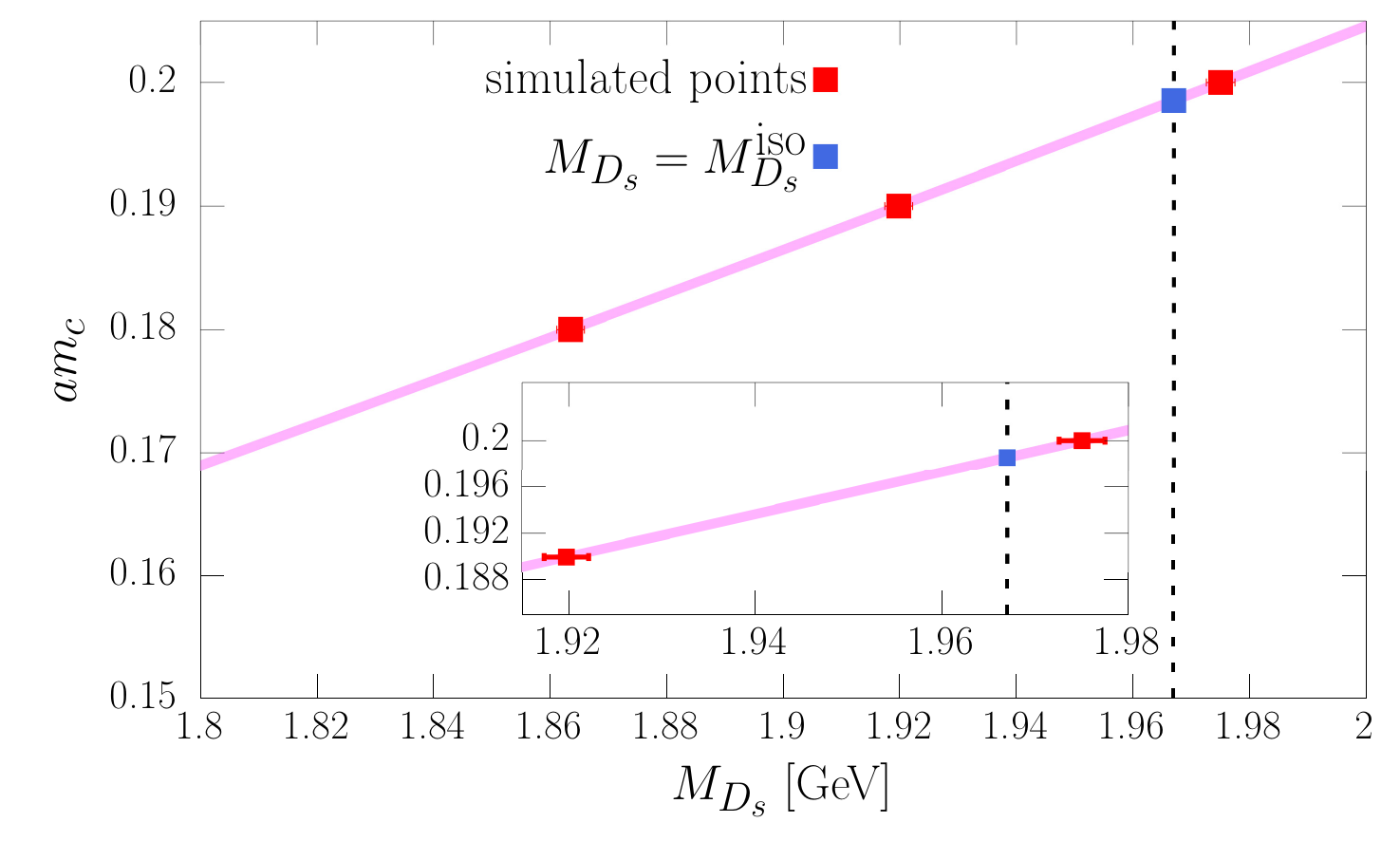}
\caption{\small\it Left: effective mass of the $D_{s}$ meson for the three simulated values of the valence charm quark mass $am_{c}^{\rm val} = 0.18, 0.19$ and $0.20$ on the cC211.060.112 ensemble. Right: interpolation of the charm quark mass $am_{c}$ to the physical point defined by $M_{D_{s}}=M_{D_{s}}^{\rm iso}=1967~{\rm MeV}$, indicated in the plot by the vertical dashed line. The interpolation is performed according to the Ansatz of Eq.~(\ref{eq:amc_ansatz}) including also a quadratic term proportional to $M_{D_{s}}^{2}$. However, the results at the physical point $M_{D_{s}}=M_{D_{s}}^{\rm iso}$, remain unchanged if a simple linear interpolation to the two rightmost red data points in the figure is performed.  \label{fig:amc}} 
\end{figure}

To illustrate the quality of the effective-mass plateaus, we show in the left panel of Figure~\ref{fig:amc} the effective mass of the $D_{s}$ meson, determined on the cC211.060.112 ensemble, for the three different values of the charm quark mass and for the single value of the valence quark mass, $m_{s}^{\rm val} \sim m_{s}^{\rm iso}$, that we employed. The right panel plot shows instead the result of the final interpolation needed to impose $M_{D_{s}} = M_{D_{s}}^{\rm iso}$, which was performed according to the following Ansatz
\begin{align}
\label{eq:amc_ansatz}
m_{c} = A + BM_{D_{s}}~,
\end{align}
where $A$ and $B$ are fit parameters. We have checked, on the ensembles where three different valence charm quark masses are employed, namely the cC211.060.112, the cD211.054.96, and the cE211.044.112 ensemble, that the inclusion of a quadratic term proportional to $M_{D_{s}}^{2}$ in the interpolation, produces a negligible change in the value of $m_{c}^{(i)}$. 
\end{itemize}

Having described all the tuning steps of the iterative procedure that allow us to match the Edinburgh/FLAG isosymmetric world, Table~\ref{tab:iso_EDI_FLAG} collects the resulting values of the lattice spacing $a^{\rm iso}$,  the critical mass $am_{\rm cr}$ and the quark masses $am_{\ell,s,c}^{\rm iso}$ for the four lattice spacings used in the calculation of $a_{\mu}^{\rm HVP}(s)$ and $a_{\mu}^{\rm HVP}(c)$. The values quoted in Table\,\ref{tab:iso_EDI_FLAG} for the lattice spacing have slightly larger uncertainties than the ones given in Ref.~\cite{tau}, due to the improved analysis of the sea-quark mistuning effects. However, the slightly increased uncertainty on the lattice spacing does not have any impact on the final values and errors quoted in Ref.~\cite{tau}.

We conclude this section with a brief description of the two discretized versions of the electromagnetic current (called in the main text the TM and the OS currents) employed for the evaluation of $a_{\mu}^{\rm HVP}(s)$ and $a_{\mu}^{\rm HVP}(c)$.  To define the TM bilinear operators, starting from the action in Eq.~(\ref{eq:Sfull}),
we introduce additional valence quark fields with no feedback on the gauge effective action. This amounts to i) including, for each quark field $q_{f}$, $f=\{u,d,s,c\}$ appearing in 
Eq.~(\ref{eq:OS_valence_quark_action}), a replica field $q'_{f}$ with the same soft mass, $m'_f = m_f$, but opposite value of the Wilson parameter, $r'_f=-r_f$ and hence opposite critical mass, $m_{\rm cr}(r'_f) = -m_{\rm cr}(r_f) $; and 
ii) adding the corresponding ghost field $\phi'_{f}$, in order to remove any contributions of these extra fields to the fermionic determinants, i.e. to the lattice gauge effective action. In summary, in order to define the TM bilinears we add to the action $S$ of Eq.~(\ref{eq:Sfull}) the action term~\footnote{Note the different sign, as compared to 
Eq.~(\ref{eq:OS_valence_quark_action}), in front of the critical Wilson term with coefficient $i\gamma_5$, which is due to $r'_f = -r_f$.}
\begin{align}
\label{eq:action_replica}
S^{\textrm{rep}} &= \sum_{f}\sum_{x} \bar{q}'_f\left\{
\gamma_\mu \bar{\nabla}_\mu[U]+ir_f\gamma_5 \left(
W^\mathrm{cl}[U]+m_\mathrm{cr} 
\right) +m_f
\right\}q'_f + \nonumber \\
&+ \sum_{f}
\sum_x \bar{\phi}'_f\left\{
\gamma_\mu \bar{\nabla}_\mu[U] +ir_f\gamma_5 \left(
W^\mathrm{cl}[U]+m_\mathrm{cr} 
\right) +m_f
\right\}\phi'_f \; , 
\end{align}
i.e.\ we work with the mixed lattice action $S^{\textrm{mixed}}=S+ S^{\textrm{rep}}$.
The TM quark bilinears are then defined as
\begin{align}
J^{\rm TM}_{\Gamma}(x) = \bar{q}_{f}(x) \Gamma q'_{f}(x)~,
\end{align}
with $\Gamma$ a generic Dirac matrix, and in particular the TM electromagnetic current  is given by
\begin{align}
J_{f}^{\mu,\rm{TM}}(x) = Z_{A} q_{\mathrm{em},f} \, \bar{q}_{f}(x) \gamma^{\mu}  q'_{f}(x)~, \qquad f=\{ s,c\}~,
\end{align}
where $Z_{A}$ is the renormalization constant of the axial current. To define the corresponding OS version, we do not need in principle to introduce additional replica valence fields, since the OS bilinears are constructed in terms of the quark fields $q_{f}$ entering the OS quark action defined in Eq.~(\ref
{eq:OS_valence_quark_action}). However, in order to single out the connected part of a two-point correlation function of OS bilinears, we find it convenient to add to the action $S$ of Eq.~(\ref{eq:Sfull}) the valence quark action term of Eq.~(\ref{eq:action_replica}) with the replacement $r'_{f} \to -r'_{f} = -r_{f}$. In terms of the quark fields $q_{f}$ and of the replica valence fields $q'_{f}$, with $r'_{f}=r_{f}$, the OS electromagnetic current is thus given by
\begin{align}
J_{f}^{\mu,\rm{OS}}(x) = Z_{V} q_{\mathrm{em},f} \, \bar{q}_{f}(x) \gamma^{\mu}  q'_{f}(x)~, \qquad f=\{s,c\}~,
\end{align}
where $Z_{V}$ is the renormalization constant of the vector current (see Appendix~\ref{sec:renormalization}). The vector correlators $V_{f}^{\rm TM}$ and $V_{f}^{\rm OS}$ entering in the rh.s.\ of Eq.~(\ref{eq:defalatt}) are then defined as
\begin{align}
   \label{eq:VV_correlators}
   V_{f}^{\rm{TM}}(t)  \; & \equiv  \; \frac{1}{3} \sum_{\bf x} \sum_{i=1,2,3} \left\langle J_{f}^{i, \rm{TM}}(x) [J_{f}^{i, \rm{TM}}]^\dagger(0) \right\rangle  ~ , ~ \nonumber \\[2mm]  
    V_{f}^{\rm{OS}}(t)  \; & \equiv  \; \frac{1}{3} \sum_{\bf x} \sum_{i=1,2,3} \left\langle J_{f}^{i, \rm{OS}}(x) [J_{f}^{i, \rm{OS}}]^\dagger(0) \right\rangle  ~ , 
\end{align}
Two-point correlation functions constructed in terms of the TM or OS bilinears, such as $V_{f}^{\rm TM}(t)$ and $V_{f}^{\rm OS
}(t)$, produce equivalent results in the continuum limit~\cite{Frezzotti:2004wz}. At non-zero values of the lattice spacing $a$, they however differ by $\mathcal{O}(a^{2})$ UV cutoff effects and, as explained in the main text, we exploit this fact to perform joint fits of the results obtained with the two regularizations while enforcing a common continuum-limit.

\section{Evaluating the HVP at the isoQCD point}
\label{sec:mass_interpolation_amu}

To evaluate the strange and charm HVP, along with the corresponding SD, LD and W contributions, at the Edinburgh/FLAG isoQCD point (discussed in Appendix~\ref{sec:masses}), we exploited all gauge ensembles in the upper part of Table~\ref{tab:simudetails}. For each ensemble, we performed the inversions of the vector TM and OS correlators of Eq.~(\ref{eq:VV_correlators}), for
different valence quark masses $m_{s}^{\rm val}$ and $m_{c}^{\rm val}$,  while fixing sea-quark masses and critical mass to $m_{f}^{\rm sim}$ and $m_{\rm cr}^{\rm sim}$, respectively. For each $\beta$, we carefully chose two values for $m_{s}^{\rm val}$ and three values for $m_{c}^{\rm val}$, in such a way that they
are always rather close to $m_{s}^{\rm iso}$ and $m_{c}^{\rm iso}$. 

Adopting the notation of Eq.~(\ref{eq:def_O_4args}), we first performed the required valence quark-mass interpolation of both $a_{\mu}^{\rm HVP,w}(s)(m_{f}^{\rm sim}, m_{\rm cr}^{\rm sim} | m_{s}^{\rm val} , m_{\rm cr}^{\rm sim})$ and $a_{\mu}^{\rm HVP,w}(c)(m_{f}^{\rm sim}, m_{\rm cr}^{\rm sim} | m_{c}^{\rm val} , m_{\rm cr}^{\rm sim})$, using the following Ansatz
\begin{align}
   \label{eq:int_amu_ansatz}
     a_{\mu}^{\rm HVP,w}(c)(m_{f}^{\rm sim}, m_{\rm cr}^{\rm sim} | m_{c}^{\rm val} , m_{\rm cr}^{\rm sim}) &= 
    a_{\mu}^{\rm HVP,w}(c)(m_{f}^{\rm sim}, m_{\rm cr}^{\rm sim} | m_{c}^{\rm iso} , m_{\rm cr}^{\rm sim}) \cdot \left[ 1 + A_c^w \left( m_c^{\rm val} - m_c^{\rm iso} \right)
    + B_c^w \left( m_c^{\rm val} - m_c^{\rm iso} \right)^{2} \right] \nonumber \\[12pt]
a_{\mu}^{\rm HVP,w}(s)(m_{f}^{\rm sim}, m_{\rm cr}^{\rm sim} | m_{s}^{\rm val} , m_{\rm cr}^{\rm sim}) &= 
    a_{\mu}^{\rm HVP,w}(s)(m_{f}^{\rm sim}, m_{\rm cr}^{\rm sim} | m_{s}^{\rm iso} , m_{\rm cr}^{\rm sim})\cdot \left[ 1 + A_s^w \left( m_s^{\rm val} - m_s^{\rm iso} \right) \right] ~,
\end{align}
where $A_{s}^{w}, A_{c}^{w}$ and $B_{c}^{w}$, as well as $a_{\mu}^{\rm HVP,w}(s)(m_{f}^{\rm sim}, m_{\rm cr}^{\rm sim} | m_{s}^{\rm iso} , m_{\rm cr}^{\rm sim})$ and $a_{\mu}^{\rm HVP,w}(c)(m_{f}^{\rm sim}, m_{\rm cr}^{\rm sim} | m_{c}^{\rm iso} , m_{\rm cr}^{\rm sim})$, are fit parameters. 

After interpolating $m_{s}^{\rm val} \to m_{s}^{\rm iso}$, and $m_{c}^{\rm val} \to m_{c}^{\rm iso}$, we applied the corrections needed to incorporate the effects of the fine-tuning of the critical mass and of the sea-quark mass parameters. As discussed in Appendix~\ref{sec:masses}, leading-order reweighting has been used to evaluate the sea-quark corrections, ultimately allowing us to determine the HVP at our isosymmetric point of reference by using Eq.~(\ref{eq:final_O}) with $m_{f}^{\rm val} = m_{f}^{\rm iso}$ and
\begin{align}
O = \left\{ a_{\mu}^{\rm HVP,w}(s), a_{\mu}^{\rm HVP,w}(c)              \right\}~.
\end{align}
Appendix~\ref{sec:mistunings} provides a detailed discussion of the evaluation of the sea-quark mass and critical mass derivatives appearing in the r.h.s. of Eq.~(\ref{eq:final_O}).

\section{Estimating effects of mistunings of simulation parameters}
\label{sec:mistunings}

In this section, we give some details on the calculation of the sea-quark mass corrections to the physical observables $O$ relevant for the present analysis, which we performed employing the leading-order reweighting discussed in Appendix~\ref{sec:masses}, i.e. making use of the formula in Eq.~(\ref{eq:der_sea}).  

\begin{figure}
\includegraphics[scale=0.28]{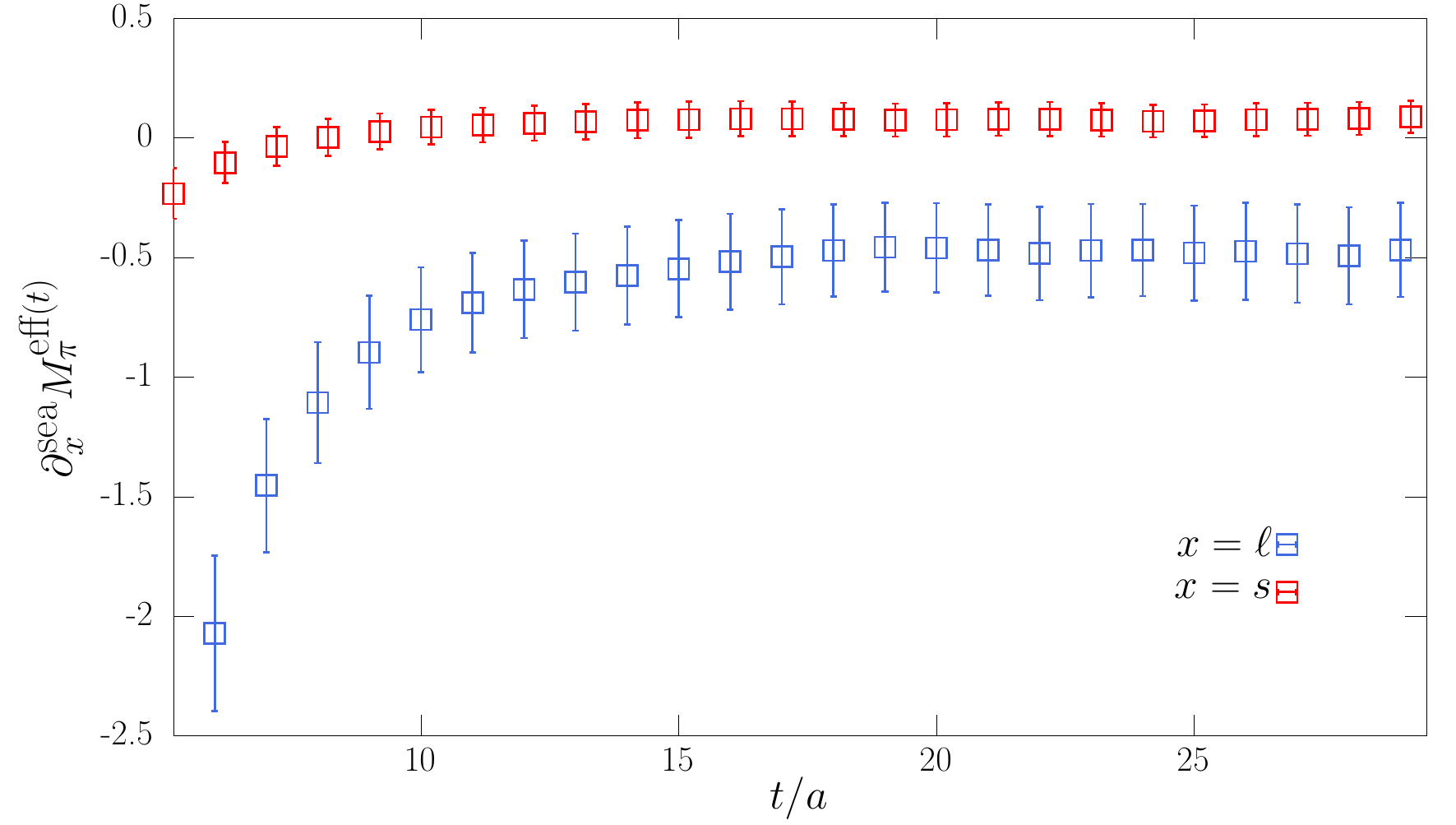}
\includegraphics[scale=0.28]{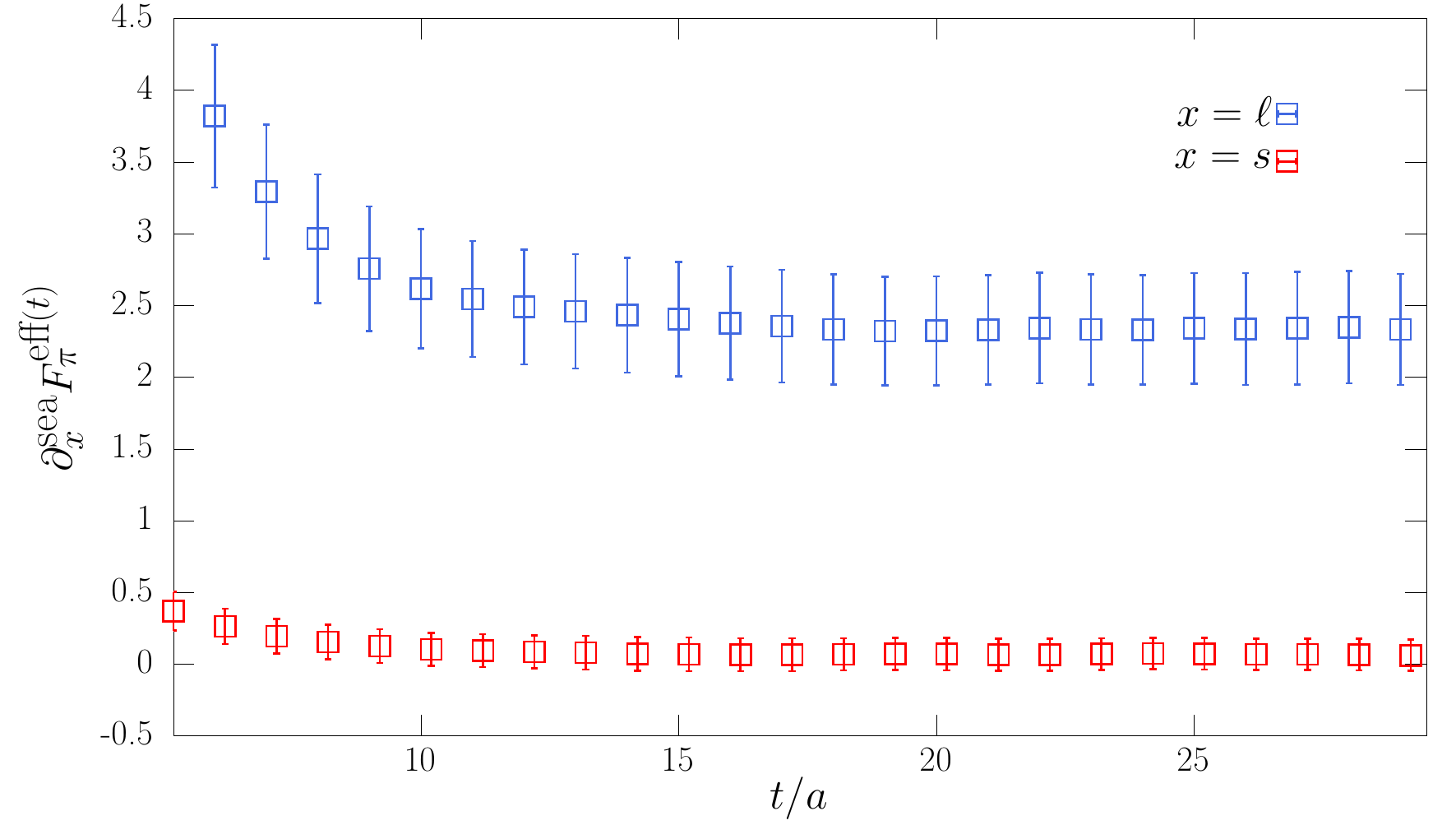}
\caption{\small\it The light and strange sea-quark mass derivatives of the effective pion mass (left) and of the effective pion decay constant (right), as a function of the Euclidean time on the cB211.072.64 ensemble. \label{fig:sea_der_mpi_fpi}}
\end{figure}

\begin{figure}
\includegraphics[scale=0.28]{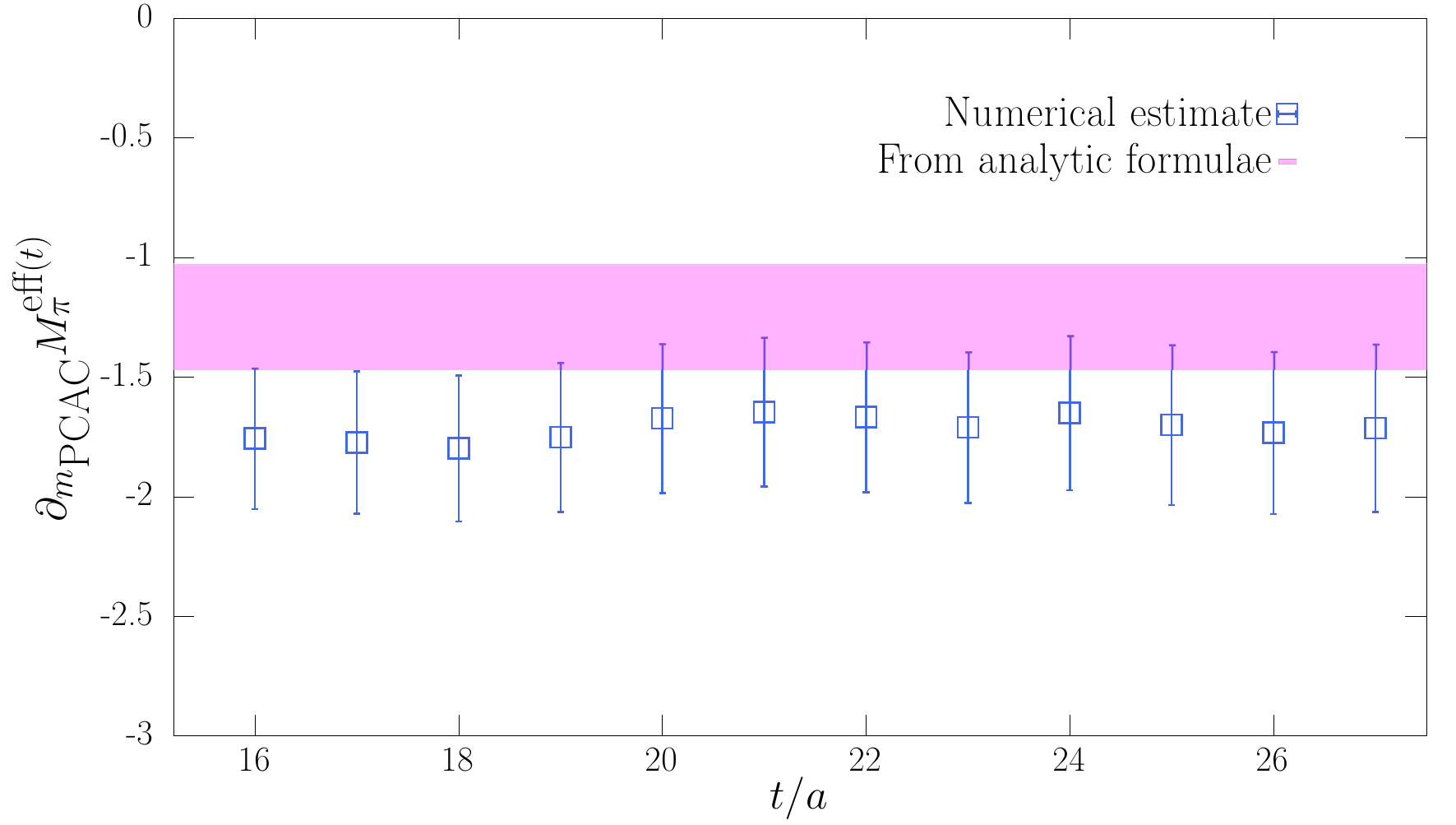}
\includegraphics[scale=0.28]{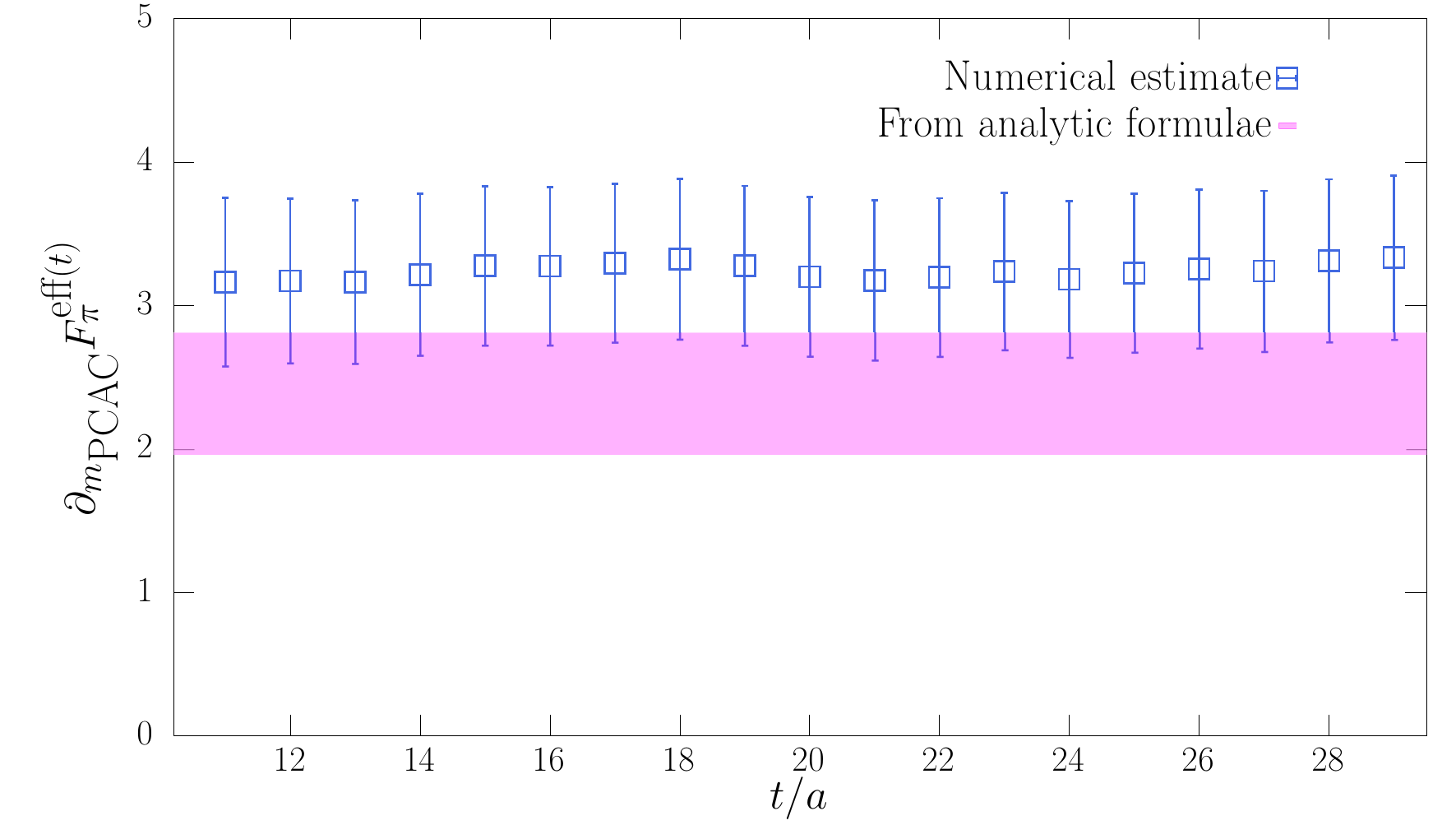}
\caption{\small\it The $m_{\rm PCAC}$-derivative of the effective pion mass (left) and of the effective pion decay constant (right) on the cC211.060.80 ensemble. The magenta bands correspond to the predictions of the analytic formulae of Eq.~(\ref{eq:an_corr_mpcac_fpi_mpi}).  \label{fig:sea_mcr_der_mpi_fpi} }
\end{figure}

\begin{figure}
\includegraphics[scale=0.28]{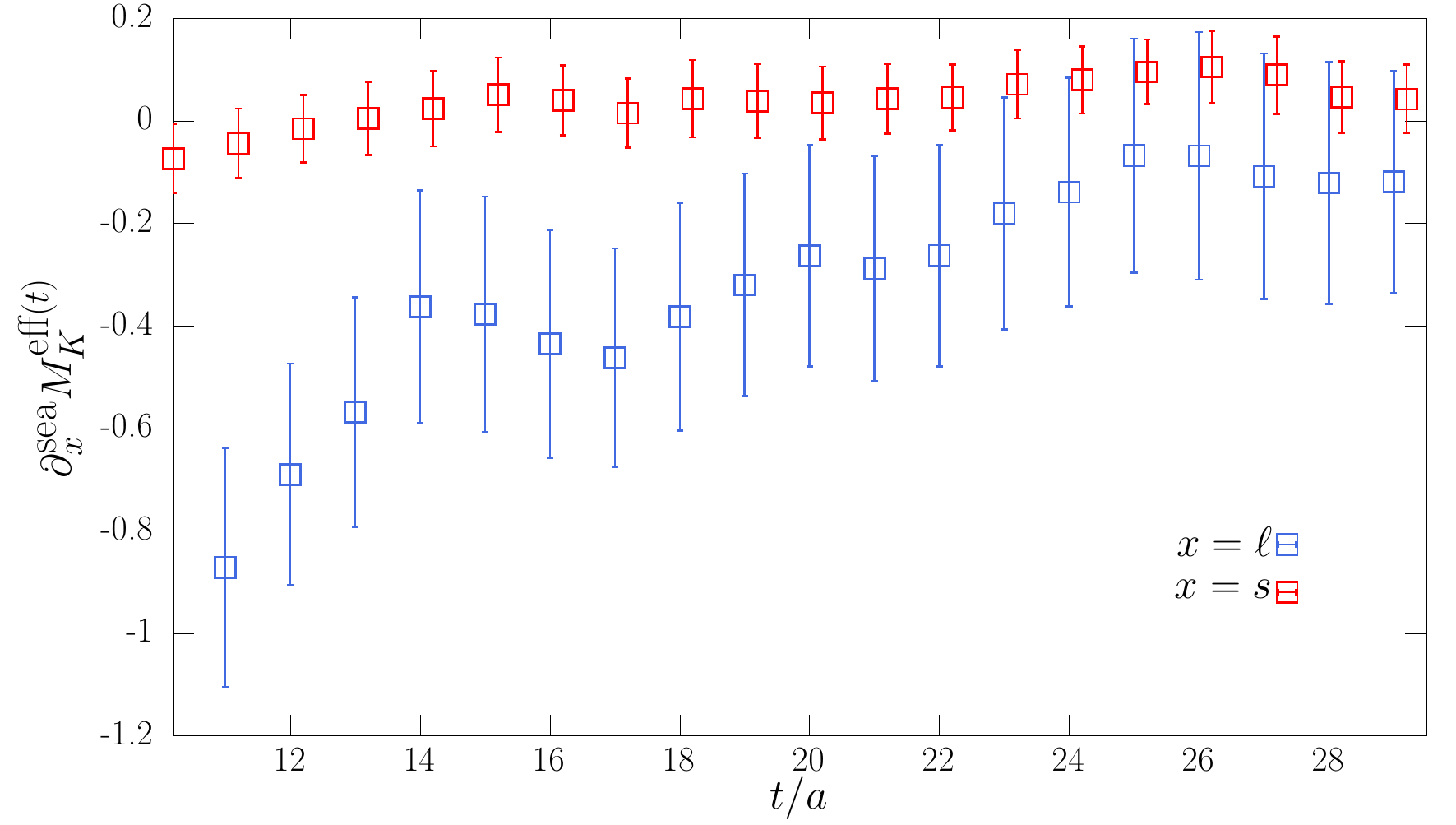}
\includegraphics[scale=0.28]{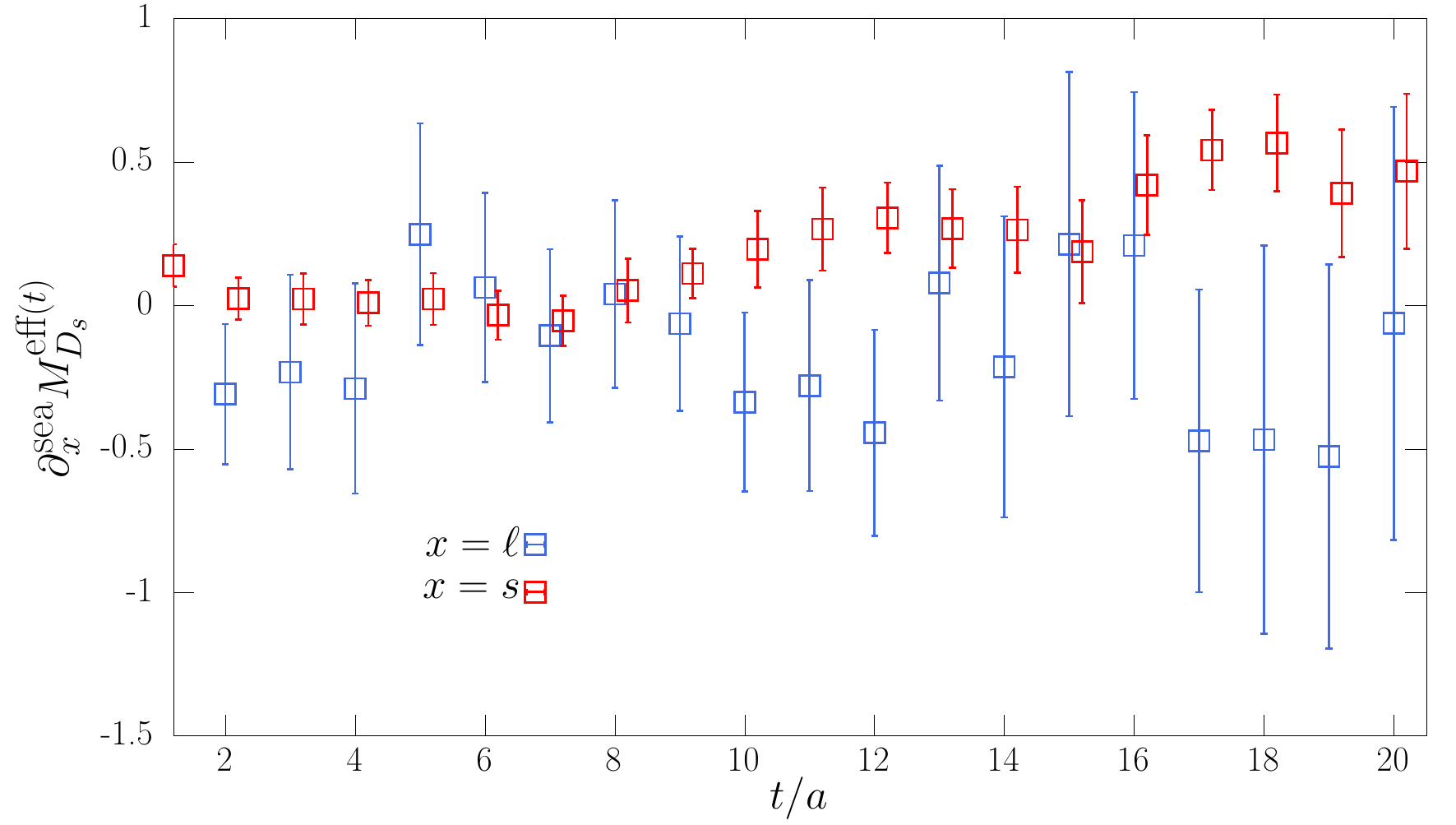}
\caption{\small\it The light and strange sea-quark mass derivatives of the effective pion mass (left) and of the effective $D_{s}$-meson mass (right), as a function of the Euclidean time on the cB211.072.64 ensemble. \label{fig:sea_der_mk_mds}}
\end{figure}

The observables $O$, of which we discuss the sea-quark mass corrections in this appendix, are:
\begin{align}
O = \left\{ M_{\pi} , F_{\pi}, M_{K}, M_{D_{s}}, a_{\mu}^{\rm HVP}(s), a_{\mu}^{\rm HVP}(c) \right\}~.
\end{align}
We start from the observables used to determine the quark masses $m_{f}^{\rm iso}$ and to set the scale, namely $M_{\pi}, F_{\pi}, M_{K}$ and $M_{D_{s}}$. 
In Figure~\ref{fig:sea_der_mpi_fpi} we show the results for the light and strange sea-quark mass derivatives of the effective pion mass and decay constant, $\partial_{\ell,s}^{\rm sea}M_{\pi}^{\rm eff}(t)$ and $\partial_{\ell,s}^{\rm sea}F_{\pi}^{\rm eff}(t)$, as obtained on the cB211.072.64 ensemble. 
As the figure shows, the quark-mass derivative, as expected,
strongly decreases as the quark mass increases (i.e. going from light to strange). The light-quark derivative is in absolute value larger for $F_{\pi}$ than for $M_{\pi}$, by a factor of about 5. The largest mistuning $\Delta m_{\ell} = m_{\ell} - m_{f}^{\rm sim}$ occurs on the cB211.072.64 and cD211.054.96 ensembles, for which $\Delta m_{\ell} \simeq 0.16~{\rm MeV}$, which then gives rise to a correction to $F_{\pi}$ of about $0.3-0.4~{\rm MeV}$, and smaller than $0.1~{\rm MeV}$ for $M_{\pi}$. The signal-to-noise in the strange-quark mass derivative is sensibly smaller than in the light-quark one. The typical size of the derivatives are $|\partial_{s}^{\rm sea} M_{\pi} | \sim 0.08$ and  $|\partial_{s}^{\rm sea} F_{\pi} | \sim 0.12$. The largest mistuning $\Delta m_{s} = m_{s} - m_{s}^{\rm sim}$ occurs on the cC211.060.80 ensemble where $\Delta m_{s} \sim -1.7~{\rm MeV}$, which produces on this ensemble a  correction to $M_{\pi}$ of about $0.13~{\rm MeV}$ and of about $0.2~{\rm MeV}$ to $F_{\pi}$, i.e. at the level of $0.1\%$ and $0.15\%$, respectively. For the charm quark, as already mentioned in footnote~\ref{charm_der_note}, the sea-quark derivatives turn out to be too noisy to provide a useful determination of the corresponding mistuning correction. For all observables $O$, we include in the analysis an estimate of the derivative $\partial_{c}^{\rm sea} O$, assuming the following approximate scaling of the sea quark mass derivative with the quark mass
\begin{align}
\label{eq:charm_der_est}
\partial_{c}^{\rm sea} O \; \simeq \;  \frac{m_{s}}{m_{c}} \, \partial_{s}^{\rm sea} O~.
\end{align}
The $1/m_{c}$ suppression of the derivative with respect to the sea quark mass of a generic observable $O$ is expected in the (approximatively well realized) limit $m_c \gg m_s \sim \Lambda_{QCD}$ on each gauge background at finite lattice spacing from its analytic expression\,Eq.~(\ref{eq:der_sea_mf}). Actually an even stronger suppression, as $1/m_{c}^2$, is predicted in the same limit up to
lattice artifacts by perturbation theory. As for the critical mass corrections to $F_{\pi}$ and $M_{\pi}$, as detailed in Appendix~\ref{sec:masses}, we rely on the analytic formulae of Eq.~(\ref{eq:an_corr_mpcac_fpi_mpi}). In Figure~\ref{fig:sea_mcr_der_mpi_fpi}, we however compare on the cC211.060.80 ensemble, for which the difference $\Delta m_{\rm cr} = m_{\rm cr} - m_{\rm cr}^{\rm sim}$ is maximal, the results obtained using Eq.~(\ref{eq:an_corr_mpcac_fpi_mpi}) with that obtained by evaluating numerically the second term in Eq.~(\ref{eq:final_O}) with $O= M_{\pi}, F_{\pi}$. Specifically, in the figure we show the results for
\begin{align}
\label{eq:def_mpcac_der}
\partial_{m_{\rm PCAC}} O \equiv \frac{\partial_{\rm cr} O}{\partial_{\rm cr} m_{\rm PCAC}} \; , \qquad O=\left\{ M_{\pi}, F_{\pi} \right\}~,
\end{align}
where $\partial_{\rm cr} \equiv \partial_{\rm cr}^{\rm sea} + \partial_{\rm cr}^{\rm val}$ (see Eq.~(\ref{eq:final_O})). \footnote{The valence-quark mass derivative $\partial_{\rm cr}^{\rm val}$ contribution to both terms in the numerator and denominator of Eq.~(\ref{eq:def_mpcac_der}), which we estimated on the cB211.072.64 ensemble, turns out to be negligible within errors w.r.t. to the sea-quark contribution, and has been neglected.} As the figure shows, the result of the numerical estimates and that of the analytic formulae of Eq.~(\ref{eq:an_corr_mpcac_fpi_mpi}) are in fairly good agreement.
We now move to the case of $M_{K}$ and $M_{D_{s}}$. In Figure~\ref{fig:sea_der_mk_mds} we show the results for the light and strange sea-quark mass derivatives of the effective mass of the kaon and of the $D_{s}$ meson, as obtained on the cB211.072.64 ensemble.
All derivatives turn out to be very small and compatible with zero within a few standard deviations. For the kaon mass $M_{K}$, the typical magnitude of the derivatives are: $|\partial_{\ell} M_K | \sim 0.2 $,   $|\partial_{s} M_K | \sim 0.1 $.  For the $D_{s}$-meson they are: $|\partial_{\ell} M_{D_{s}} | \sim 0.5 $,   $| \partial_{s} M_{D_{s}} | \sim 0.3 $.  The largest mistunings that we have are $\Delta m_{\ell} \simeq 0.16~{\rm MeV}$, $\Delta m_{s} \simeq -1.7~{\rm MeV}$. The largest correction to the kaon mass (including those from the fine-tuning of the critical mass, and from charm sea-quark mass mistuning estimated employing Eq.~(\ref{eq:charm_der_est})) is of about $0.2~{\rm MeV}$, while it is of about $0.5~{\rm MeV}$ for $M_{D_{s}}$. In both cases, the corrections are smaller than our statistical uncertainties, hence completely negligible.

Finally, we discuss the mistuning corrections on $a_{\mu}^{\rm HVP}(s)$ and $a_{\mu}^{\rm HVP}(c)$. In this case, there are two different types of corrections. The first is an indirect one that it is caused by the feedback on the HVP due to a change of the lattice spacing (through which we define the physical time $t$ in the integrand of Eq.~(\ref{eq:amu_HVP})) and of the isoQCD quark masses $m_{f}^{\rm iso}$. These corrections can be applied from scratch by evaluating the HVP using the values of the lattice spacing $a^{\rm iso}$ and of the quark masses $m_{f}^{\rm iso}$ given in Table~\ref{tab:iso_EDI_FLAG}. The second source of corrections, which we discuss now, arises from the 
sea-quark mass and critical mass corrections to the vector correlators $V_{s}^{\rm{reg}}(t)$ and $V_{c}^{\rm{reg}}(t)$ of Eq.~(\ref{eq:VV_correlators}), and we define the following partial
derivatives of $a_{\mu}^{\rm HVP}(s)$ and $a_{\mu}^{\rm HVP}(c)$
w.r.t.\ to the $x$-flavour sea quark mass and the (sea plus valence)
critical quark mass at fixed lattice spacing $a=a^{\rm iso}$:
\begin{align}
\label{eq:sea_part_der}
\partial^{\rm sea-part}_{x}a_{\mu}^{\rm HVP}(f) &= 2\alpha_{\rm em}^{2} a^{3}\sum_{n=1}^{T/(2a)} w(n)\, n^{2} K(a m_{\mu} n)\, \partial_{x}^{\rm sea-part}~V_{f}^{\rm reg}(na)~,\qquad  f=s,c~,\qquad  x=\ell,s,c,  \nonumber \\[8pt]
\partial^{\rm part}_{\rm cr}a_{\mu}^{\rm HVP}(f) &= 2\alpha_{\rm em}^{2} a^{3}\sum_{n=1}^{T/(2a)} w(n)\, n^{2} K(a m_{\mu} n) \left[ \partial_{\rm cr}^{\rm sea} + \partial_{\rm cr}^{\rm val}\right]~V_{f}^{\rm reg}(na)~,\qquad  f=s,c~.
\end{align}
where $n=t/a =1,\ldots, T/(2a)$ is the Euclidean time in lattice units. %
The partial derivatives of the SD, W and LD contributions can be defined analogously.

\begin{figure}
\includegraphics[scale=0.28]{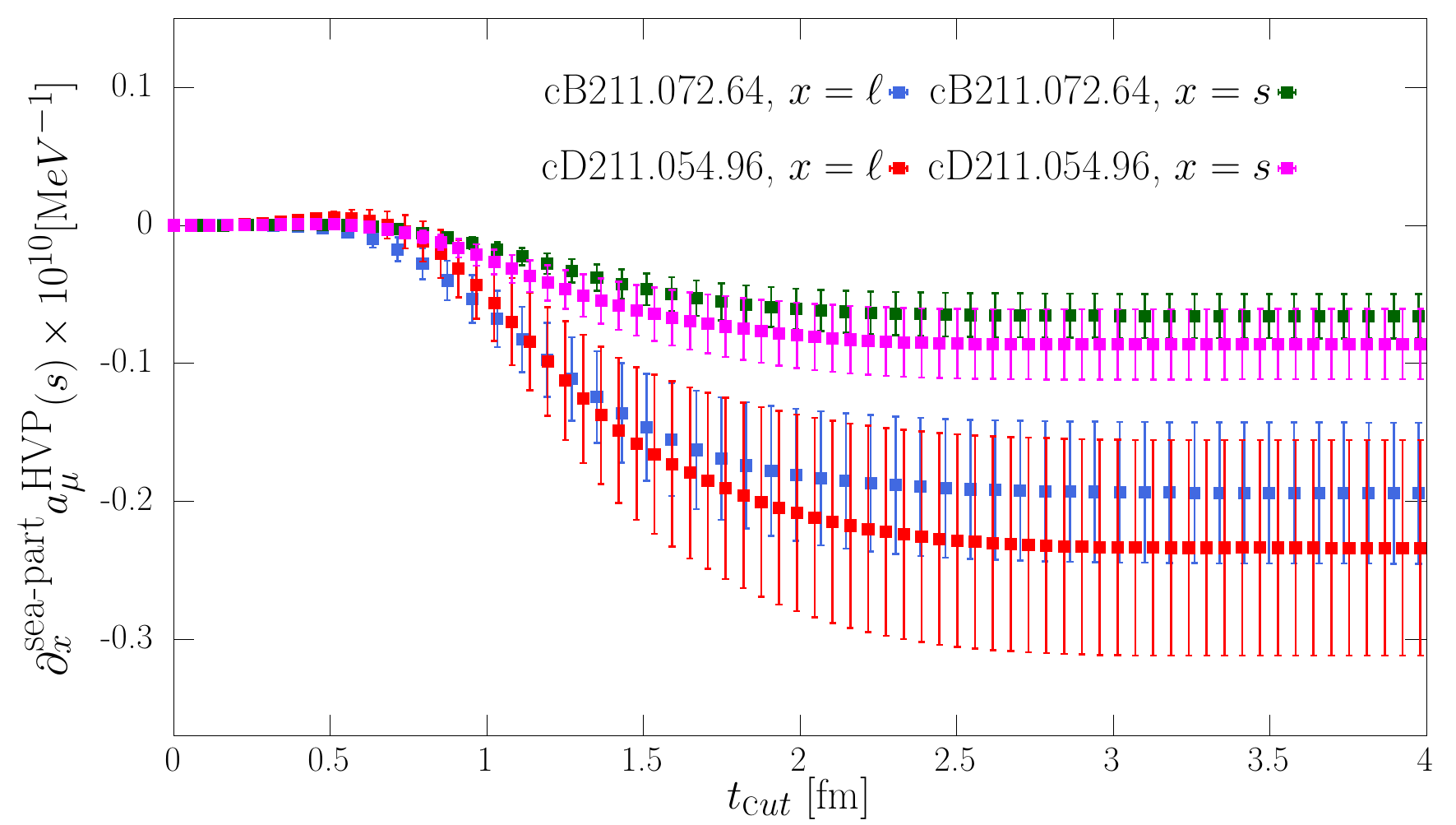}
\includegraphics[scale=0.28]{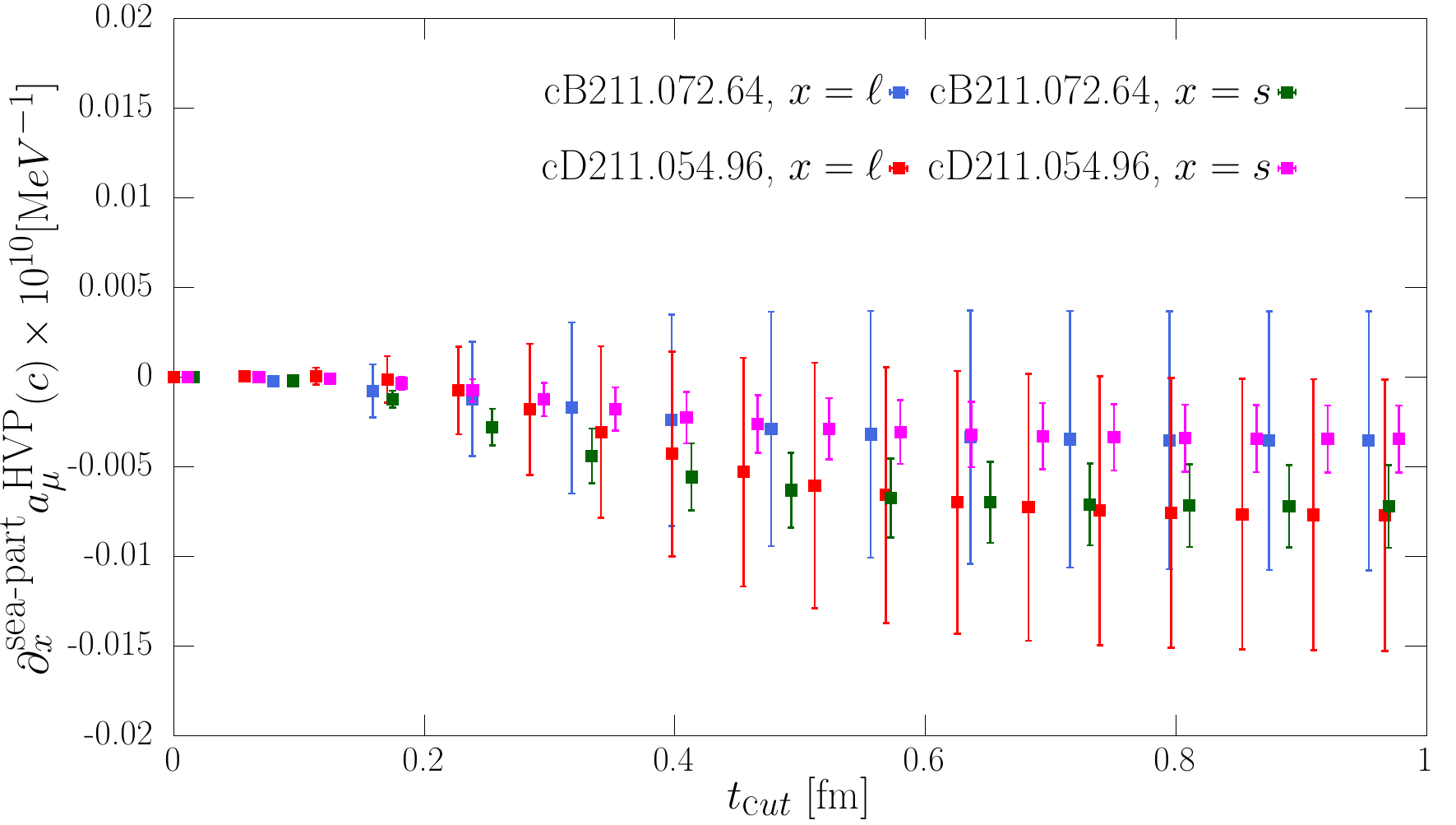}
\caption{\small\it The light and strange (partial) sea-quark mass derivatives of $a_{\mu}^{\rm HVP}(s)$ (left) and of $a_{\mu}^{\rm HVP}(c)$, as a function of the upper limit $t_{\rm cut}/a$ on the sum in Eq.~(\ref{eq:sea_part_der}), as obtained on the cB211.072.64 and cD211.054.96 ensembles. \label{fig:sea_part_der_amu}}
\end{figure}

In Figure~\ref{fig:sea_part_der_amu}, we present the results for the light and strange (partial) sea-quark mass derivatives of $a_{\mu}^{\rm HVP}(s)$ and $a_{\mu}^{\rm HVP}(c)$ as obtained on the cB211.072.64 and cD211.054.96 ensembles. The results correspond to $\rm{reg} = $TM, but within uncertainties 
no dependence of the derivatives on the regularization has been observed.
As the figure shows the derivatives are sensibly larger for $a_{\mu}^{\rm HVP}(s)$ than for $a_{\mu}^{\rm HVP}(c)$. The typical size of the light sea-quark mass derivative is for the strange HVP $|\partial^{\rm sea-part}_{\ell}\, a_{\mu}^{\rm HVP}(s)| \sim 0.2-0.3~\rm{MeV}^{-1}$ while it is of about $|\partial^{\rm sea-part}_{\ell}\, a_{\mu}^{\rm HVP}(c)| \sim 0-0.015~\rm{MeV}^{-1}$ for the charm HVP. Considering that the largest light sea-quark mistuning is $\Delta m_{\ell} \sim 0.16~{\rm MeV}$, the corresponding corrections to the strange HVP, $a_{\mu}^{\rm HVP}(s)$, turn out to be smaller than $0.1\%$ (although this is only slightly smaller than our statistical errors on $a_{\mu}^{\rm HVP}(s)$), and completely negligible for $a_{\mu}^{\rm HVP}(c)$. The corrections due to the fine-tuning of the critical mass are of a similar magnitude to those resulting from light sea-quark mass mistuning effects.

\begin{figure}
    \centering
    \includegraphics[width=0.48\linewidth]{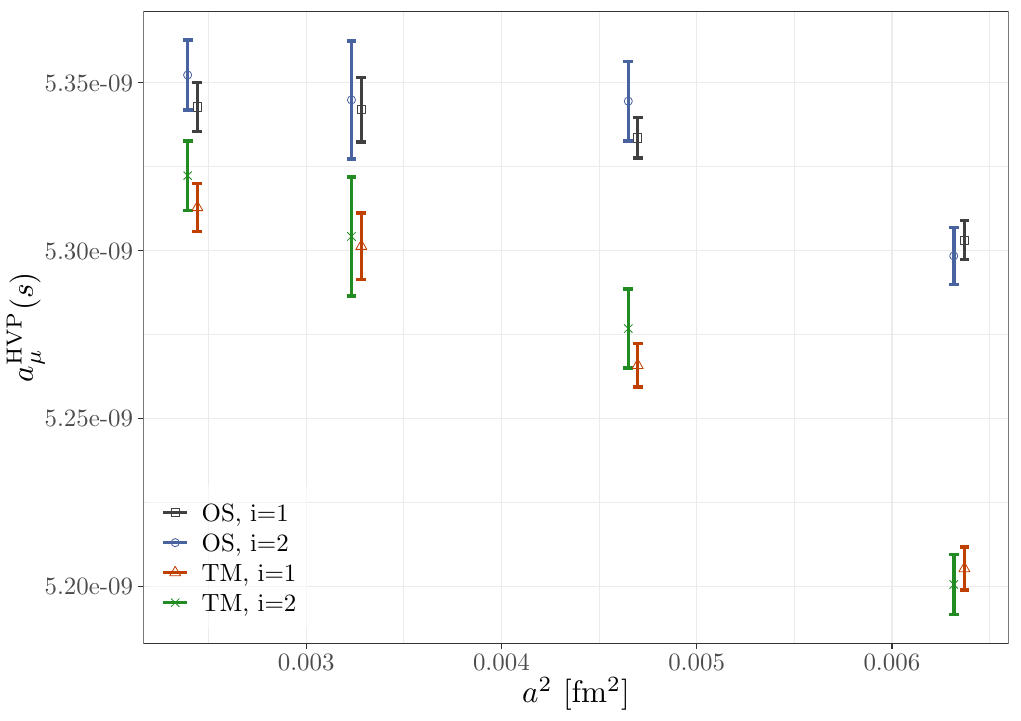}
    \includegraphics[width=0.48\linewidth]{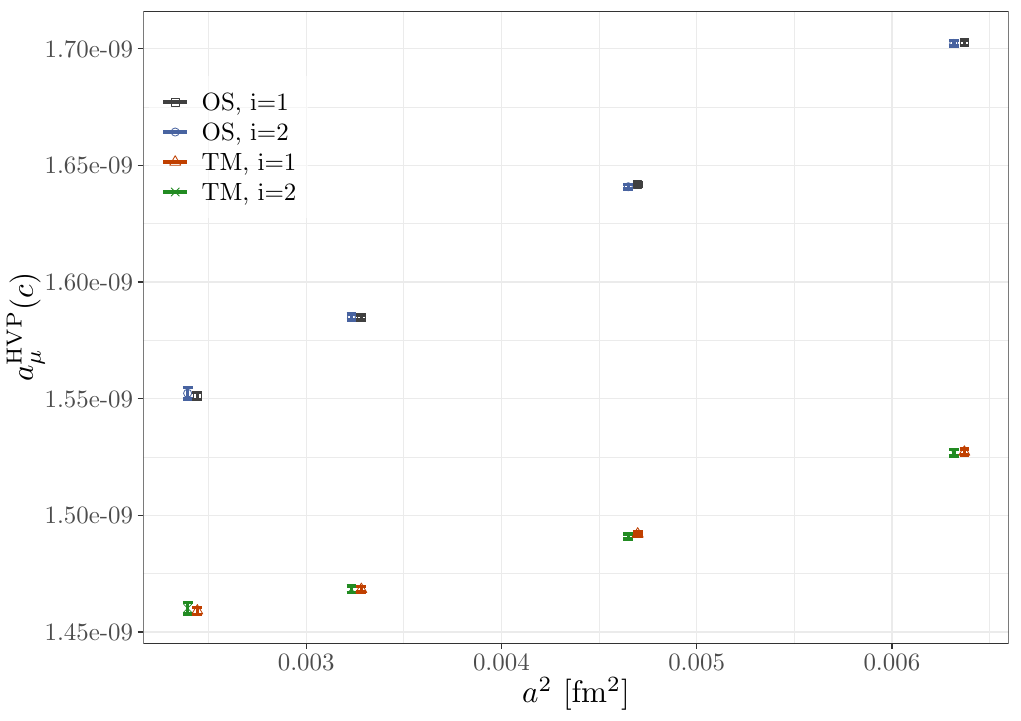}
    \caption{\small\it Comparison between the values of $a_{\mu}^{\rm HVP}(s)$ (left) and $a_{\mu}^{\rm HVP}(c)$ (right) obtained before (black for the OS and red for the TM regularization) and after (blue for the OS and green for the TM regularization) applying the corrections due to the mistuning of sea-quark masses and critical mass. }
    \label{fig:amu_tuning}
\end{figure}

As for the strange sea-quark mass derivatives, they are of order $|\partial^{\rm sea-part}_{s}\, a_{\mu}^{\rm HVP}(s)| \sim 0.05-0.1~{\rm MeV}^{-1}$ and $|\partial^{\rm sea-part}_{c}\, a_{\mu}^{\rm HVP}(s)| \sim 0.005-0.01~{\rm MeV}^{-1}$. The largest strange sea-quark mass mistuning is $\Delta m_{s} \sim -1.7~{\rm MeV}$. For the charm HVP, $a_{\mu}^{\rm HVP}(c)$, the size of these corrections is still below (on some ensembles however comparable to) our statistical uncertainties. For the strange HVP, \(a_{\mu}^{\rm HVP}(s)\), the contribution is significantly larger—on some ensembles, up to twice the size of our statistical uncertainties in \(a_{\mu}^{\rm HVP}(s)\). After accounting for the charm sea-quark mass mistuning effect, estimated using Eq.~(\ref{eq:charm_der_est}), the errors on \(a_{\mu}^{\rm HVP}(s)\) (as well as on \(a_{\mu}^{\rm HVP, W}\)) increased. This almost completely offset the improvements in the intermediate strange window resulting from the addition of a new lattice spacing ensemble (cE211.044.112) and the larger statistics used in this calculation compared to our previous results in Ref.~\cite{ExtendedTwistedMass:2022jpw}. In Figure~\ref{fig:amu_tuning} we show, for each $\beta$, the comparison between the values of $a_{\mu}^{\rm HVP}(s)$ and $a_{\mu}^{\rm HVP}(c)$ obtained before and after applying the corrections due to the mistuning of sea-quark masses and critical mass.

We conclude this section with a comment on the corrections to the scale-invariant RCs $Z_{V}$ and $Z_{A}$ (which enter the determination of the strange and charm HVP), due to sea-quark mass and critical mass mistuning effects. For these two quantities the corrections turned out to be extremely tiny, reaching at most, on the coarser ensembles, the level of $0.005$-$0.01\%$. For $Z_{V}$, which has an astonishing precision $< 0.002\%$, the correction is however larger than the statistical uncertainty. The central values and errors of $Z_{V}$ and $Z_{A}$, that we provide in Appendix~\ref{sec:renormalization}, are inclusive of the corrections due to mistuning effects.

\section{Hadronic determination of \texorpdfstring{$Z_{V}$}{ZV} and \texorpdfstring{$Z_{A}$}{ZA}}
\label{sec:renormalization}

In order to reach a high precision determination of the two scale-invariant RCs $Z_{V}$ and $Z_{A}$ we employ the hadronic method, already adopted in Ref.\,\cite{ExtendedTwistedMass:2022jpw}, based on the Ward identity (WI) and universality of renormalized correlation functions, combined with a high statistics determination of the relevant bare correlators. This allows us to obtain on the ensembles of Table\,\ref{tab:iso_EDI_FLAG}, after correcting for sea-quark mass and critical mass mistuning effects (see Appendix~\ref{sec:mistunings}), an accuracy of $\simeq 0.03\%$ for $Z_A$ and of $\simeq 0.01 \%$ for $Z_V$, thus reaching the desired accuracy. We collect in Table\,\ref{tab:RCs} the values of $Z_{A}$ and $Z_{V}$ used in this work for each of the ETMC ensembles of Table\,\ref{tab:iso_EDI_FLAG}.

\begin{table}[t!]
\centering
\begin{tabular}{ lcc }
ensemble & $Z_{V}$ & $Z_{A}$ \\[4pt]
\colrule
\\
\textrm{B64} & ~ $0.706354(54)$  ~  & ~ $0.74296(19)$  ~  \\
\textrm{B96} &  ~  $0.706406(52)$  ~  &  ~  $0.74261(19)$   ~  \\
\textrm{C80} &  ~  $0.725440(33)$  ~  &  ~  $0.75814(13)$  ~ \\
\textrm{C112} &  ~  $0.725458(31)$  ~  &  ~  $0.75824(15)$  ~ \\
\textrm{D96} & ~  $0.744132(31)$  ~  &  ~  $0.77367(10)$  ~   \\
\textrm{E112} &   ~  $0.758238(18)$  ~  &  ~  $0.78548(9)$  ~  \\[8pt]
\hline
\end{tabular}
 \caption{\it \small The values of $Z_{V}$ and $Z_{A}$ used in this work for each of the ETMC ensembles of Table\,\ref{tab:iso_EDI_FLAG}, determined by employing the WI-based hadronic method described in Appendix B of Ref.\,\cite{ExtendedTwistedMass:2022jpw}.}
    \label{tab:RCs} 

\end{table}

\bibliography{biblio}
\bibliographystyle{JHEP}

\end{document}